\documentclass[twocolumn,preprintnumbers,amsmath,amssymb,superscriptaddress]{revtex4}
\usepackage{mathrsfs}
\usepackage{slashbox}
\usepackage{adjustbox}
\usepackage{amsfonts}
\usepackage{revsymb4-1}
\usepackage{graphicx}
\usepackage[dvipsnames]{xcolor}
\usepackage{booktabs}
\usepackage{cancel}
\usepackage{soul}
\usepackage{tikz}
\usepackage{hyperref}
\usepackage{newtxtext,newtxmath,amsmath}
\hypersetup{
 citecolor=red, linkcolor=blue, filecolor=blue, urlcolor=cyan,linktocpage  = true, colorlinks=true}

\usetikzlibrary{calc}

\graphicspath{{figs/}}
\begin{document}

\newcommand{\ba}{\begin{array}}
\newcommand{\ea}{\end{array}}

\newcommand{\req}[1]{Eq.~(\ref{#1})}
\newcommand{\res}[1]{Section~\ref{#1}}
\newcommand{\reapp}[1]{Appendix~\ref{#1}}
\newcommand{\rep}[1]{\cite{#1}}
\newcommand{\refig}[1]{Fig.~\ref{#1}}
\newcommand{\ret}[1]{Table~\ref{#1}}

\newcommand{\dif}{{\rm d}}
\newcommand{\Tr}{{\rm Tr}}
\newcommand{\tr}{{\rm tr}}
\newcommand{\Pf}{{\rm Pf}}
\newcommand{\Id}{\mathbf{1}}
\newcommand{\pexp}{\mbox{P} \, \mbox{exp}}
\newcommand{\Per}[1]{\mathbf{S}_{#1}}
\newcommand{\eps}{\varepsilon}
\newcommand{\ci}[1]{\boldsymbol{#1}}
\renewcommand{\Re}[1]{{\mathcal Re}\left\{#1\right\}}
\renewcommand{\Im}[1]{{\mathcal Im}\left\{#1\right\}}
\newcommand{\Dslash}{\relax{\kern+.25em / \kern-.70em D}}

\newcommand{\dl}{\stackrel{\leftarrow}{D}}
\newcommand{\dr}{\stackrel{\rightarrow}{D}}
\newcommand{\dcl}{\stackrel{\leftarrow}{\cD}}
\newcommand{\dcr}{\stackrel{\rightarrow}{\cD}}
\newcommand{\pdl}{\stackrel{\leftarrow}{\partial}}
\newcommand{\pdr}{\stackrel{\rightarrow}{\partial}}

\newcommand{\Dwd}{D_{\rm\scriptstyle w}}
\newcommand{\Dtwd}{D_{\rm\scriptstyle tw}}
\newcommand{\arcsinh}{\rm arcsinh}
\newcommand{\arccosh}{\rm arccosh}
\newcommand{\arctanh}{\rm arctanh}
\newcommand{\fm}{{\rm fm}}
\newcommand{\MeV}{{\rm MeV}}
\newcommand{\GeV}{{\rm GeV}}

\newcommand{\Real}{\relax{\mathsf{\Gamma\kern-.35em R}}}
\newcommand{\Int}{\relax{\mathsf{Z\kern-.40em Z}}}
\newcommand{\Ztwo}{Z_2}
\newcommand{\Zf}{Z_4}
\newcommand{\ZN}{Z_{\NC}}
\newcommand{\ZTN}{Z_{2\NC}}
\newcommand{\UO}{\mbox{U}(1)}
\newcommand{\SUT}{\mbox{SU}(2)}
\newcommand{\UT}{\mbox{U}(2)}
\newcommand{\SUt}{\mbox{SU}(3)}
\newcommand{\Ut}{\mbox{U}(3)}
\newcommand{\SUN}{\mbox{SU}(\NC)}
\newcommand{\UN}{\mbox{U}(\NC)}
\newcommand{\SUQ}{\mbox{SU}(Q)}
\newcommand{\UQ}{\mbox{U}(Q)}
\newcommand{\SU}{\mbox{SU}}
\newcommand{\CF}{C_{\rm F}}

\newcommand{\half}{{\scriptstyle{{1\over 2}}}}
\newcommand{\third}{{\scriptstyle{{1\over 3}}}}
\newcommand{\quart}{{\scriptstyle{{1\over 4}}}}
\newcommand{\sixth}{{\scriptstyle{{1\over 6}}}}
\newcommand{\eighth}{{\scriptstyle{{1\over 8}}}}
\newcommand{\sixteenth}{{\scriptstyle{{1\over 16}}}}
\newcommand{\twth}{{\scriptstyle{{1\over 12}}}}
\newcommand{\thalf}{{\scriptstyle{{3\over 2}}}}
\newcommand{\fhalf}{{\scriptstyle{{5\over 2}}}}
\newcommand{\tthird}{{\scriptstyle{{2\over 3}}}}
\newcommand{\fthird}{{\scriptstyle{{5\over 3}}}}
\newcommand{\ihalf}{{\scriptstyle{{i\over 2}}}}
\newcommand{\iquart}{{\scriptstyle{{i\over 4}}}}
\newcommand{\oneovN}{{\scriptstyle{{1\over \NC}}}}

\newcommand{\NC}{N}
\newcommand{\NF}{N_\mathrm{\scriptstyle f}}
\newcommand{\NSS}{{\cal N}}

\newcommand{\MS}{{\rm MS}}
\newcommand{\MSbar}{{\overline{\rm MS}}}
\newcommand{\SF}{{\rm SF}}
\newcommand{\GF}{{\rm GF}}

\newcommand{\gbar}{\kern1pt\overline{\kern-1pt g\kern-0pt}\kern1pt}
\newcommand{\mbar}{\kern2pt\overline{\kern-1pt m\kern-1pt}\kern1pt}
\newcommand{\obar}[1]{\kern3pt\overline{\kern-2pt #1\kern-0pt}\kern1pt}
\newcommand{\gren}{g_{\rm R}}
\newcommand{\mren}[1]{m_{{\rm R} #1}}
\newcommand{\muren}[1]{\mu_{{\rm R} #1}}
\newcommand{\oren}[1]{#1_{\rm R}}
\newcommand{\lQCD}{\Lambda_{\rm\scriptscriptstyle QCD}}
\newcommand{\MW}{M_{\rm\scriptscriptstyle W}}
\newcommand{\mrgi}{M}
\newcommand{\orgi}[1]{\hat #1}
\newcommand{\msubt}{m_{\rm q}}
\newcommand{\mcrit}{m_{\rm cr}}
\newcommand{\hopc}{\kappa_{\rm c}}
\newcommand{\muq}{\mu_{\rm q}}

\newcommand{\fX}{f_{\rm\scriptscriptstyle X}}
\newcommand{\fP}{f_{\rm\scriptscriptstyle P}}
\newcommand{\fS}{f_{\rm\scriptscriptstyle S}}
\newcommand{\fA}{f_{\rm\scriptscriptstyle A}}
\newcommand{\fV}{f_{\rm\scriptscriptstyle V}}
\newcommand{\kX}{k_{\rm\scriptscriptstyle X}}
\newcommand{\kA}{k_{\rm\scriptscriptstyle A}}
\newcommand{\kV}{k_{\rm\scriptscriptstyle V}}
\newcommand{\kT}{k_{\rm\scriptscriptstyle T}}
\newcommand{\kTt}{k_{\rm\scriptscriptstyle \widetilde T}}
\newcommand{\hT}{h_{\rm\scriptscriptstyle T}}
\newcommand{\gX}{g_{\rm\scriptscriptstyle X}}
\newcommand{\gP}{g_{\rm\scriptscriptstyle P}}
\newcommand{\gS}{g_{\rm\scriptscriptstyle S}}
\newcommand{\gA}{g_{\rm\scriptscriptstyle A}}
\newcommand{\gV}{g_{\rm\scriptscriptstyle V}}
\newcommand{\kVt}{k_{\rm\scriptscriptstyle \widetilde V}}
\newcommand{\gVt}{g_{\rm\scriptscriptstyle \widetilde V}}
\newcommand{\lVt}{l_{\rm\scriptscriptstyle \widetilde V}}
\newcommand{\lX}{l_{\rm\scriptscriptstyle X}}
\newcommand{\lA}{l_{\rm\scriptscriptstyle A}}
\newcommand{\lV}{l_{\rm\scriptscriptstyle V}}
\newcommand{\lT}{l_{\rm\scriptscriptstyle T}}
\newcommand{\lTt}{l_{\rm\scriptscriptstyle\widetilde{T}}}

\newcommand{\Zm}{Z_{\rm m}}
\newcommand{\ZP}{Z_{\rm\scriptscriptstyle P}}
\newcommand{\ZS}{Z_{\rm\scriptscriptstyle S}}
\newcommand{\ZA}{Z_{\rm\scriptscriptstyle A}}
\newcommand{\ZV}{Z_{\rm\scriptscriptstyle V}}
\newcommand{\ZVt}{Z_{\rm\scriptscriptstyle \widetilde V}}
\newcommand{\ZT}{Z_{\rm\scriptscriptstyle T}}
\newcommand{\zf}{z_{\rm f}}
\newcommand{\ZPSF}{Z_{\rm\scriptscriptstyle P}^{\rm\scriptscriptstyle SF}}
\newcommand{\ZPchiSF}{Z_{\rm\scriptscriptstyle P}^{\rm\scriptscriptstyle \chi SF}}
\newcommand{\ZSchiSF}{Z_{\rm\scriptscriptstyle S}^{\rm\scriptscriptstyle \chi SF}}
\newcommand{\ZTSF}{Z_{\rm\scriptscriptstyle T}^{\rm\scriptscriptstyle SF}}
\newcommand{\ZTSFf}{Z_{\rm\scriptscriptstyle T}^{\rm\scriptscriptstyle SF-f}}
\newcommand{\ZTSFk}{Z_{\rm\scriptscriptstyle T}^{\rm\scriptscriptstyle SF-k}}
\newcommand{\ZTchiSFud}{Z_{\rm\scriptscriptstyle T}^{{\rm\scriptscriptstyle \chi SF},ud}}
\newcommand{\ZTchiSFuup}{Z_{\rm\scriptscriptstyle T}^{{\rm\scriptscriptstyle \chi SF},uu^\prime}}
\newcommand{\sigmaP}{\sigma_{\rm\scriptscriptstyle P}}
\newcommand{\sigmaS}{\sigma_{\rm\scriptscriptstyle S}}
\newcommand{\sigmaT}{\sigma_{\rm\scriptscriptstyle T}}
\newcommand{\SigmaP}{\Sigma_{\rm\scriptscriptstyle P}}
\newcommand{\SigmaS}{\Sigma_{\rm\scriptscriptstyle S}}
\newcommand{\SigmaPS}{\Sigma_{\rm\scriptscriptstyle P/S}}
\newcommand{\SigmaSP}{\Sigma_{\rm\scriptscriptstyle S/P}}
\newcommand{\SigmaT}{\Sigma_{\rm\scriptscriptstyle T}}
\newcommand{\SigmaPSF}{\Sigma_{\rm\scriptscriptstyle P}^{\rm\scriptscriptstyle SF}}
\newcommand{\SigmaPchiSF}{\Sigma_{\rm\scriptscriptstyle P}^{\rm\scriptscriptstyle \chi SF}}
\newcommand{\SigmaSchiSF}{\Sigma_{\rm\scriptscriptstyle S}^{\rm\scriptscriptstyle \chi SF}}
\newcommand{\SigmaTSF}{\Sigma_{\rm\scriptscriptstyle T}^{\rm\scriptscriptstyle SF}}
\newcommand{\SigmaTchiSFud}{\Sigma_{\rm\scriptscriptstyle T}^{{\rm\scriptscriptstyle \chi SF},ud}}
\newcommand{\SigmaTchiSFuup}{\Sigma_{\rm\scriptscriptstyle T}^{{\rm\scriptscriptstyle \chi SF},uu^\prime}}

\newcommand{\gammaP}{\gamma_{\rm\scriptscriptstyle P}}
\newcommand{\gammaT}{\gamma_{\rm\scriptscriptstyle T}}
\newcommand{\deltaT}{\delta_{\rm\scriptscriptstyle T}}
\newcommand{\rhoT}{\rho_{\rm\scriptscriptstyle T}}

\newcommand{\lmax}{L_{\rm max}}
\newcommand{\lmin}{L_{\rm min}}

\newcommand{\Oa}{\mbox{O}(a)}
\newcommand{\Oasq}{\mbox{O}(a^2)}
\newcommand{\Ogsqa}{\mbox{O}(g_0^2 a)}
\newcommand{\Ogqa}{\mbox{O}(g_0^4 a)}
\newcommand{\icsw}{c_{\rm sw}}
\newcommand{\ict}{c_{\rm t}}
\newcommand{\icttil}{\tilde c_{\rm t}}
\newcommand{\icP}{c_{\rm\scriptscriptstyle P}}
\newcommand{\icA}{c_{\rm\scriptscriptstyle A}}
\newcommand{\icT}{c_{\rm\scriptscriptstyle T}}
\newcommand{\icV}{c_{\rm\scriptscriptstyle V}}
\newcommand{\ibP}{b_{\rm\scriptscriptstyle P}}
\newcommand{\ibA}{b_{\rm\scriptscriptstyle A}}
\newcommand{\ibT}{b_{\rm\scriptscriptstyle T}}
\newcommand{\ibV}{b_{\rm\scriptscriptstyle V}}
\newcommand{\ibm}{b_{\rm m}}
\newcommand{\ibmtil}{\tilde{b}_{\rm m}}
\newcommand{\ibmu}{b_\mu}
\newcommand{\uSF}{u_{\rm\scriptscriptstyle SF}}
\newcommand{\uGF}{u_{\rm\scriptscriptstyle GF}}

\newcommand{\abar}{\kern1pt\overline{\kern-1pt a\kern-0.5pt}\kern1pt}
\newcommand{\abarhalf}{{\scriptstyle \frac{\abar}{2}}}
\newcommand{\psil}{\psi_{\mbox{\tiny L}}}

\newcommand{\bA}{{\mathbf{A}}}
\newcommand{\bB}{{\mathbf{B}}}
\newcommand{\bC}{{\mathbf{C}}}
\newcommand{\bD}{{\mathbf{D}}}
\newcommand{\bE}{{\mathbf{E}}}
\newcommand{\bF}{{\mathbf{F}}}
\newcommand{\bG}{{\mathbf{G}}}
\newcommand{\bH}{{\mathbf{H}}}
\newcommand{\bI}{{\mathbf{I}}}
\newcommand{\bJ}{{\mathbf{J}}}
\newcommand{\bK}{{\mathbf{K}}}
\newcommand{\bL}{{\mathbf{L}}}
\newcommand{\bM}{{\mathbf{M}}}
\newcommand{\bN}{{\mathbf{N}}}
\newcommand{\bO}{{\mathbf{O}}}
\newcommand{\bP}{{\mathbf{P}}}
\newcommand{\bQ}{{\mathbf{Q}}}
\newcommand{\bR}{{\mathbf{R}}}
\newcommand{\bS}{{\mathbf{S}}}
\newcommand{\bT}{{\mathbf{T}}}
\newcommand{\bU}{{\mathbf{U}}}
\newcommand{\bV}{{\mathbf{V}}}
\newcommand{\bW}{{\mathbf{W}}}
\newcommand{\bX}{{\mathbf{X}}}
\newcommand{\bY}{{\mathbf{Y}}}
\newcommand{\bZ}{{\mathbf{Z}}}
\newcommand{\cA}{{\cal A}}
\newcommand{\cB}{{\cal B}}
\newcommand{\cC}{{\cal C}}
\newcommand{\cD}{{\cal D}}
\newcommand{\cE}{{\cal E}}
\newcommand{\cF}{{\cal F}}
\newcommand{\cG}{{\cal G}}
\newcommand{\cH}{{\cal H}}
\newcommand{\cI}{{\cal I}}
\newcommand{\cJ}{{\cal J}}
\newcommand{\cK}{{\cal K}}
\newcommand{\cL}{{\cal L}}
\newcommand{\cM}{{\cal M}}
\newcommand{\cN}{{\cal N}}
\newcommand{\cO}{{\cal O}}
\newcommand{\cP}{{\cal P}}
\newcommand{\cQ}{{\cal Q}}
\newcommand{\cR}{{\cal R}}
\newcommand{\cS}{{\cal S}}
\newcommand{\cT}{{\cal T}}
\newcommand{\cU}{{\cal U}}
\newcommand{\cV}{{\cal V}}
\newcommand{\cW}{{\cal W}}
\newcommand{\cX}{{\cal X}}
\newcommand{\cY}{{\cal Y}}
\newcommand{\cZ}{{\cal Z}}

\renewcommand{\vr}{\mathbf{r}}
\newcommand{\vx}{\mathbf{x}}
\newcommand{\vy}{\mathbf{y}}
\newcommand{\vz}{\mathbf{z}}
\newcommand{\vt}{\mathbf{t}}
\newcommand{\vp}{\mathbf{p}}
\newcommand{\vq}{\mathbf{q}}
\newcommand{\vk}{\mathbf{k}}
\newcommand{\vl}{\mathbf{l}}
\newcommand{\vn}{\mathbf{n}}
\newcommand{\vu}{\mathbf{u}}
\newcommand{\vw}{\mathbf{w}}
\newcommand{\vzero}{\mathbf{0}}
\newcommand{\ring}[1]{\mathaccent"7017 #1}

\title{
\begin{flushright}
{\rm CERN-TH-2023-210}
\end{flushright}
Nonperturbative running of the tensor operator for $\NF=3$ QCD
       from the chirally rotated Schr\"odinger Functional}
 
\author{\begin{center}
     \includegraphics[height=2.0\baselineskip]{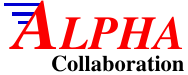}
       \end{center} Isabel~Campos~Plasencia}
\affiliation{Instituto de F\'isica de Cantabria IFCA-CSIC, Avda. de los Castros s/n, 39005 Santander, Spain}
\author{Mattia~Dalla~Brida}
\affiliation{Theoretical Physics Department, CERN, CH-1211 Geneva 23, Switzerland}
\author{Giulia~Maria~de~Divitiis}
\affiliation{Dipartimento di Fisica, Universit\`a di Roma ``Tor Vergata'', Via della Ricerca Scientifica 1, 00133 Rome, Italy}
\affiliation{INFN, Sezione di Tor Vergata, Via della Ricerca Scientifica 1, 00133 Rome, Italy}
\author{Andrew~Lytle}
\affiliation{Department of Physics, 
        University of Illinois at Urbana-Champaign,
        Urbana, Illinois, 61801, USA}
\author{Mauro~Papinutto}
\affiliation{Dipartimento di Fisica,  Universit\`a di Roma La Sapienza, and INFN, Sezione di Roma, Piazzale A.~Moro 2, 00185 Rome, Italy}
\author{Ludovica~Pirelli}
\affiliation{Dipartimento di Fisica, Universit\`a di Roma ``Tor Vergata'', Via della Ricerca Scientifica 1, 00133 Rome, Italy}
\affiliation{INFN, Sezione di Tor Vergata, Via della Ricerca Scientifica 1, 00133 Rome, Italy}
\author{Anastassios~Vladikas}
\affiliation{INFN, Sezione di Tor Vergata, Via della Ricerca Scientifica 1, 00133 Rome, Italy}
\date{\today}

\begin{abstract}
We study the Renormalisation Group (RG) running of the non-singlet
tensor operator,
for $\NF=3$ QCD with Wilson fermions in a mixed action setup,
with standard Schr\"odinger Functional (SF) boundary conditions 
for sea quarks and  chirally rotated Schr\"odinger Functional 
($\chi$SF) boundary conditions for valence quarks.
Based on a recursive finite-size scaling technique we compute 
non-perturbatively the tensor step-scaling function for an energy range between
a hadronic scale and an electroweak scale, above which perturbation
theory may be safely applied. Our result is expressed as the RG-running factor
$T^{\mathrm{RGI}}/[ T(\mu_{\mathrm{had}})]_{\scriptstyle \rm R}$,
where the numerator is the scale independent (Renormalisation Group Invariant - RGI) 
tensor operator and 
the denominator is its renormalised counterpart at a hadronic scale 
$\mu_{\mathrm{had}} = 233(8)$~MeV in a given scheme.
We determine the step-scaling function in four distinct renormalisation schemes.
We also compute the renormalisation parameters of these schemes at
$\mu_{\mathrm{had}}$ which, combined with the RG-running factor,
gives the scheme-independent quantity $Z^{\mathrm{RGI}}_{\mathrm T}(g_0^2)$ 
in four schemes and for a range of bare gauge couplings in which large volume
hadronic matrix element simulations are performed by the CLS consortium in
$\NF=2+1$ QCD.
All four results are compatible and also agree with a recent determination
based on a unitary setup for Wilson quarks 
with Schr\"odinger Functional boundary conditions~\cite{Chimirri:2023ovl}.
This provides a strong universality test. 
\\ \end{abstract}
\maketitle

\section{Introduction} 
\label{sec:intro}

The hadronic matrix elements of the non-singlet tensor bilinear quark operator
are an essential ingredient to several processes, such as
rare heavy meson and $\beta$-decays and quantities like
the neutron electric dipole moment; see 
e.g. refs.~\cite{Buchalla:2008jp,Antonelli:2009ws,Blake:2016olu,Bhattacharya:2015esa,Bhattacharya:2016zcn,Abramczyk:2017oxr}.  

In the present work we concentrate on the non-perturbative renormalisation and renormalisation group (RG)
running of the tensor composite field. As in our previous publication on the
RG-running of the pseudoscalar density~\cite{Plasencia:2021eon}, we follow the ALPHA finite size scaling approach, applied in $\NF=3$ massless QCD. We compute the step scaling function (SSF) of the tensor composite field for renormalisation scales $\mu$ which cover a wide energy range 
$\lQCD \lessapprox \mu \equiv 1/L \lessapprox \MW$. Here $L$ denotes the 
lattice physical extension; note that it ranges from very small values of about $L \approx 10^{-3}$fm to $L \lessapprox 1$fm.
We use the gauge ensembles of ref.~\cite{Campos:2018ahf},
generated for the nonperturbative determination of the 
RG-running of the quark mass in $\NF=3$ massless QCD with SF boundaries. On these ensembles we compute correlation functions with the dimension-3 tensor 
composite field in the bulk and chirally rotated boundary conditions~\cite{Sint:2010eh}. This is known as the $\chi$SF renormalisation scheme.
So we have a mixed action approach, as sea and valence quarks have different regularisations
(SF and $\chi$SF respectively). 

Our setup is similar to the traditional SF scheme used in ref.~\cite{Pena:2017hct} (for $\NF=0,2$) and more recently in ref.~\cite{Chimirri:2023ovl} (for $\NF=3$), but has certain advantages related to the property of ``automatic" improvement of $\chi$SF, 
as discussed in Sect.~\ref{sec:theory}. Our main result consists of estimates of the ratio $T^{\mathrm{RGI}}/[ T(\mu_{\mathrm{had}})]_{\scriptstyle \rm R}$, where $T^{\mathrm{RGI}}$ is the scale independent (Renormalisation Group Invariant - RGI) tensor operator and $[ T(\mu_{\mathrm{had}})]_{\scriptstyle \rm R}$ is its renormalised counterpart, computed at a hadronic scale 
$\mu_{\mathrm{had}} = 233(8)$~MeV in a given scheme. We have obtained four such estimates for four renormalisation schemes of the $\chi$SF variety. A second result, again obtained for the four schemes in question, is the tensor renormalisation parameter $\ZT(g_0^2,a\mu_{\mathrm{had}})$, computed in the range of bare gauge couplings where CLS simulations of low-energy hadronic matrix elements in $\NF = 2+1$ QCD are performed~\cite{Bruno:2014jqa,Bruno:2016plf,Bali:2016umi,Mohler:2017wnb,RQCD:2022xux}.

This paper is organised as follows: Sect.~\ref{sec:theory} is an overview of the fundamental properties
of the $\chi$SF lattice regularisation with Wilson quarks and the basic definitions and properties of the 
quantities of interest. Sect.~\ref{sec:simulation} describes the details of our numerical simulations.
In this section we also show results for the ratio of correlation functions with different tensor components
in the bulk which goes to unity in the continuum limit. This is in agreement with theoretical expectations,
confirming restoration of chiral symmetry in the continuum limit of the $\chi$SF setup.
For reasons explained in Sects.~\ref{sec:theory} and \ref{sec:simulation}, computations are split in two energy ranges,
a high one $\lQCD \lessapprox \mu \lessapprox \mu_0/2$ (with $\mu_0/2 \sim 2$~GeV) and a low one
$\mu_0/2 \lessapprox \mu \lessapprox 100$~GeV. In Sect.~\ref{sec:results-he} we collect our results of the high-energy 
RG running factor $T^{\mathrm{RGI}}/[ T(\mu_0/2)]_{\scriptstyle \rm R}$. 
In Sect.~\ref{sec:results-le} we collect our results of the low-energy one
$[ T(\mu_0/2)]_{\scriptstyle \rm R}/[ T(\mu_{\mathrm{had}})]_{\scriptstyle \rm R}$.
These are combined in Sect.~\ref{sec:results-final}, in order to obtain the total running factor
$T^{\mathrm{RGI}}/[ T(\mu_{\mathrm{had}})]_{\scriptstyle \rm R}$. 
Finally, in Sect.~\ref{sec:results-final} we compute the renormalisation parameter $\ZT(g_0^2,a\mu_{\mathrm{had}})$
for a range of bare gauge couplings.
Several details of our analysis are treated
separately in the Appendices. In Appendix~\ref{app:comparison} we compare our results to those
of ref.~\cite{Chimirri:2023ovl}.

Our work has been presented in preliminary form in refs.~\cite{deDivitiis:2021ugi,CamposPlasencia:2022epo}.
For the determination of the renormalisation parameter of the tensor in the RI'/(S)MOM scheme in three-flavour QCD, see ref.~\cite{RQCD:2020kuu}.

\section{Theoretical Considerations}
\label{sec:theory}
We cover briefly those aspects of the $\chi$SF regularisation of refs.~\cite{Sint:2010eh,Brida:2016rmy} which are most relevant to our work. At the formal level, the massless QCD action is invariant under general flavour and chiral transformations.
In particular, it is invariant under the change of variables,
\begin{equation}
\label{eq:ferm-rots}
\psi = R(\pi/2) \, \psi^\prime \,\, , \qquad  \bar \psi = \bar \psi^\prime \,  R(\pi/2) \,\, ,
\end{equation}
where $\psi, \bar \psi$ and $\psi^\prime, \bar \psi^\prime$ are doublets in isospin space
(e.g.\ $\psi = (\psi_u \,\, \psi_d)^T$), related through the above chiral non-singlet transformations 
with $R(\alpha) = \exp(i \alpha \gamma_5 \tau^3/2)$.
In the SF-QCD setup, lattices have finite physical volume $L^3 \times T$ (in the present work $T=L$),
with fields obeying Dirichlet (periodic) boundary conditions in time (space).
The former are defined at $x_0 = 0$ and $x_0 = T$ as follows:
\begin{eqnarray}
P_+ \psi(x) \big \vert_{x_0 = 0} = 0 \,\, , & \qquad & P_- \psi(x) \big \vert_{x_0 = T} = 0 \,\, ,
\nonumber \\
\bar \psi(x) P_-  \big \vert_{x_0 = 0} = 0 \,\, , & \qquad & \bar  \psi(x) P_+ \big \vert_{x_0 = T} = 0 \,\, ,
\end{eqnarray}
with projectors $P_\pm = (1 \pm \gamma_0)/2$. The chiral rotations (\ref{eq:ferm-rots}) map the above
conditions onto the $\chi$SF boundary conditions
\begin{eqnarray}
\tilde Q_+ \psi^\prime(x) \big \vert_{x_0 = 0} = 0 \,\, , & \qquad & \tilde Q_- \psi^\prime(x) \big \vert_{x_0 = T} = 0 \,\, ,
\nonumber \\
\bar \psi^\prime(x) \tilde Q_+  \big \vert_{x_0 = 0} = 0 \,\, , & \qquad & \bar  \psi^\prime(x) \tilde Q_- \big \vert_{x_0 = T} = 0 \,\, ,
\end{eqnarray}
with projectors $\tilde Q_\pm = (1 \pm i \gamma_0 \gamma_5 \tau^3)/2$. Thus SF-QCD and $\chi$SF-QCD 
are equivalent theories, since one is obtained from the other by the redefinition of fermionic fields~(\ref{eq:ferm-rots}).

Given the equivalence of the two theories, it is hardly surprising that they share all symmetries: the well known SF-QCD symmetries,
once transcribed in terms of fields $\psi^\prime$ and $\bar \psi^\prime$, are those of $\chi$SF-QCD. Flavour symmetry in
its standard SF-QCD version (e.g.\ Eq.~(2.15) of ref.~\cite{Brida:2016rmy}) takes the form of Eqs.~(2.16) and (2.17) 
of~\cite{Brida:2016rmy}; parity $\cP$ (Eq.~(2.18) of~\cite{Brida:2016rmy}) becomes $\cP_5$ (Eq.~(2.19) of~\cite{Brida:2016rmy}) 
in $\chi$SF-QCD. Charge conjugation is form-invariant in the two versions. We note in passing that the parity operator $\cP_5$ commutes with the boundary projectors $\tilde Q_\pm$.

\begin{table*}
\begin{align*}
&
\qquad \text{\scriptsize flavours} \qquad \qquad  \scriptstyle (f_1, f_2) = (u, u^\prime), (d, d^\prime), (u, d), (d, u) \\
&
  \qquad \text{\scriptsize bulk operators}  \qquad  \scriptstyle X=V_0,A_{0},S,P \quad 
 \scriptstyle Y_{k}=V_{k},A_k,T_{k0},\widetilde{T}_{k0}\\
 & \begin{array}{lll} 
 \toprule
 &\text{SF}&\color{red}\text{$\chi$SF}\\
 &\scriptstyle f_{\rm X}(x_{0})=-\frac12 {a^{3} \over L^3} \sum_{\mathbf{x}} \big\langle X^{ud}(x)\mathcal{O}_{5}^{du}\big\rangle
 &\scriptstyle g_{\rm X}(x_0) = -\frac12 {a^{3} \over L^3} \sum_{\mathbf{x}} \big\langle X^{f_1f_2}(x){\color{red}{\cal Q}_{5}^{f_2f_1}}\big\rangle\\
  &\scriptstyle k_{\rm Y}(x_{0})=-\frac16 {a^{3} \over L^3} \sum_{k=1}^{3} \sum_{\mathbf{x}} \big\langle Y_{k}^{ud}(x)\mathcal{O}_{k}^{du}\big\rangle
 &\scriptstyle l_{\rm Y}(x_0) = -\frac16 {a^{3} \over L^3} \sum_{k=1}^3 \sum_{\mathbf{x}} \big\langle Y_k^{f_1f_2}(x){\color{red}{\cal Q}_{k}^{f_2f_1}}\big\rangle \\ \\
  &\scriptstyle f_{1}= - \frac12 \big\langle {\cal O}_{5}^{ud} {\cal O}_{5}^{\prime du}\big\rangle
&\scriptstyle g_{1} =
 - \frac12 \big\langle {\color{red}{\cal Q}_{5}^{f_1f_2} {\cal Q}_{5}^{'f_2f_1}}\big\rangle\\
  &\scriptstyle k_{1}= - \frac16 \sum_{k=1}^3\big\langle {\cal O}_{k}^{u d} {\cal O}_{k}^{\prime d u}\big\rangle
 &\scriptstyle l_{1}= - \frac16 \sum_{k=1}^3
           \big\langle {\color{red}{\cal Q}_{k}^{f_1f_2} {\cal Q}_{k}^{'f_2f_1}}\big\rangle\\ \\
& \scriptstyle \mathcal{O}_{5}^{f_{1}f_{2}} =a^{6}\sum_{\mathbf{y,z}}\overline{\zeta}_{d}(\mathbf{y})P_{+}\gamma_{5}\zeta_{u}(\mathbf{z})
&  \scriptstyle {\color{red}\mathcal{Q}_{5}^{uu'}} =a^{6} \sum_{\mathbf{y,z}}\overline{\zeta}_{u}(\mathbf{y})\gamma_{0}\gamma_{5}{\color{red} Q_{-}}\zeta_{u'}(\mathbf{z})\\
 & \scriptstyle \mathcal{O}_{k}^{f_{1}f_{2}} = a^{6} \sum_{\mathbf{y,z}}\overline{\zeta}_{f_{1}}(\mathbf{y})P_{+}\gamma_{k}\zeta_{f_{2}}(\mathbf{z})
 & \scriptstyle {\color{red} \mathcal{Q}_{k}^{uu'}} =a^{6} \sum_{\mathbf{y,z}}\overline{\zeta}_{u}(\mathbf{y})\gamma_{k}{ \color{red} Q_{-}}\zeta_{u'}(\mathbf{z})\\
& \vdots & \color{red} \vdots\\
& \scriptstyle P_\pm \equiv \tfrac{1}{2}(1\pm \gamma_0) & \scriptstyle \color{red} {Q_\pm \equiv \tfrac{1}{2}(1\pm i\gamma_0\gamma_5)} \\
 \toprule
\end{array}
\end{align*}
\caption{List of correlation functions, some operators defined on time-boundaries, and projectors in SF and $\chi$SF setups. Operators $\cO_5^{f_1 f_2}, \cO_k^{f_1 f_2}, \cQ_5^{f_1 f_2}, \cQ_k^{f_1 f_2}$ are defined at $x_0 = 0$, whilst $\cO_5^{\prime f_1 f_2}, \cO_k^{\prime f_1 f_2}, \cQ_5^{\prime f_1 f_2}, \cQ_k^{\prime f_1 f_2}$ (not listed here) are defined at $x_0 = T$. A complete list of SF and $\chi$SF time-boundary operators is provided in Appendix A of ref.~\cite{Brida:2016rmy}. Quantities specific to the $\chi$SF setup are highlighted in red.}
\label{tab:sfchisf}
\end{table*}
Similar considerations apply to correlation functions. 
Following ref.~\cite{Brida:2016rmy}, we introduce, in $\chi$SF-QCD, a second flavour doublet 
$(\psi_{u^\prime} \,\, \psi_{d^\prime})^T$, with exactly the same properties as the original one.
In Table~\ref{tab:sfchisf} we gather all pertinent definitions of correlation functions, operators defined on time-boundaries, and projectors. 
We are specifically interested in correlation functions with antisymmetric tensor operators in the bulk:
\begin{eqnarray}
\label{eq:Tmunu}
T_{k0}^{f_1 f_2}(x) &\equiv& i\bar \psi^{f_1} (x) \; \sigma_{k0} \; \; \psi^{f_2}(x) \, \\
\widetilde T_{k0}^{f_1 f_2} &\equiv& -\frac12 \epsilon_{k0ij} T^{f_1 f_2}_{ij} \,,
\label{eq:Ttmunu}
\end{eqnarray}
with $\sigma_{k0} \equiv \dfrac{i}{2} [\gamma_k,\gamma_0]$ and $k=1,2,3$.
The allowed combinations of flavour indices are
$(f_1, f_2) = (u, u^\prime), (d, d^\prime), (u, d), (d, u)$ (so that no disconnected diagrams arise). 
See ref.~\cite{Brida:2016rmy} for more detailed explanations. The present work is based on the formal (continuum)
relations between the following correlation functions:
\begin{subequations}
 \begin{alignat}{4}
    \kT  &=&\,i \lTt^{uu'} &=&\,-i \lTt^{dd'} &=&\,   \lT^{ud} &= \,\phantom{i} \lT^{du} \, ,
   \label{eq:dictkt}\\
    \fA  &=&\, \gA^{uu'} &=&\, \gA^{dd'} &=&\, -i \gV^{ud} &= \, i \gV^{du} \, ,
   \label{eq:dictfA}\\
    \kV  &=&\, \lV^{uu'} &=&\, \lV^{dd'} &=&\, -i \lA^{ud} &= \, i \lA^{du} \, .   
   \label{eq:dictfV}\\
     k_1  &=&\,  l_1^{uu'} &=&\, l_1^{dd'} &=&\,  l_1^{ud} &= \, l_1^{du} \, ,
  \label{eq:dictk1}\\
     f_1 \, &=& \, g_1^{uu^\prime} \, &=& \, g_1^{dd^\prime} \, &=&  \, g_1^{ud} \, &=  \, g_1^{du} \, .
   \label{eq:dictf1}
 \end{alignat}
\end{subequations}

The above properties, though trivial at the formal level, have non-trivial consequences once the lattice
regularisation with Wilson fermions ($\chi$SF-LQCD) is introduced. (Of the three lattice $\chi$SF-QCD versions
proposed in ref.~\cite{Sint:2010eh}, we use that of ref.~\cite{Brida:2016rmy}; see Sec.~3.1 of the latter
work for the definition of the action etc.). The Wilson term and boundary terms in $\chi$SF-LQCD
induce the breaking of the rotated flavour symmetry (i.e.\ Eqs.~(2.16) and (2.17) of~\cite{Brida:2016rmy})
and parity $\cP_5$ (i.e.\ Eq.~(2.19) of~\cite{Brida:2016rmy}). However, a symmetry argument analogous to that introduced
in twisted-mass QCD~\cite{Frezzotti:2003ni} holds in the present case~\cite{Sint:2007ug,Sint:2010eh}, with the
result that $\cP_5$-even correlation functions of the $\chi$SF-LQCD theory, once renormalised, are
$O(a)$-improved in the bulk. An important additional ingredient of the lattice formulation consists in the introduction of
boundary terms in the action. One such term is an improvement counterterm, which cancels boundary
$O(a)$-effects, once its coefficient $d_s(g_0^2)$ is properly tuned. Moreover, the aforementioned symmetry-breaking pattern of $\chi$SF-LQCD necessitates the introduction of an additional $O(a^0)$ boundary operator with coefficient $\zf(g_0^2)$,
which must be appropriately tuned,  in order for the rotated flavour and $\cP_5$ symmetries to be recovered
in the continuum. In Sect.~\ref{sec:simulation} we discuss the $\zf$ and $d_s$ values we have used in the present work and the pertinent subtleties associated with these choices.

Symanzik $O(a)$ improvement is achieved in $\chi$SF~\cite{Sint:2010eh,Sint:2010xy,DallaBrida:2018tpn}: the ${\cal P}_5$-even  correlation functions (generically denoted as $g_{\scriptstyle \rm even}$)
receive  corrections only at second order in the lattice spacing, whereas the ${\cal P}_5$-odd ones (denoted as $g_{\scriptstyle \rm odd}$) are pure lattice artefacts:
\begin{align}
\label{eq:geven}
&g_{\scriptstyle \rm even} = g_{\scriptstyle \rm even}^{\rm continuum} + O(a^2) \, , \\
&g_{\scriptstyle \rm odd} = O(a)\,.
\label{eq:godd}
\end{align}
This  property  turns out to be advantageous for the tensor operator, as we will se below (cf. eq.~(\ref{eq:Timp})).

Before applying this property to the tensor operator, we recall that in a standard Wilson fermion setup $T_{\mu \nu}$ is $O(a)$-improved by the addition of the following Symanzik counterterm~\cite{Bhattacharya:2005rb}:
\begin{align}
T_{\mu \nu}^{{\scriptstyle \rm I}, f_1 f_2} =T_{\mu \nu}^{f_1 f_2} + { c_{\rm T}(g_0^2)\; a\,(\tilde{\partial}_\mu V_\nu^{f_1 f_2} - \tilde{\partial}_\nu V_\mu^{f_1 f_2})}\, ,
\end{align}
where $\tilde{\partial}_{\mu}$ denotes the symmetric lattice derivative. (Obviously, the above equation, combined with eq.~(\ref{eq:Ttmunu}), gives
$\widetilde T_{\mu \nu}^{{\scriptstyle \rm I}, f_1 f_2} = \widetilde T_{\mu \nu}^{f_1 f_2} - c_{\rm T}(g_0^2)\; a\,\epsilon_{\mu\nu\rho\sigma} \tilde{\partial}_\rho V_{\sigma}^{f_1 f_2}$.) The insertion of this counterterm in $\kT$ gives rise to 
\begin{align}
  \kT^{\rm I}=\kT + c_{\rm T}(g_0^2)\; a\,\tilde{\partial}_{0}\kV \,.
\end{align}
Note that the term involving a $\tilde{\partial}_{j} V_0(x)$ insertion in the correlation function vanishes under the summation over all space points ${\bf x}$.
The coefficient $c_{\rm T}$ is known to 1-loop in perturbation theory~\cite{Sint:1997dj}. It has been used in a study of SF step-scaling functions of the tensor operator in $\NF=2$ QCD~\cite{Pena:2017hct}. Using a Ward identity, non-perturbative estimates of $c_{\rm T}$ in a SF setup for $\NF=3$ QCD have recently appeared for a range of bare couplings relevant to large volume simulations~\cite{Chimirri:2023ovl}.

In a $\chi$SF setup, $c_{\rm T}$ is irrelevant for the improvement of $\lT^{ud}$, since the vector correlation function $\lV^{ud}$ is $O(a)$, being  ${\cal P}_5$-odd; cf. eqs.~(\ref{eq:geven},\ref{eq:godd}).
Therefore the Symanzik correction, being $O(a^2)$, may be dropped:
\begin{align}
\label{eq:Timp}
&\qquad \lT^{{\scriptstyle \rm I}, ud}=\lT^{ud} + \cancel{ c_{\rm T}(g_0^2)\; a\,\tilde{\partial}_{0}\lV^{ud} }\,, \\
&\qquad \lTt^{{\scriptstyle \rm I}, uu^\prime}=\lTt^{uu^\prime} \,.
\end{align}
In the second equation, the $c_{\rm T}$ counterterm drops out, being the integral of functions of the form $\partial_i V_j$ over a spatial volume with periodic boundaries.

The same argument holds for all other correlation functions $\lT^{f_1 f_2}$ and $\lTt^{f_1 f_2}$ appearing in eq.~(\ref{eq:dictkt}).
As a side remark we point out that standard parity $\cP$, combined with flavour exchanges $u \leftrightarrow d$ and  $u^\prime \leftrightarrow d^\prime$ is an exact 
symmetry of $\chi$SF-LQCD~\cite{Sint:2010eh,Brida:2016rmy}. This ensures that  $\lT^{ud} = \lT^{du}$ and
$\lTt^{uu^\prime} = - \lTt^{dd^\prime}$ are exact lattice relations. For this reason we have not used the correlation functions $ \lT^{du}$ and $\lTt^{dd^\prime}$, as they do not convey any new information.

In refs.~\cite{Pena:2017hct,Chimirri:2023ovl}, two SF renormalisation conditions for the tensor operator are imposed by generalising in analogy to those of ref.~\cite{Capitani:1998mq}:
\begin{eqnarray}
\dfrac{\ZTSFf{\scriptstyle(g_0^2,L/a)} \kT\scriptstyle{(T/2)}}{\sqrt{f_1}} &=& \Bigg [ \dfrac{\kT\scriptstyle{(T/2)}}{\sqrt{f_1}} \Bigg ]^{\rm \scriptstyle t.l.} \label{eq:ZT-SFf} \,\, ,\\
\dfrac{\ZTSFk{\scriptstyle(g_0^2,L/a)} \kT\scriptstyle{(T/2)}}{\sqrt{k_1}} &=& \Bigg [ \dfrac{\kT\scriptstyle{(T/2)}}{\sqrt{k_1}} \Bigg ]^{\rm \scriptstyle t.l.} \label{eq:ZT-SFk} \,\, ,
\end{eqnarray}
where the superscripts $\rm \scriptstyle t.l.$ on the r.h.s. stand for ``tree level''. These tree level quantities are computed at non-vanishing $a/L$ and vanishing bare quark mass (i.e. the hopping parameter $\kappa = 1/8)$. The denominators cancel the renormalisation of the boundary fields contained in the numerators. From the above renormalisation conditions the renormalisation constants $\ZTSFf,\ZTSFk$ are determined in two SF renormalisation schemes~\cite{Pena:2017hct,Chimirri:2023ovl}.

So far we have discussed the SF renormalisation schemes~(\ref{eq:ZT-SFf}),(\ref{eq:ZT-SFk}); we now generalise these even further. Based on the formal (continuum) relations~(\ref{eq:dictkt})-(\ref{eq:dictf1}), we can impose the $\chi$SF renormalisation conditions
\resizebox{0.85\linewidth}{!}{
\begin{minipage}{\linewidth}
\begin{eqnarray}
\dfrac{\ZTchiSFud{\scriptstyle(g_0^2,L/a)} \lT^{ud}{\scriptstyle(T/2)}}{[g_1^{ud}]^{\alpha} [l_1^{ud}]^{\beta} [-i \gVt^{ud}]^{\gamma}[\lVt^{uu^\prime}]^{\delta}} &=& \Bigg [ \dfrac{\lT^{ud}{\scriptstyle(T/2)}}{[g_1^{ud}]^{\alpha} \,\, [l_1^{ud}]^{\beta} \,\, [-i \gVt^{ud}]^{\gamma} [\lVt^{uu^\prime}]^{\delta}} \Bigg ]^{\rm \scriptstyle t.l.} , \nonumber \\
\label{eq:ZT_chiSFdef3} \\
\dfrac{\ZTchiSFuup{\scriptstyle(g_0^2,L/a)} i \lTt^{uu^\prime}{\scriptstyle(T/2)}}{[g_1^{uu^\prime}]^{\alpha} [l_1^{uu^\prime}]^{\beta} [-i \gVt^{ud}]^{\gamma}[\lVt^{uu^\prime}]^{\delta}} &=& \Bigg [ \dfrac{i \lTt^{uu^\prime}{\scriptstyle(T/2)}}{[g_1^{uu^\prime}]^{\alpha} [l_1^{uu^\prime}]^{\beta} [-i \gVt^{ud}]^{\gamma}[\lVt^{uu^\prime}]^{\delta}} \Bigg ]^{\rm \scriptstyle t.l.} . \nonumber \\
\label{eq:ZT_chiSFdef4}
\end{eqnarray}
\end{minipage}}
\\
The correlation functions $\gVt^{uu^\prime}, \lVt^{uu^\prime}$ are as defined in Tab.~\ref{tab:sfchisf}, with the operator insertions $X^{f_1 f_2} = \widetilde V^{f_1 f_2}_0$ and $Y^{f_1 f_2} = \widetilde V^{f_1 f_2}_k$, where $\widetilde V^{f_1 f_2}_\mu$ is the conserved current (point-split for Wilson fermions) with $\ZVt = 1$. The exponents $\alpha, \beta, \gamma, \delta$ are real numbers satisfying $2(\alpha + \beta) + \gamma +\delta =1$. This ensures that the renormalisation factors of the boundary fields cancel in the ratios (\ref{eq:ZT_chiSFdef3}) and (\ref{eq:ZT_chiSFdef4}). For fixed $(\alpha, \beta, \gamma, \delta)$ the above equations correspond to the same continuum renormalisation condition, with different lattice realisations, from which the same continuum 
step-scaling function is obtained. Each choice of $(\alpha, \beta, \gamma, \delta)$ defines a different scheme, if only by modifying the denominator of the ratios of correlation functions. The introduction of products of boundary-to-boundary and boundary-to-bulk correlation function in $\chi$SF renormalisation schemes was introduced in the context of four-fermion operators in ref.~\cite{Mainar:2016uwb}.

To simplify matters we only obtain results for the following four schemes:
\begin{align}
(\alpha, \beta, \gamma, \delta) = & (1/2,0,0,0), (0,1/2,0,0), \nonumber \\
& (0,0,1,0), (0,0,0,1) \, .
\end{align}
Henceforth, we will refer to these schemes by the Greek letter corresponding to the non-zero entry of each of the above rows; e.g.
scheme-$\alpha$, scheme-$\gamma$ etc.
Schemes-$\alpha$ and -$\beta$ are defined by imposing the same renormalisation conditions as those of 
eqs.~(\ref{eq:ZT-SFf}) and (\ref{eq:ZT-SFk}) respectively; these are the f- and k-schemes of ref.~\cite{Chimirri:2023ovl}.
The difference lies in the lattice setup: Whereas in the f- and k-
schemes of ref.~\cite{Chimirri:2023ovl} the bare correlation functions are defined on lattices with SF boundary conditions, in this work 
we use $\chi$SF boundaries for the valence quarks (see Section~\ref{sec:simulation} for details). Thus the anomalous
dimensions of our schemes-$\alpha$ and -$\beta$ are respectively equal to those of the f- anf k-schemes of ref.~\cite{Chimirri:2023ovl}, up to discretisation effects. The introduction of schemes-$\gamma$ and -$\delta$ is appealing only in a $\chi$SF setup, since in the SF setup $\fV = 0$, while $\kVt$ requires the extra counterterm  $c_{\rm V}$.

It is well known that the anomalous dimension $\gammaT$ of the tensor operator is scheme dependent; in perturbation theory this  dependence sets in at NLO. Thus at high energies some schemes convergence better to their perturbative behaviour than others. Moreover, both the statistical accuracy and the size of the discretisation errors afflicting our non-perturbative results 
depend on the renormalisation scheme. Having more schemes at our disposal may potentially 
provides extra handles for a better control of these effects. 

\begin{figure*}
\includegraphics[width=0.5\textwidth]{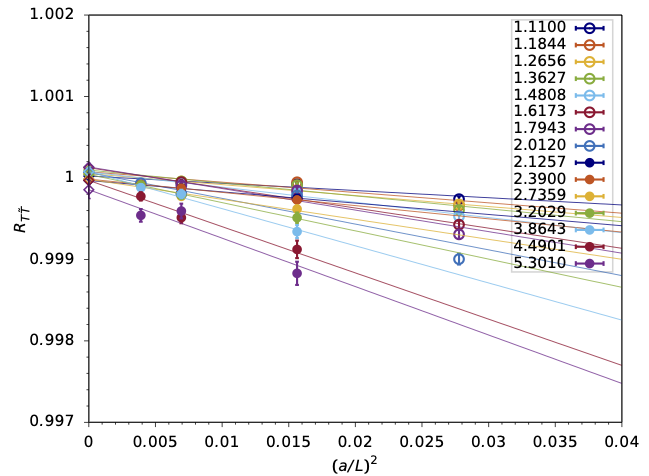}
\caption{Ratio $R^l_{\scriptstyle T \widetilde T}$ as function of the squared lattice spacing. Data of the same color refers to the same renormalized coupling value $u$,
with open circles indicating couplings in the high energy region (SF) and 
filled circles those at low energies (GF). The diamond symbols show
the results of the linear continuum  extrapolation at fixed values of the 
coupling $u$.}
\label{fig:rttt}
\end{figure*}

By definition $\widetilde T_{\mu\nu}$ is proportional to $T_{\rho\sigma}$ (with indices $\rho,\sigma$ complementary to $\mu,\nu$). Therefore a common renormalisation  factor $\ZT$ renormalises both operators. Moreover, the ratios (\ref{eq:ZT_chiSFdef3})-(\ref{eq:ZT_chiSFdef4}) are related in the continuum limit through eqs.~(\ref{eq:dictkt})-(\ref{eq:dictf1}). From these considerations  we deduce that the renormalisation constants thus defined only differ by discretisation effects and hence
\begin{equation}
\label{eq:Rl}
R^l_{\scriptstyle T \widetilde T} \, \equiv \, \dfrac{i \lTt^{uu^\prime}(x_0)}{\lT^{ud}(x_0)} \, = \, 1 + O(a^2) \,\, .
\end{equation}
As shown in Fig.~\ref{fig:rttt}, 
the ratio $R^l_{\scriptstyle T \widetilde T}$
is numerically different from unity by a small deviation of
per thousand, and extrapolates to 1 with an accuracy 
of $\sim 0.1$ \textperthousand.
This shows that the correlation function $\lT^{uu^\prime}$
is numerically  extremely close to $\lTt^{ud}$ and that it provides redundant information 
within our statistical errors. In this article we have therefore chosen to present 
only the results derived from $\lT^{ud}$ and the renormalisation scheme~(\ref{eq:ZT_chiSFdef3});
the renormalisation condition~(\ref{eq:ZT_chiSFdef4}) will not be used.

The definition of the lattice step-scaling function (SSF) is fairly standard:
\begin{equation}
\SigmaTchiSFud(g_0^2,a/L) = \dfrac{\ZTchiSFud(g_0^2,2L/a)}{\ZTchiSFud(g_0^2,L/a)} \,\, , \label{eq:Ssf-TchiSFud}
\end{equation}
Henceforth the superscripts $\chi$SF,$ud$ will be dropped; instead indices -$\alpha$, $\beta$, -$\gamma$, and -$\delta$ 
will be introduced where necessary, in order to indicate the scheme we are referring to.
The above SSF has a well defined continuum limit:
\begin{equation}
\sigmaT(u) \,\, = \,\, \lim_{a \rightarrow 0} \SigmaT(g_0^2,a/L) \Big \vert_{\gbar^2(L) = u} \,\, .
\end{equation}
The squared renormalised finite-volume coupling $\gbar^2(L) = u$ is meant to be held fixed while the continuum limit is taken. 
Although $\SigmaT(g_0^2, a/L)$ is computed at specific values of the bare coupling and lattice volume, we are interested in its behaviour with varying renormalised coupling; hence we will also use the notation $\SigmaT(u, a/L)$.
In terms of the renormalised continuum tensor operator $[T_{\mu\nu}^{ud}(\mu)]_{\mathrm R}$, defined at a scale $\mu=1/L$, which corresponds to a renormalised coupling $\gbar^2(\mu) = u$, the continuum step-scaling function is given by the ratio
\begin{equation}
\label{sigmaP-mbar}
\sigmaT(u) \,\, = \,\,\dfrac{[T_{\mu\nu}^{ud}(\mu/2)]_{\scriptstyle \rm R}}{[T_{\mu\nu}^{ud}(\mu)]_{\scriptstyle \rm R}}\Bigg \vert_{\gbar^2(\mu) = u} \,\,  .
\end{equation}
In what follows, we will simplify the notation by dropping subscripts ($\mu,\nu$) and superscripts  ($u,d$) in $T_{\mu\nu}^{ud}$; cf. eq.~(\ref{Rk}). The continuum SSF can also be expressed in terms of the tensor anomalous dimension $\gammaT$ and the Callan-Symanzik $\beta$-function as
\begin{equation}
\sigmaT(u)=\exp \Big [ \int_{\sqrt{u}}^{\sqrt{\sigma(u)}}dx\frac{\gammaT(x)}{\beta(x)} \Big ] \,\, ,
\label{eq:sigma-integr}
\end{equation}
where $\sigma(u) = \gbar^2(\mu/2)\vert_{\gbar^2(\mu) = u}$. For the series expansions of $\beta, \gammaT$, and $\sigmaT$ see Appendix~\ref{app:pert-coeffs}.
 
\section{Numerical Simulations} 
\label{sec:simulation}
Our numerical setup has been detailed in ref.~\cite{Plasencia:2021eon}.
For completeness, we make a r\'esum\'e of the most important points. 
The lattice volumes $L^4$ in which simulations are performed
define the range of accessible energy scales $\mu = 1/L$.
We distinguish two energy regimes~\cite{Brida:2016flw,DallaBrida:2016kgh,DallaBrida:2018rfy}.
In the high-energy range ($\mu_0/2 \sim 2~{\rm GeV} \lessapprox \mu
\lessapprox \MW$) the renormalised squared coupling $\gbar^2_{\mathrm{SF}}$ is
the nonperturbative Schr\"odinger Functional (SF) coupling first introduced in ref.~\cite{Luscher:1992an,Luscher:1993gh}.
In the low-energy one ($\lQCD \lessapprox \mu \lessapprox  \mu_0/2$) 
$\gbar^2_{\mathrm{GF}}$ is the squared gradient flow (GF) coupling defined in
ref.~\cite{Fritzsch:2013je,DallaBrida:2016kgh}. This allows for a very precise computation of the 
Callan-Symanzik $\beta$-function and ultimately of $\lQCD$; see ref.~\cite{Bruno:2017gxd}.
In refs.~\cite{Brida:2016flw,DallaBrida:2016kgh,DallaBrida:2018rfy},
the switching scale $\mu_0/2$ is implicitly defined by the value of the coupling 
$\gbar_{\SF}^2(\mu_0)$; see eqs.~(\ref{eq:gbar-iter})-(\ref{uSF-mu0/2}) below.
The matching between schemes was subsequently specified by determining
the value of the GF coupling at the same scale; see eq.~(\ref{uGF-mu0/2}) below.
In physical units this corresponds to a switching scale $\mu_0/2 \sim 2 {\rm GeV}$; 
see ref.~\cite{Bruno:2017gxd}.

Our gauge ensembles are practically those described in 
ref.~\cite{Campos:2018ahf}: different lattice regularisations were adopted in each energy
regime. At high energies, simulations were carried out 
using the plaquette gauge action~\cite{Wilson:1974sk}
and the clover fermion action~\cite{Sheikholeslami:1985ij} 
with the nonperturbative value of  $\icsw$~\cite{Yamada:2004ja} and
the one-~\cite{Sint:1997jx} and two-loop~\cite{Bode:1998hd}
values of $\icttil$ and $\ict$ respectively. At low energies the tree-level 
Symanzik-improved (L\"uscher-Weisz) gauge action was
used~\cite{Luscher:1985zq}. The fermion action was the $O(a)$-improved 
clover~\cite{Sheikholeslami:1985ij}, with the nonperturbative value 
of the improvement coefficient $\icsw$~\cite{Bulava:2013cta} and 
one-loop values of $\icttil$~\cite{VilasecaPrivate,DallaBrida:2016kgh} and $\ict$~\cite{Aoki:1998qd}.

The simulations and bare parameter tunings have been described in ref.~\cite{Campos:2018ahf}:
At high energies (SF range) we have the 8 values of the
squared renormalised coupling $\uSF$ listed in Appendix~\ref{app:Z-Sigma-he}. For each of these couplings, 
corresponding to a fixed renormalisation scale $\mu = 1/L$, the inverse 
bare coupling $\beta = 6/g_0^2$ was tuned appropriately for
$L/a = 6,8,12$. At the strongest coupling $\uSF = 2.012$
an extra finer lattice with $L/a = 16$
was simulated. At low energies (GF range), simulations were carried out for the 7 values of the squared renormalised coupling $\uGF$ listed in Appendix~\ref{app:Z-Sigma-le}.

The inverse bare coupling $\beta = 6/g_0^2$ 
was chosen so that $\uGF$ remains {\it approximately} constant for the three lattice volumes $L/a = 8,12,16$.
In both the high- and the low-energy ranges, gauge ensembles were generated at 
each $(\beta,L/a)$ and $(\beta,2L/a)$ combination. At fixed $(\beta,L/a)$ the 
hopping parameter $\kappa$ was tuned to its critical value $\hopc$ in standard Alpha fashion~\cite{Campos:2018ahf}.
Note that fermion fields have SF boundary conditions in space-time with a periodicity angle $\theta =1/2$.
All correlation functions were computed in the $Q=0$ topological sector of the theory. (These choices were also
tacitly adopted in our previous paper~\cite{Plasencia:2021eon}, concerning the RG-running of the quark mass.) 

This concludes the recapitulation of the simulations generating $\NF=3$ sea quark ensembles.
In ref.~\cite{Chimirri:2023ovl} these ensembles were combined with the SF renormalisation
conditions (\ref{eq:ZT-SFf}) and (\ref{eq:ZT-SFk}) in order to determine the non-perturbative renormalisation and 
RG-running of the tensor composite field in a unitary setup. We adopt a different approach:
As in ref.~\cite{Plasencia:2021eon}, valence quark propagators are obtained in a mixed-action
framework by inverting the Dirac-Wilson operator with a Clover term in the bulk and $\chi$SF boundary conditions in the whole energy range.
As mentioned earlier, an important ingredient of the $\chi$SF lattice formulation~\cite{Sint:2010eh} consists in the introduction of two
boundary terms in the action, namely $\zf$ and $d_s$.
In ref.~\cite{Plasencia:2021eon} $\zf$ was determined as a function of $\beta$, so as 
to ensure that chiral and flavour symmetries are recovered in the continuum and thus the theory belongs to the same
universality class as other lattice regularisations. 

Concerning the second boundary term $d_s$, we had argued in ref.~\cite{Plasencia:2021eon} that it is safe
to use the tree-level value $d_{s}^{(0)} =0.5$ for the SSF $\SigmaP$. We use the same tree-level estimate here
for $\SigmaT$. It is worth recalling that in some sense the $\chi$SF 
counter-term $d_s$ ``replaces" the SF one $\icttil$. The latter counter-term, known at one-loop in PT~\cite{Sint:1997jx},
introduces barely noticeable systematic effects in the pseudoscalar renormalisation parameter $\ZP$~\cite{Campos:2018ahf},
but is a measurable source of systematic uncertainty for the tensor parameter $\ZT$; see ref~\cite{Chimirri:2023ovl}. An explanation
for this is provided by perturbation theory: The dependence of $\ZP$ on $\icttil$ is of higher order in the coupling than naively expected due to some ``accidental" cancellation between different terms, only valid at one-loop. Such a cancellation is absent in the perturbative estimate of $\ZT$. It is plausible that this is the reason for the stronger dependence
of the non-perturbative SF results of $\ZT$ on $\icttil$, seen in ref.~\cite{Chimirri:2023ovl}. A similar situation also characterises
the perturbative $\ZP$ and $\ZT$ results in $\chi$SF, obtained in ref.~\cite{Brida:2016rmy}: Due to accidental cancellations,
dependence of $\ZP$ on $d_s$ is absent at one-loop order, while being present in $\ZT$. However, the one-loop dependence of $\ZT$ on $d_s$ is very weak. Given this last property, we have used the tree-level value for $d_s$ in our analysis.

In this work we always use the $\Gamma$-method error analysis of refs.~\cite{Wolff:2003sm,Schaefer:2010hu}. The $\Gamma$-method-based package used for the error propagation was developed in refs.~\cite{Ramos:2018vgu,Ramos:2020scv}.
The code to compute the $\chi$SF correlation functions is built on openQCD 1.0 and previously  used in ref.~\cite{DallaBrida:2018tpn}. 

\section{Tensor running at high energies}
\label{sec:results-he}
We now turn to the computation of the step-scaling functions themselves,
which are the main input for the determination of the nonperturbative
running of the tensor operator. On each pair of ($L$, $2L$) ensembles,
and for each scheme-$\alpha$, -$\beta$, -$\gamma$, -$\delta$,
we compute $\SigmaTchiSFud$, defined in Eq.~(\ref{eq:Ssf-TchiSFud}); 
henceforth the superscripts $\chi$SF and $ud$ will be dropped:
\begin{equation}
\label{eq:SigmaTrat}
\SigmaT(u, a/L) = \frac{\ZT(u, 2L/a)}{\ZT(u, L/a)} \,.
\end{equation}
Occasionally a superscript $\alpha$, $\beta$, $\gamma$, $\delta$ will indicate the
relevant $\chi$SF-scheme.

In both the high energy and low energy regimes we work at three
different lattice spacings, except for the largest coupling in the high energy
range (the switching point, $u=2.0120$) where we use four lattice spacings.
We note here
that the values of $\SigmaT$ at different couplings 
are statistically uncorrelated. 
In the spirit of ref.~\cite{Luscher:1993gh} and as done in
refs.~\cite{Campos:2018ahf,Plasencia:2021eon}, we also
define the ratio of the lattice SSF $\SigmaT(u, a/L)$, computed in 1-loop lattice perturbation 
theory, to the 1-loop quantity in the $L/a=\infty$ limit, in order to determine
the numerical effect of the lattice artefacts appearing at this order:
\begin{gather}
\label{eq:R_T_PT}
\frac{\SigmaT^{\rm\scriptscriptstyle 1-loop}(u, a/L)}{\sigmaT^{\rm\scriptscriptstyle 1-loop}(u)}
\,\, =  \,\, 1 + u \deltaT(a/L) \,\,, \\
\deltaT(a/L) = \gammaT^{(0)} \ln(2) \epsilon_{\rm\scriptscriptstyle{T}}(a/L) \,\, .
\label{eq:delta}
\end{gather}
In the above, $\gammaT^{(0)}$ is the universal anomalous dimension coefficient for the tensor 
field; see eq.~(\ref{eq:gamma0}). 
The lattice artefacts of $O(g_0^2 a^n)$ may be subtracted from $\SigmaT$, 
computed nonperturbatively, according to
\begin{gather}
\label{eq:SigmaT_sub}
\SigmaT^{\text{sub}}(u, a/L) \equiv 
\frac{\SigmaT(u, a/L)}{1 + u \deltaT(a/L)} \,\, .
\end{gather}
The remaining discretisation errors in $\SigmaT^{\text{sub}}$ are $O(g_0^4 a^2)$. 

However there is a subtlety in the present work, where we use the subtraction coefficients 
$\epsilon_{\rm\scriptscriptstyle{T}}(a/L)$ obtained in~\cite{Brida:2016rmy,VilasecaPrivate}
and quoted in Table~\ref{tab:cT}. These have been calculated perturbatively 
using the asymptotic values of the critical quark mass $\mcrit(a/L=0)$ and $\zf(a/L=0)$, rather than the 
$\mcrit(a/L)$ and $\zf(a/L)$ corresponding to a finite $a/L$. Thus in practice an
$O(g_0^2 a^2)$ effect could be present, which will be discussed below.

\begin{table}
\label{tab:cT}
\begin{tabular}{c|cccc}
\hline 
L/a & \multicolumn{1}{c}{$\epsilon_{{\mathrm T}}^{\alpha}$} & \multicolumn{1}{c}{$\epsilon_{{\mathrm T}}^{\beta}$} & \multicolumn{1}{c}{$\epsilon_{{\mathrm T}}^{\gamma}$} & $\epsilon_{{\mathrm T}}^{\delta}$\tabularnewline
\hline 
6 & -0.2594 & -0.2024 & -0.5377 & 0.1544\tabularnewline
8 & -0.1112 & -0.0928 & -0.2572 & 0.0682\tabularnewline
10 & -0.0576 & -0.0515 & -0.1475 & 0.0423\tabularnewline
12 & -0.0337 & -0.0322 & -0.0945 & 0.0296\tabularnewline
14 & -0.0215 & -0.0218 & -0.0652 & 0.0220\tabularnewline
16 & -0.0145 & -0.0156 & -0.0473 & 0.0170\tabularnewline
18 & -0.0102 & -0.0116 & -0.0358 & 0.0135\tabularnewline
20 & -0.0074 & -0.0089 & -0.0278 & 0.0110\tabularnewline
22 & -0.0055 & -0.0070 & -0.0222 & 0.0091\tabularnewline
24 & -0.0042 & -0.0056 & -0.0181 & 0.0077\tabularnewline
\end{tabular}
\caption{Subtraction coefficients $\epsilon_{\mathrm T}(a/L)$ for the four schemes 
$\alpha$, $\beta$, $\gamma$, $\delta$, used to remove discretisation effects from the non-perturbative step scaling functions  $\SigmaT$ 
up to $O(g_0^2)$ as given in 
eqs.~(\ref{eq:R_T_PT}) and (\ref{eq:delta}).
}
\end{table}

Our results for the $Z_{\mathrm T}$'s, $\SigmaT$'s, and $\SigmaT^{\mathrm{sub}}$'s for
all values of $u$, $L/a$, $\beta$ and $\hopc$, in all four schemes, are gathered in Appendix~\ref{app:Z-Sigma-he}.

As in ref.~\cite{Plasencia:2021eon}, the $\SigmaT$ dependence on the gauge coupling and 
number of lattice sites may be expressed as a double power series:
\begin{equation} 
\label{eq:sigmaT_fit}
\SigmaT(u, a/L) = 1 + \sum_{i=1, j=0} \, b_{ij} u^i (a/L)^j \, .
\end{equation}
The continuum step-scaling function $\sigmaT$ is then given by
$\sigmaT(u) = \lim_{a \to 0} \SigmaT(u, a/L)$.
Note that upon subtracting leading $O(u)$ discretisation errors from our data 
(cf.\ Eq.~(\ref{eq:SigmaT_sub})), we could also set the terms $b_{1j>0}$ to zero. 
For the reasons explained above (possible residual $O(g_0^2 a^2)$ due to $\epsilon_{\mathrm T}$),
we will refrain from setting $b_{1j>0}=0$.
We exclude the $O(a)$ contributions from eq.~(27) (i.e. $b_{i1}=0$) because of the presence of a Clover term in the action(s) and the fact that the boundary improvement described at the end of Section III is deemed adequate.
Since the gauge
ensembles have been generated for sea-quarks with SF boundary
conditions, powers of $(a/L)^3$ and higher cannot be excluded. Such odd powers are
also present, due to the  $\chi$SF boundary terms. Moreover, we have ignored the presence
of powers of $\ln(a/L)$ and their resummation, resulting to discretisation effects of the form $a^2 [\alpha_{\mathrm s}^{\MSbar}(1/a)]^{\gamma}$,
with $\gamma$ generically indicating the anomalous dimension of a Symanzik counter-term; cf. refs.~\cite{Husung:2019ytz,Husung:2022kvi,Husung:2021mfl}. 

In ref.~\cite{Campos:2018ahf} several fit and extrapolation
methods have been proposed in order to compute the continuum SSF
$\sigmaP(u)$ and the anomalous dimension $\gammaP(u)$ of the pseudoscalar density
from the lattice SSF $\SigmaP(u)$.
We apply similar methods in order to obtain $\sigmaT(u)$ and $\gammaT(u)$
as continuum functions from the  $\SigmaT(u)$ data. 
The ultimate goal is the computation of
the running factor $T^{RGI}/[T(\mu_0/2)]_{\mathrm R}$, 
i.e.\ the ratio of the renormalisation group invariant (RGI) 
tensor $T^{\mathrm{RGI}}$ to the renormalised tensor field 
$[T(\mu_0/2)]_{\mathrm R}$ in a $\chi$SF scheme. 

\subsection{$u$-by-$u$ fits}
The first method,
denoted here as $\sigmaT$:$u$-{\it by}-$u$, is the original Alpha 
proposal~\cite{Luscher:1993gh}, in which $\SigmaT$ is written as a linear 
function in $(a/L)^2$:
\begin{equation}
\SigmaT(u,a/L)=\sigmaT(u)+\rhoT(u)\,(a/L)^{2} \,\, .
\label{eq:lin-extr} 
\end{equation}
With $u$ held constant, $\sigmaT$ and $\rhoT$ are linear fit parameters. 

We show results from these fits in Appendix~\ref{app:sigma-rho-he}, 
both for $\SigmaT(u,a/L)$ and $\SigmaT^{\text{sub}}(u, a/L)$. 
The quality of the fits is satisfactory, in that
$\chi^2/{\rm d.o.f.} \lessapprox 1$ in most of the 32 cases analysed 
(8 couplings for each of 4 schemes).
For the $\SigmaT$ fits, $\chi^2/{\rm d.o.f.}$ is substantially greater than 1 only for three $u$'s;
two in scheme-$\gamma$ and one in scheme-$\delta$.
For the $\SigmaT^{\text{sub}}$ fits, $\chi^2/{\rm d.o.f.}$ is substantially greater than 1 
only for three $u$'s; one in scheme-$\beta$ and two in scheme-$\gamma$.
Comparing $\chi^2/{\rm d.o.f.}$ from fitting $\SigmaT$ to that from $\SigmaT^{\text{sub}}$, we see
that the former is almost always bigger, with only a handful of exceptions: $u$=1.11 (schemes-$\alpha, \beta, \gamma$)
and $u$ = 1.2656 (schemes-$\gamma, -\delta$). As seen in the Tables of Appendix~\ref{app:sigma-rho-he},
the dependence of $\SigmaT$ on $(a/L)^2$ (i.e. the various slopes $\rhoT(u)$) is quite varied for each $u$ and for
each scheme, with $\rhoT^{\text{sub}}(u) < \rhoT(u)$. Nevertheless, the extrapolated $\sigmaT(u)$'s are compatible to the
$\sigmaT^{\text{sub}}(u)$'s. Henceforth we will only analyse and
quote results from the $\SigmaT^{\text{sub}}$ data, dropping the subscript
``sub" from now on.

The next step in this procedure consists in fitting the eight extrapolated $\sigmaT(u)$ results by the polynomial
\begin{equation}
\sigmaT (u) \,\, = \,\, 1  \, + \, \sum_{i=1}^{n_\sigma} b_{i0}u^{i} \,\, ,
\label{eq:sigma-ubyu}
\end{equation}
with $b_{10}$ and $b_{20}$ fixed by perturbation theory; cf. eqs~(\ref{eq:sigma-b10b20}).
We perform fits with $n_\sigma =3$ (one fitting parameter $b_{30}$), $n_\sigma =4$ (two fitting parameters, 
$b_{30}$ and $b_{40}$) and $n_\sigma =5$ (three fitting parameters, $b_{30}$, $b_{40}$, and $b_{50}$). 
For schemes-$\alpha$ and -$\beta$, $\chi^2/{\rm d.o.f.}$ increases as $n_\sigma$ grows, from $ \sim 1.1$ to a maximum of $\sim 2.3~$. The situation improves for schemes-$\gamma$ and -$\delta$, with $\chi^2/{\rm d.o.f.} \lessapprox 0.9$ (except for the $\chi^2/{\rm d.o.f.}$ of scheme-$\gamma$, $\sim 1.3~$ for $n_\sigma =3)$.
As the errors of the fit parameter $b_{30}$ grows by about an order 
of magnitude when $n_\sigma =5$ in all schemes, this fit is discarded. The $b_{30}$ results for $n_\sigma =4$ are shown in
eq.(\ref{eq:b30}).

The resulting continuous function for $\sigmaT(u )$ is readily calculated for the coupling values provided by
the recursion
\begin{equation}
\label{eq:gbar-iter}
\gbar^2_{\mathrm{SF}}(\mu_0) \, = \, 2.012 \,\,\, , \qquad u_k \, = \, \gbar^2_{\mathrm{SF}}(2^k\mu_0) \,\,\, ,
\end{equation}
and the RG evolution can be determined by the renormalised operator ratios at different scales
\begin{equation} 
\label{Rk}
R^{(k)} = \dfrac{[ T(2^k \mu_0) ]_{\scriptstyle \rm R}}{[ T(\mu_0/2) ]_{\scriptstyle \rm R}} = 
\prod_{n=0}^{k} [\sigmaT(u_n)]^{-1} \,\, .
\end{equation}
Recall that~\cite{DallaBrida:2016kgh}
\begin{equation} 
\label{uSF-mu0/2}
u_{\mathrm{SF}}(\mu_0/2) \, = \, \gbar^2_{\mathrm{SF}}(\mu_0/2) \, = \, \sigma(2.012) \, = \, 2.452(11) \,\, .
\end{equation}

Finally, we can construct the running factor that takes a renormalised tensor
composite field in our chosen $\chi$SF scheme at the scale $\mu_0/2$ to the
renormalisation group invariant tensor $T^{\mathrm{RGI}}$:
\begin{equation}
\label{eq:Tbar/TRGI}
\frac{T^{\mathrm{RGI}}}{[ T(\mu_0/2) ]_{\scriptstyle \rm R}} = \frac{T^{\mathrm{RGI}}}{[ T(2^k\mu_0) ]_{\scriptstyle \rm R}} \, \,\frac{[ T(2^k\mu_0) ]_{\scriptstyle \rm R}}{[ T(\mu_0/2) ]_{\scriptstyle \rm R}} \, .
\end{equation}
The first factor on the r.h.s. can be calculated in perturbation theory from 
\begin{multline}
\frac{T^{\mathrm{RGI}}}{[ T(2^k\mu_0) ]_{\scriptstyle \rm R}}  = 
\label{eq:T-barT-2k}
\Big [ \dfrac{\gbar^2_{\SF}(2^k \mu_0)}{4\pi} \Big ]^{-\gammaT^{(0)}/2 b_0} \times \\
\exp \Bigl\{ -\int_0^{\gbar_{\SF}(2^k \mu_0)} dx 
\Bigl[\frac{\gammaT(x)}{\beta(x)} - \frac{\gammaT^{(0)}}{b_0 x} \Bigr]
\Bigr\} \,,
\end{multline}
with $\gammaT$ and $\beta$ given by eqs.~(\ref{eq:beta})-(\ref{eq:gamma1}).
The second factor on the r.h.s. of eq.~(\ref{eq:Tbar/TRGI}) is given in eq.~(\ref{Rk}).
In ref.~\cite{Plasencia:2021eon}, we had verified that the analogous factor for the running of the quark mass  
can be reliably computed for very high $k$-values, which surpass the high-end
of the energy range covered by our simulations. We have checked that the same stability
is displayed in the case at hand.

A second procedure, labelled as $\gammaT$:$u$-{\it by}-$u$, starts off just like $\sigmaT$:$u$-{\it by}-$u$;
at constant $u$, we fit the data points $\SigmaT(u,a/L)$ with eq. (\ref{eq:lin-extr}), 
obtaining $\sigmaT(u)$. Then the continuum values $\sigmaT(u)$ are fitted with
\begin{equation}
\sigmaT(u) \,\, = \,\, \exp \Big [ \int_{\sqrt{u}}^{\sqrt{\sigma(u)}} dx \, \frac{\gammaT(x)}{\beta(x)} \Big ] \,\, ,
\label{eq:gamma-ubyu}
\end{equation}
where in the integrand we use eqs.~(\ref{eq:beta},\ref{eq:betaPTa}-\ref{eq:betaPTd}) for the Callan-Symanzik $\beta$-function and
eqs.~(\ref{eq:gamma},\ref{eq:gamma0},\ref{eq:gamma1}) for the anomalous dimension $\gammaT$.
The series 
\begin{equation}
\gammaT(g)=-g^{2} \sum_{n=0}^{n_\gamma} \gammaT^{(n)}g^{2n} \,\, ,
\label{eq:gamma}
\end{equation}
is truncated at $n_\gamma \in [2, \cdots, 5]$, with $\gammaT^{(k)}$ for $k \geq 2$ 
free fit parameter(s). The fits have $\chi^2/{\rm d.o.f.} < 0.5$. 
These are significantly smaller values that those obtained when fitting
$\sigmaT(u)$ with the $\sigmaT$:$u$-{\it by}-$u$ procedure. 
The $n_\gamma=2$ results (one fitting parameter $\gammaT^{(2)}$) have very small errors, while
$n_\gamma=3$ results (two fitting parameters $\gammaT^{(2)}$ and $\gammaT^{(3)}$) are predictably 
larger. For $n_\gamma=4,5$ the errors increase overwhelmingly, so these cases are discarded.
As an example, the fit parameter $\gammaT^{(2)}$ for all four schemes is given in eq.~(\ref{eq:gammaT2}).

Having thus obtained an estimate for the anomalous dimension $\gammaT(u)$, we arrive at another determination of the renormalised tensor ratios, using the expression
\begin{equation}
\label{eq:Rk2}
R^{(k)} \,\, = \,\, \dfrac{[ T(2^k \mu_0) ]_{\scriptstyle \rm R}}{[ T(\mu_0/2) ]_{\scriptstyle \rm R}}  \,\, = \,\,
\exp \Big [ - \int_{\sqrt{u_k}}^{\sqrt{u_{-1}}} dx \, \frac{\gammaT(x)}{\beta(x)} \Big ] \,\, ,
\end{equation}
with the couplings determined through eq.~(\ref{eq:gbar-iter}).

Since our fit functions have the correct asymptotic behaviour built in for $g^2 \rightarrow 0$,  the running factor of eq.~(\ref{eq:Tbar/TRGI}) can now be computed directly from
\begin{multline}
\frac{T^{\mathrm{RGI}}}{[ T(\mu_0/2) ]_{\scriptstyle \rm R}}  = 
\Big [ \dfrac{\gbar^2_{\SF}(\mu_0/2)}{4\pi} \Big ]^{-\gammaT^{(0)}/2 b_0} \times \\
\exp \Bigl\{ -\int_0^{\gbar_{\SF}(\mu_0/2)} dx 
\Bigl[\frac{\gammaT(x)}{\beta(x)} - \frac{\gammaT^{(0)}}{b_0 x} \Bigr]
\Bigr\} \,.
\label{eq:T-barT-np}
\end{multline}

\subsection{Global fits}
A third procedure is a variant of the method $\sigmaP$:{\it global}, applied in ref.~\cite{Campos:2018ahf}, which we here
denote as $\sigmaT$:{\it global}. We start by rewriting eq.~(\ref{eq:sigmaT_fit}) as a more general, if truncated, power
series
\begin{equation} 
\label{eq:sigmaT_fit2}
\SigmaT(u, a/L) = 1 + \sum_{i=1}^{n_\sigma}  \sum_{j=0}^{j_{\mathrm{max}}} \, b_{ij} u^i (a/L)^{j} \, ,
\end{equation}
with $b_{i1} = 0$, so as to exclude linear discretisation effects. As previously discussed, we also allow for the possibility $b_{i3} \neq 0$,
in order to check for $O(a/L)^3$ contaminations. The coefficients $b_{10}, b_{20}$ are fixed by perturbation theory; cf. eq.(\ref{eq:sigma-b10b20}).
The complete dataset is globally fit by the above expression and $\sigmaT$ is obtained by the $b_{i0}$ terms of the above
series. 
These $\sigmaT$ results are then used as in 
eqs.~(\ref{Rk}), (\ref{eq:Tbar/TRGI}), and (\ref{eq:T-barT-2k}). We investigate the stability of 
$T^{\mathrm{RGI}}/[ T(\mu_0/2) ]_{\scriptstyle \rm R}$ 
(cf. eq.~(\ref{eq:Tbar/TRGI})) and its errors
for various choices of $n_\sigma$ and $j_{\mathrm{max}}$ in the fit~(\ref{eq:sigmaT_fit2}).

We denote by $[n_\sigma, j_{\mathrm{max}}, (L/a)_{\mathrm{min}}]$ a fit in which series~(\ref{eq:sigmaT_fit2}) is truncated
at $n_\sigma, j_{\mathrm{max}}$ and the data corresponding to $(L/a)_{\mathrm{min}}$ are included
in the fit. In practice we will consider two cases: $(L/a)_{\mathrm{min}}=6$ (i.e. all datapoints included) and $(L/a)_{\mathrm{min}}=8$
(i.e. $L/a=6$ datapoints excluded). In the $\sigmaT$:$u$-{\it by}-$u$ analysis, $n_\sigma =5$ fits were discarded. Also here we only analyse $n_\sigma =3,4$. Fits with $[3, 3, 8], [4, 3, 8]$ are immediately discarded, as their errors increase beyond control. The same is observed for $[3, 4, 8], [4, 4, 8]$; note that for the last two fits we allow for $j_{\mathrm{max}} =4$ (i.e. an $O((a/L)^4)$ discretisation effect), while suppressing the $O((a/L)^3)$ term (i.e. we set $b_{i3}=0$). In conclusion, the $\sigmaT$:{\it global} procedure is applied with the following parameters: 
\begin{itemize}
\item 
with $j_{\mathrm{max}} = 2$, we vary $n_\sigma=3,4$ and  $(L/a)_{\mathrm{min}}=6,8$;
\item
with  $j_{\mathrm{max}} = 3$, we vary $n_\sigma=3,4$ and  fix $(L/a)_{\mathrm{min}}=6$;
\item
with  $j_{\mathrm{max}} = 4$ (and no $(a/L)^3$ allowed!), we vary $n_\sigma=3,4$ and  fix $(L/a)_{\mathrm{min}}=6$.
\end{itemize}
In all these cases, $\chi^2/{\rm d.o.f.}$ is much smaller than 1. We will discuss these results below.

Finally, our fourth procedure, denoted as $\gammaT$:{\it global}, is a variant of $\gammaP$:{\it global} of ref.~\cite{Campos:2018ahf}. It
is based on a combination of eqs.~(\ref{eq:gamma-ubyu}) and (\ref{eq:sigmaT_fit2}):
\begin{equation}
\label{eq:SigmaT=exp+sums} 
\SigmaT \, = \, \exp \Big [ \int_{\sqrt{u}}^{\sqrt{\sigma(u)}} dx \, \frac{\gammaT(x)}{\beta(x)} \Big ] \,\,
+ \,\, \sum_{i=1}^{n_\sigma^\prime}  \sum_{j=2}^{j_{\mathrm{max}}} \, b_{ij} u^i (a/L)^{j} \,\, .
\end{equation}
The Callan-Symanzik coefficients $b_j (i=0, \cdots, 3)$, and the anomalous dimension coefficients $\gammaT^{(j)} (j=0,1)$ are fixed as discussed in the $\gammaT$:$u$-{\it by}-$u$ procedure. The power series of $\gammaT$ terminates at $g^{2n_{\gamma}}$. Each fit is now labelled by $[n_\gamma, n_\sigma^\prime, j_{\mathrm{max}}, (L/a)_{\mathrm{min}}]$.

\begin{figure*}
\includegraphics[width=0.5\textwidth]{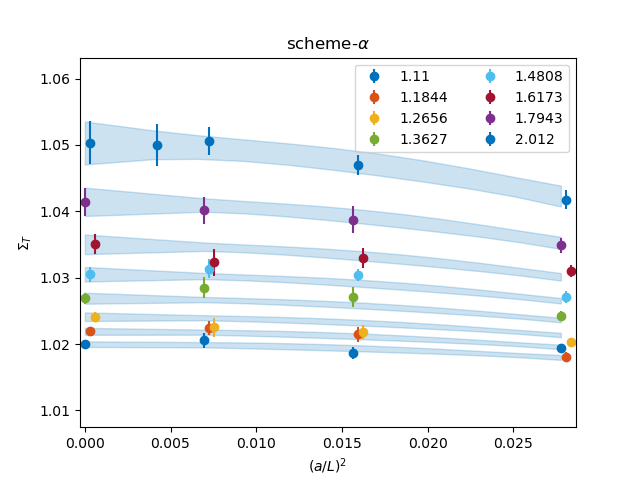}~\includegraphics[width=0.5\textwidth]{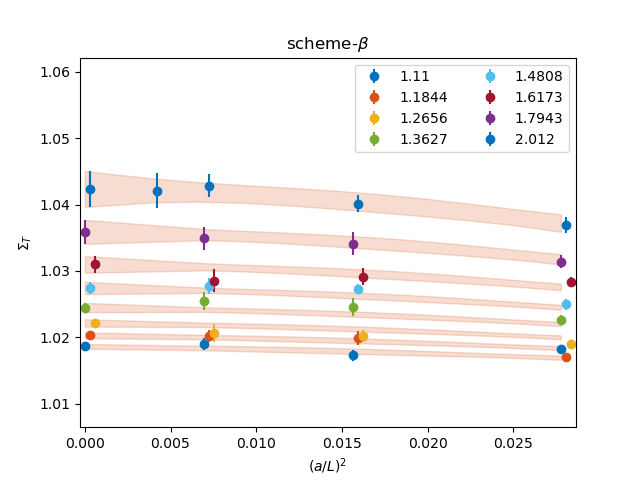}
\includegraphics[width=0.5\textwidth]{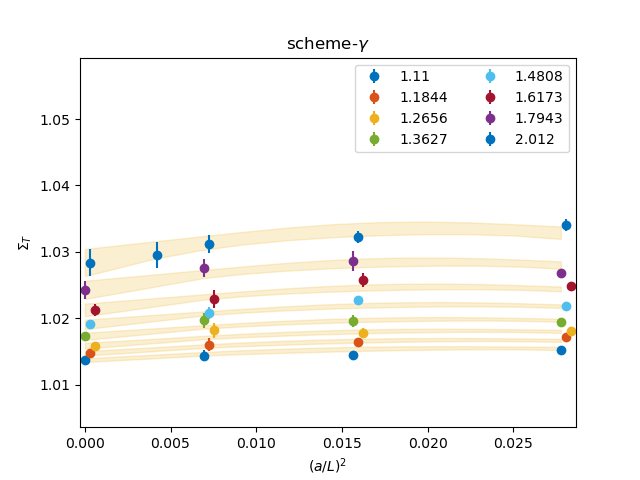}~\includegraphics[width=0.5\textwidth]{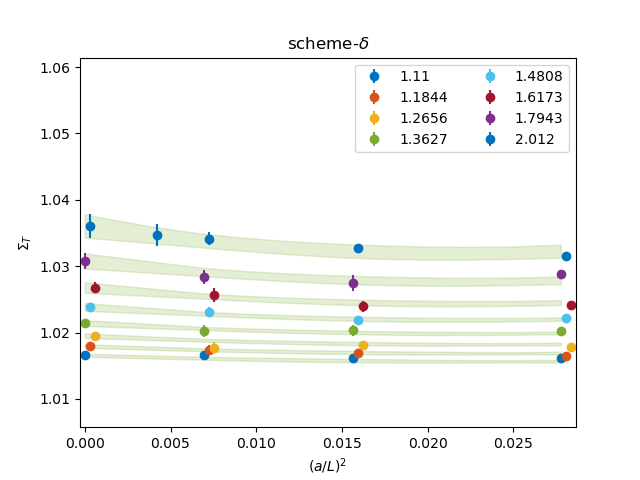}
\caption{Example of the $\gammaT$:{\it global} fit of the step-scaling data $\SigmaT(u, a/L)$, with $[n_\gamma, n_\sigma^\prime, j_{\mathrm{max}}, (L/a)_{\mathrm{min}}] = [2,2,4,6]$, in the high-energy
regime (see Sect.~\ref{sec:results-he} for fit details). The filled circles give the raw data while the bands are
the results returned from the fit at the respective $u$ values. 
The data points of the same colour are at a fixed value of the renormalised squared coupling $u$, indicated in the inset. Some points are slightly shifted in x for better readability.
\label{fig:sigmaT-global-SF}}
\end{figure*}

In the $\gammaT$:$u$-{\it by}-$u$ analysis, $n_\gamma =4,5$ fits were discarded. For this reason, also here we only analyse $n_\gamma =2,3$. We vary  $n_\sigma^\prime$ in the range $[2,4]$, finding that its value has no impact on the results. 
Moreover, when $j_{\mathrm{max}} = 3$ or $ j_{\mathrm{max}} = 4$ (with the cubic power again excluded in the presence of the quartic one) and $(L/a)_{\mathrm{min}} =8$, the errors increase beyond control and these fits are discarded. In conclusion, the third procedure
will be applied with the following parameters:
 \begin{itemize}
\item 
with $j_{\mathrm{max}} = 2$, we vary $n_\gamma=2,3$ and  $(L/a)_{\mathrm{min}}=6,8$;
\item
with  $j_{\mathrm{max}} = 3$, we vary $n_\gamma=2,3$ and  fix $(L/a)_{\mathrm{min}}=6$;
\item
with  $j_{\mathrm{max}} = 4$ (and no $(a/L)^3$ allowed!), we vary $n_\gamma=2,3$ and  fix $(L/a)_{\mathrm{min}}=6$.
\end{itemize}
In all these cases, $\chi^2/{\rm d.o.f.}$ is much smaller than 1. 
Similarly to what we did in the $\gammaT$:$u$-{\it by}-$u$ case, also here we determine
the running factor $T^{RGI}/[T(\mu_0/2)]_{\mathrm R}$ from Eq.~(\ref{eq:T-barT-np}).
The fit results are collected in Appendix~\ref{app:TRGI/Tmu2-he} (Table~\ref{tab:SF_ris}) (also the discarded ones, shown for comparison).
Those of scheme-$\gamma$ tend to be the least stable as we change fitting procedure and/or number of fitting parameters. Henceforth we concentrate on the other three schemes, showing results for the $\gamma$-fit only occasionally for comparison. The results of scheme-$\delta$ tend to have the smallest relative errors. The $\sigmaT$:$u$-{\it by}-$u$ and $\gammaT$:$u$-{\it by}-$u$ results are not sensitive to the choice of $n_\sigma$ and $n_\gamma$. In the two $u$-{\it by}-$u$ procedures we have allowed for discretisation effects which are at most $(a/L)^2$. Comparing these results with the ones from the two {\it global} fit procedures with $j_{\mathrm{max}} =2$, we find that they are always compatible for schemes-$\alpha,\beta,\delta$. For this reason, we will concentrate on the {\it global} fits only. When $j_{\mathrm{max}} =2$, it is preferable to discard the fits with $(L/a)_{\mathrm{min}}=6$ (i.e. the fits that include also points at the largest lattice spacing). An example of the fit quality is shown in Fig.~\ref{fig:sigmaT-global-SF}, for $\gammaT$:{\it global} fits with
$[n_\gamma, n_\sigma^\prime, j_{\mathrm{max}}, (L/a)_{\mathrm{min}}] = [2,2,4,6]$.

The bottom line is that, for each scheme, we prefer results from the two {\it global} procedures, obtained with $[j_{\mathrm{max}},(L/a)_{\mathrm{min}}]=[2,8]$ and $[j_{\mathrm{max}},(L/a)_{\mathrm{min}}]=[3\, \rm{or} \,4,6]$. 
The power series truncations $n_\sigma, n_\sigma^\prime$, and $n_\gamma$ vary as shown in Appendix~\ref{app:TRGI/Tmu2-he} (Table~\ref{tab:SF_ris}). We also display the fit results in Fig.~\ref{fig:TRGI/Tmu2-global}. The comparison of the plots shows that:
\begin{itemize}
\item Results in scheme-$\alpha$ tend to be stable as we vary the number of the fit parameters, with errors that tend to be larger than those in the other schemes.
\item Results in the scheme-$\beta$ are the most stable as we vary the number of the fit parameters, with errors that are smaller that those in scheme-$\alpha$, comparable to those in scheme-$\gamma$, and larger that those in scheme-$\delta$.
\item Results in scheme-$\gamma$ show the largest spread of the mean values; we point out that the spread would appear even bigger had we included the previously discarded fits. Errors are comparable to those in scheme-$\beta$.
\item Results in scheme-$\delta$ have the smallest errors. Although the mean values from different fits are compatible within errors, 
results which include $L/a=6$ datapoints tend to be systematically lower than those which do not. In scheme-$\gamma$, results which include $L/a=6$ datapoints tend to be systematically higher than those which do not; however in this case this systematic effect is less prominent.
\end{itemize}

\subsection{Final values}
\begin{figure*}
\includegraphics[width=0.5\textwidth]{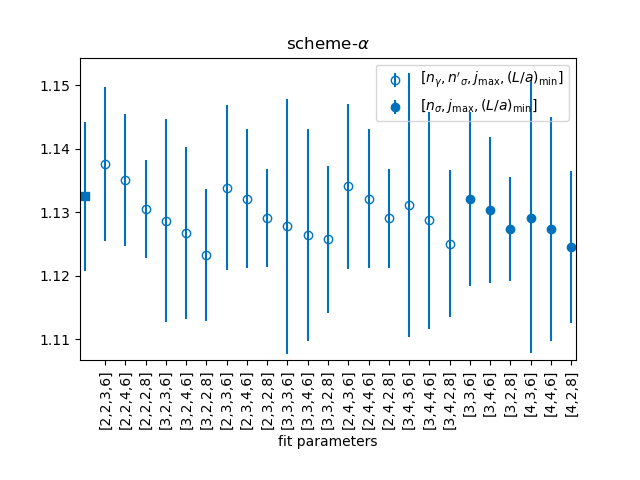}~\includegraphics[width=0.5\textwidth]{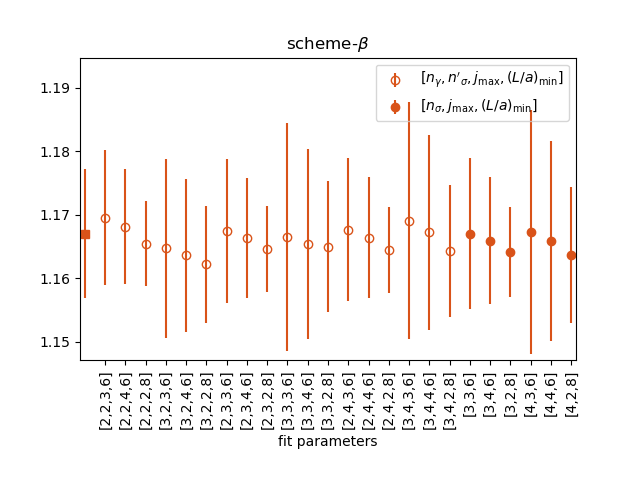}
\includegraphics[width=0.5\textwidth]{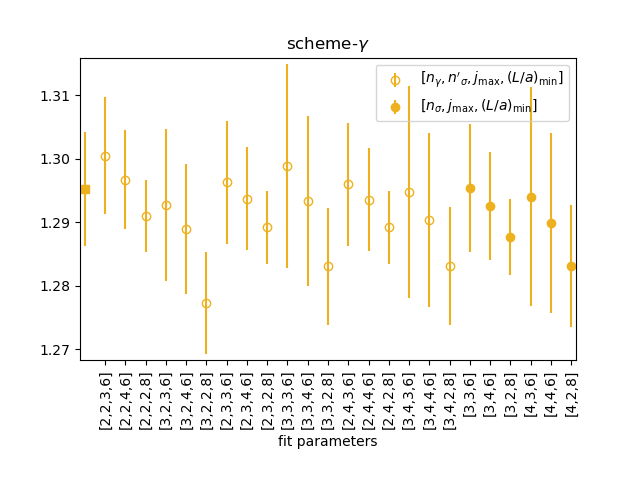}~\includegraphics[width=0.5\textwidth]{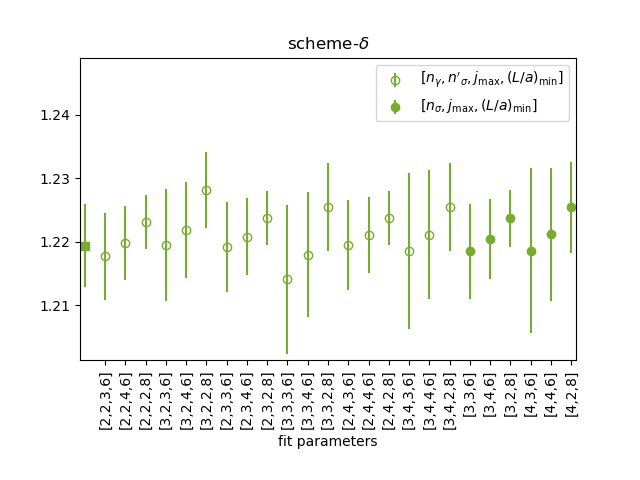}
\caption{Comparison of different fit results for $T^{\mathrm{RGI}}/[ T(\mu_0/2) ]_{\scriptstyle \rm R}$. The open circles
stand for $\gammaT$:{\it global} fits; the close circles  for $\sigmaT$:{\it global}. The fit parameters for each point,
explained in the plot insets, are shown in the abscissa. The filled square is the average of all data by the AIC method~\cite{Jay:2020jkz}.
\label{fig:TRGI/Tmu2-global}}
\end{figure*}

The correlated results of each scheme are model-averaged according to AIC (Aikake Information Criterion); see ref.~\cite{Jay:2020jkz}. 
Given $N_{\mathrm{fits}}$ results $\langle O \rangle_{n}$ from several fits ($n = 1, \cdots N_{\mathrm{fits}}$),
the AIC average of an observable $O$ is given by
\begin{equation}
\langle O \rangle=\underset{n=1}{\overset{N_\mathrm{fits}}{\sum}}w_{n}^{\mathrm{AIC}} \langle O \rangle_{n} \,\, ,
\label{eq:AIC average}
\end{equation}
where for the $n^{\mathrm{th}}$ fit the AIC weight is:
\begin{equation}
w_{n}^{\mathrm{AIC}}\sim N\exp\Big[-\frac{1}{2}(\chi_{n}^{2}+2N_{n}^{\mathrm{param}}+2N_{n}^{\mathrm{cut}})\Big] \,\, ,
\label{eq:w AIC}
\end{equation} 
with $N^{\mathrm{param}}$ the number of fit parameters and $N^{\mathrm{cut}}$ the number of datapoints excluded in the fit.
Inverse-variance weighted averaging produces similar results with slightly smaller errors. The AIC results from the 24 datapoints of Fig.~\ref{fig:TRGI/Tmu2-global} are:
\begin{equation}
\dfrac{T^{\mathrm{RGI}}}{[T(\mu_{0}/2)]_{\mathrm R}}=\begin{cases}
1.1325(117) & {\rm scheme-}\alpha\\
1.1671(102) & {\rm scheme-}\beta\\
1.2952(90) & {\rm scheme-}\gamma\\
1.2194(65)& {\rm scheme-}\delta
\end{cases}
\label{eq:final-he}
\end{equation}
In conclusion our favourite choice in the high energy range is scheme-$\beta$ because of its stability. Second acceptable options are scheme-$\alpha$, with larger, more conservative errors, and scheme-$\delta$, with smaller errors. At this stage of the analysis, scheme-$\gamma$ appears to behave worse than the others, on account of the larger spread of results displayed in Fig.~\ref{fig:TRGI/Tmu2-global}.

\section{Tensor running at low energies}
\label{sec:results-le}
Having computed the running factor to convert the renormalised mass at
the scale $\mu_0/2$ to the renormalisation group invariant mass, we now
turn to the computation of the running factor in the low-energy (GF) regime.
Recall that~\cite{DallaBrida:2016kgh}
\begin{equation} 
\label{uGF-mu0/2}
u_{\mathrm{GF}}(\mu_0/2) \, = \, \gbar^2_{\mathrm{GF}}(\mu_0/2) \, = \, 2.6723(64) \,\, .
\end{equation}

Whereas in the high-energy (SF) regime, we had lattices of extent
$L/a=6,8,12$ (and $L/a=16$ at $u=2.0120$), in the low-energy (GF) regime our lattices have
extent $L/a=8,12,16$. 
Our full results for the $Z_{\mathrm T}$'s and $\SigmaT$'s, for
all values of $u$, $L/a$, $\beta$ and $\hopc$, in all four schemes, are gathered in Appendix~\ref{app:Z-Sigma-le}. Note that $\SigmaT^{\mathrm{sub}}$'s are not useful in the non-perturbative energy range under consideration.
The values of $\gbar^2_{\mathrm{GF}}(g_0^2,a/L)$ have been computed following the procedure described in ref.~\cite{DallaBrida:2016kgh}. Since $\gbar^2_{\mathrm{GF}}$ and $\SigmaT$ are computed on the same ensembles, they are correlated observables. Correlations have been taken into account employing the methods and package of refs.~\cite{Ramos:2018vgu,Ramos:2020scv}.
The low energy hadronic scale $\mu_{\text{had}}$ is defined by
\begin{equation}
u(\mu_{\text{had}}) = 9.25 \,,
\end{equation}
corresponding to a physical scale 
$\mu_{\text{had}} = 233(8)$~MeV~\cite{Bruno:2017gxd}. Since the ratio of the
switching scale $\mu_0/2 \sim 2$~GeV to the hadronic scale $\mu_{\text{had}}$ is 
not a power of two, it is inconvenient to carry out the analysis in terms of the step
scaling function $\sigmaT(u)$, which only expresses the tensor running between
consecutive scales $\mu$ and $\mu/2$. Thus, we rule out $\sigmaT$:$u$-{\it by}-$u$
and  $\sigmaT$:{\it global} fits. Moreover, in the GF regime the bare couplings $g_0^2$ have {\it not} been 
precisely tuned to a fixed value $\gbar^2_{\mathrm{GF}}(g_0^2, a/L)$, as $a/L$ is varied. This implies
that $\gammaT$:$u$-{\it by}-$u$ fits are also not to be implemented.
It is therefore preferable to apply the process $\gammaT$:{\it global},
adapted to the current low-energy regime; i.e. without input from perturbation theory.
This means that now eq.~(\ref{eq:SigmaT=exp+sums}) is written as
\begin{eqnarray}
\label{eq:SigmaT=exp+sums2} 
\SigmaT(u,a/L) \, &=& \, \exp \Big [ \int_{\sqrt{u}}^{\sqrt{\sigma(u)}} dx \, f(x) \Big ] \nonumber \\
&+& \,\, \sum_{i=1}^{n_\sigma^\prime}  \sum_{j=2}^{j_{\mathrm{max}}} \, c_{ij} u^i (a/L)^{j} \,\, .
\end{eqnarray}
where $f(g) \equiv \gammaT(g)/\beta(g)$ is expressed as a truncated series
\begin{equation}
\label{eq:f-series}
f(g) \,\, = \,\, \frac{1}{g}\sum_{k=0}^{n_{\mathrm{f}}} f_{k} \, g^{2k} \,\,.
\end{equation}
The coefficients $f_k$ are determined by fitting all datapoints $\SigmaT(u,a/L)$ with eq.~(\ref{eq:SigmaT=exp+sums2}).
The RG-running for the tensor in the low-energy range is then given by
\begin{equation}
\label{eq:Tmu0half/Tmuhad}
\frac{[ T(\mu_0/2) ]_{\scriptstyle \rm R}}{[ T(\mu_{\mathrm{had}}) ]_{\scriptstyle \rm R}} \,\, = \,\, 
\exp \Big [ \int_{\gbar(\mu_0/2)}^{\gbar(\mu_{\mathrm{had}})} dx \, f(x) \Big ] \,\, .
\end{equation}
The tensor anomalous dimension is obtained as the product $\gammaT(g) = f(g) \beta_{\mathrm{GF}}(g)$, with $f(g)$
known from the fit (\ref{eq:SigmaT=exp+sums2}),(\ref{eq:f-series}) and the Callan-Symanzik function in the low-energy (GF)
regime, parametrised as in ref.~\cite{DallaBrida:2016kgh}:
\begin{equation}
\label{eq:betafunct-GF}
\beta_{\mathrm{GF}}(g) \,\, = \,\, - \dfrac{g^3}{\sum_{k=0}^{2} p_k g^{2k}} \,\, ,
\end{equation}
with
\begin{alignat}{4}
    p_0  \, &=& \, 16.63 &\pm 0.61 \,\, , \nonumber \\
    p_1  \, &=& \, -0.05 &\pm 0.20 \,\, , \label{eq:betafunct-GFp} \\
    p_2  \, &=& \, 0.008 &\pm 0.016 \,\, . \nonumber
 \end{alignat}
(The results for $p_0,  p_1$, and $p_2$, having been computed afresh on configuration ensembles which differ - by a few configurations - from those of  ref.~\cite{DallaBrida:2016kgh}, are compatible but not identical to those of refs.~\cite{DallaBrida:2016kgh} and~\cite{Campos:2018ahf}.)

\begin{figure*}
\includegraphics[width=0.5\textwidth]{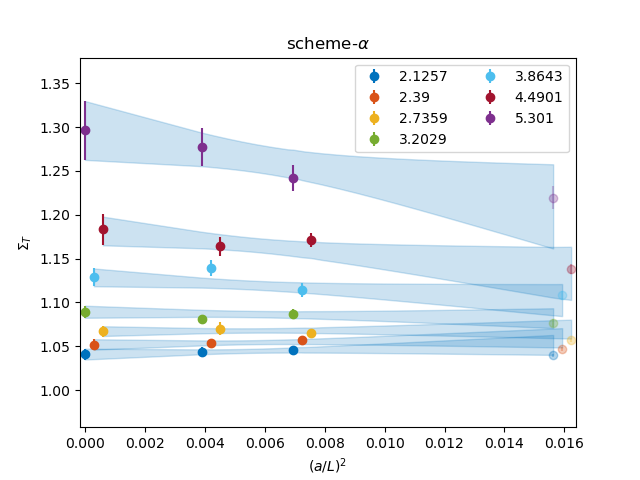}~\includegraphics[width=0.5\textwidth]{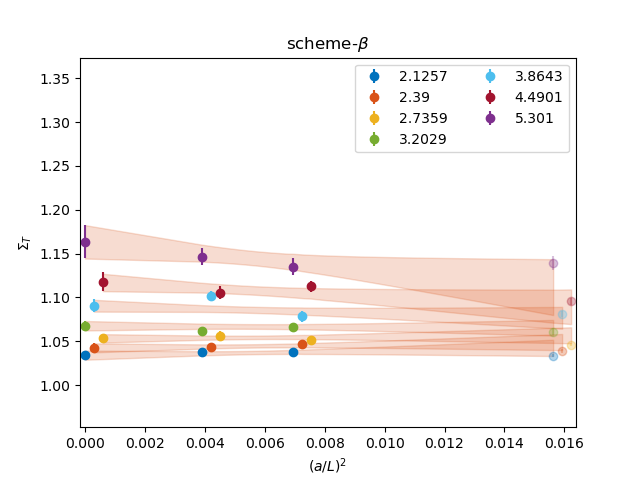}
\includegraphics[width=0.5\textwidth]{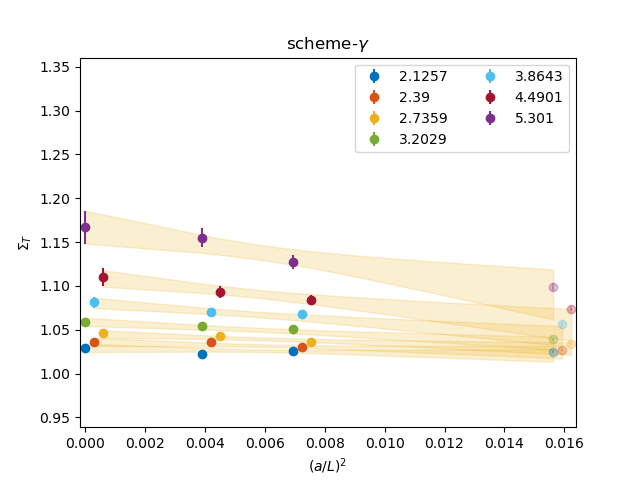}~\includegraphics[width=0.5\textwidth]{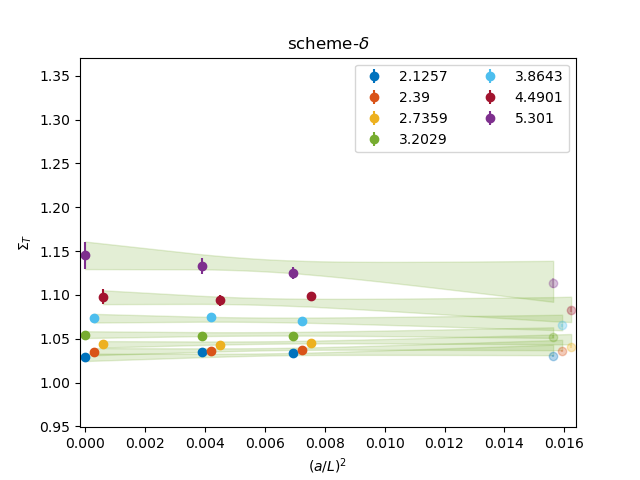}
\caption{Example of the $\gammaT$:{\it global} fit of the step-scaling data $\SigmaT(u, a/L)$, with $[n_\gamma, n_\sigma^\prime, j_{\mathrm{max}}, (L/a)_{\mathrm{min}}] = [2,2,2,12]$, in the low-energy regime (see Sect.~\ref{sec:results-le} for fit details). The filled circles give the raw data while the bands are the results returned from the fit at the respective $u$ values. The data points of the same colour are at {\it approximately} the values of $u = \gbar^2(L/a=16)$ indicated in the inset. Some points are slightly shifted in x for better readability.
\label{fig:sigmaT-global-GF}}
\end{figure*}

We denote by $[n_{\mathrm{f}}, n_\sigma^\prime, j_{\mathrm{max}}, (L/a)_{\mathrm{min}}]$ a fit in which series~(\ref{eq:f-series}) is truncated at $n_{\mathrm{f}}$, the double series of eq.~(\ref{eq:SigmaT=exp+sums2}) is truncated at  $n_\sigma^\prime, j_{\mathrm{max}}$ and the data corresponding to $(L/a)_{\mathrm{min}}$ are included in the fit. In practice we will consider two cases: $(L/a)_{\mathrm{min}}=8$ (i.e. all datapoints included) and $(L/a)_{\mathrm{min}}=12$ (i.e. $L/a=8$ datapoints excluded).
Whenever we allow for $j_{\mathrm{max}} =4$ (i.e. an $O((a/L)^4)$ discretisation effect), we suppress the $O((a/L)^3)$ term (i.e. we set $b_{i3}=0$). In conclusion, fits are performed with the following parameters: 
\begin{itemize}
\item 
with $j_{\mathrm{max}} = 2$, we vary $n_{\mathrm{f}} =2,3$, $n_\sigma^\prime = 2,3,4$ and  $(L/a)_{\mathrm{min}}=8,12$;
\item
with  $j_{\mathrm{max}} = 3$, we vary $n_{\mathrm{f}} =2,3$, $n_\sigma^\prime = 2,3,4$ and  fix $(L/a)_{\mathrm{min}}=8$;
\item
with  $j_{\mathrm{max}} = 4$ (and $(a/L)^3$ not allowed!), we vary $n_{\mathrm{f}} =2,3$, $n_\sigma^\prime = 2,3,4$ and  fix $(L/a)_{\mathrm{min}}=8$.
\end{itemize}
All $\alpha$-scheme results have $\chi^2/{\rm d.o.f.} \in [0.3,0.4]$. For the $\beta$-scheme, $\chi^2/{\rm d.o.f.} \in [0.4,0.6]$. For the $\gamma$-scheme, $\chi^2/{\rm d.o.f.} \in [0.2,0.5]$, while for the $\delta$-scheme, $\chi^2/{\rm d.o.f.} \in [0.1,0.3]$.
An example of the fit quality is shown in Fig.~\ref{fig:sigmaT-global-GF}, for $[n_\gamma, n_\sigma^\prime, j_{\mathrm{max}}, (L/a)_{\mathrm{min}}] = [2,2,2,12]$. It is a representative choice, since this fit has a big weight on the final average proposed at the end of this section; see Table~\ref{tab:GF_ris} of Appendix~\ref{app:Tmu2/Tmuhad-le}.
More details about this fit will be given in Appendix ~\ref{app:pert-coeffs}.

The results for $[ T(\mu_0/2) ]_{\scriptstyle \rm R}/[ T(\mu_{\mathrm{had}}) ]_{\scriptstyle \rm R}$ are summarised in Appendix~\ref{app:Tmu2/Tmuhad-le} and Fig.~\ref{fig:TRGI/Tmu2-muhad}. 
\begin{figure*}
\includegraphics[width=0.5\textwidth]{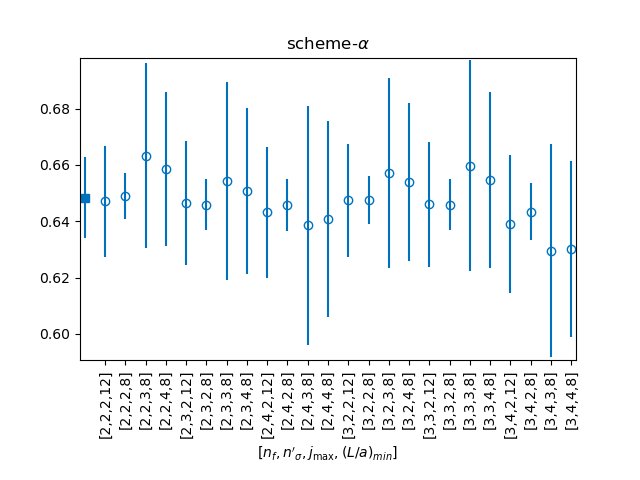}~\includegraphics[width=0.5\textwidth]{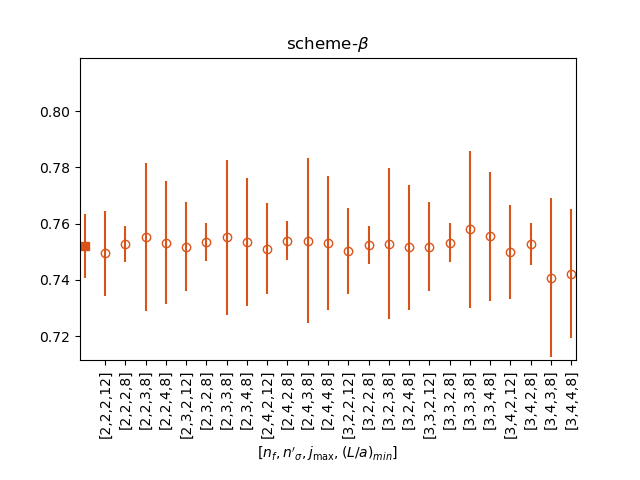}
\includegraphics[width=0.5\textwidth]{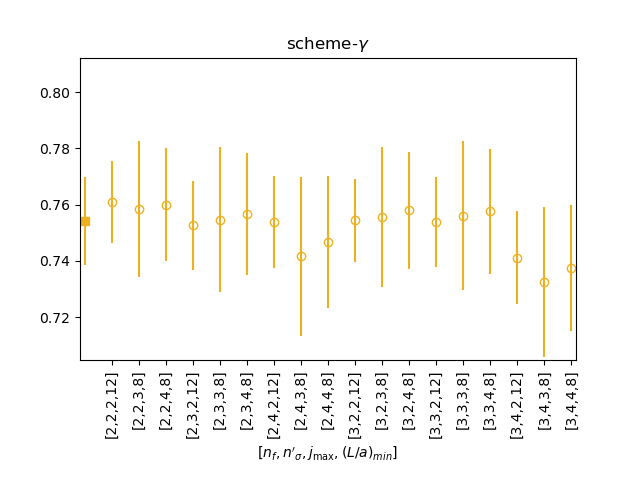}~\includegraphics[width=0.5\textwidth]{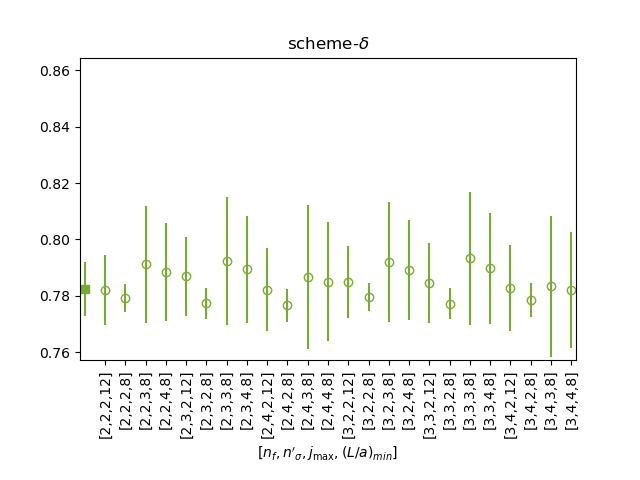}
\caption{Comparison of different fit results for $[T(\mu_0/2)]_{\scriptstyle \rm R}/[T(\mu_{\mathrm{had}})]_{\scriptstyle \rm R}$. 
The fit parameters for each point are shown in the abscissa. The filled square is the average of all data by the AIC method~\cite{Jay:2020jkz}. For each scheme we show only the fits that are included in the averages.
\label{fig:TRGI/Tmu2-muhad}}
\end{figure*}
We see that:
\begin{itemize}
\item
For each scheme, when only $(a/L)^2$ discretisation effects are allowed in eq.~(\ref{eq:SigmaT=exp+sums2}), results are stable as we vary $n_{\mathrm{f}}$ and $n_\sigma^\prime$. Unsurprisingly, the errors for $(L/a)_{\mathrm{min}}=12$ fits are larger than those for $(L/a)_{\mathrm{min}}=8$ fits.
\item
Predictably, the introduction of either $(a/L)^3$ or $(a/L)^4$ discretisation effects also results in an increase of the error in all schemes.
\item
Upon comparing results across schemes, we see that scheme-$\delta$ is the most accurate and scheme-$\alpha$ is the least accurate, in the sense that they have the smallest and biggest errors respectively.
\item
Similarly to what happens in the high-energy regime, schemes -$\alpha$, -$\beta$, and -$\delta$ show remarkable stability with the variation of the fit parameters $[n_{\mathrm{f}}, n_\sigma^\prime, j_{\mathrm{max}}, (L/a)_{\mathrm{min}}]$. For what concerns scheme-$\gamma$, we find that results with $[n_{\mathrm{f}}, n_\sigma^\prime, j_{\mathrm{max}}, (L/a)_{\mathrm{min}}]$=[*,*,2,8] are systematically shifted towards the top end of the error bars of all other results.
 For this reason, these scheme-$\gamma$ results are excluded for the final average. They are not shown in Fig.~\ref{fig:TRGI/Tmu2-muhad} but they can be found in Appendix~\ref{app:Tmu2/Tmuhad-le}. For the other schemes all the fits are included in the final estimate.
\end{itemize}
Combining the above considerations we conclude that our preferred scheme is $\beta$, with $\gamma$ and $\delta$ second best. On account of its relatively bigger errors, scheme$-\alpha$  is also an acceptable conservative option.

The correlated results of each scheme are model-averaged according to AIC~\cite{Jay:2020jkz}. Inverse-variance weighted averaging produces similar results with slightly smaller errors. The AIC results are:
\begin{equation}
\frac{[ T(\mu_0/2) ]_{\scriptstyle \rm R}}{[ T(\mu_{\mathrm{had}})]_{\scriptstyle \rm R}} = 
\begin{cases}
 0.648(14) & {\rm scheme}-\alpha \\
 0.752(11) & {\rm scheme}-\beta \\
 0.754(16) & {\rm scheme}-\gamma \\
 0.7825(96) & {\rm scheme}-\delta
\end{cases}
\label{eq:final-le}
\end{equation}

\section{Tensor running at all energy scales}
\label{sec:results-he-le}
\begin{figure*}
\includegraphics[width=0.5\textwidth]{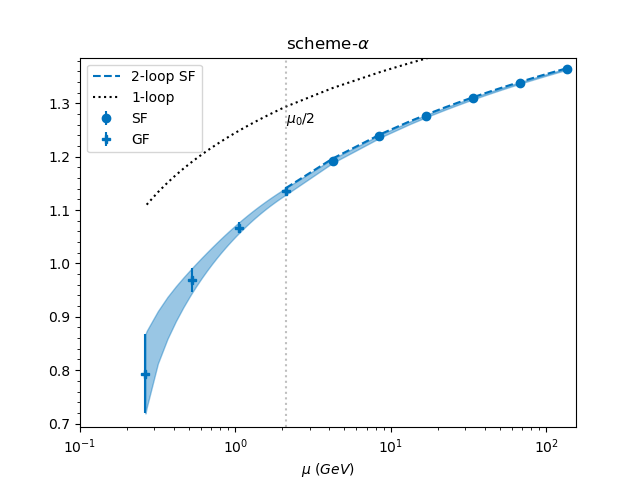}~\includegraphics[width=0.5\textwidth]{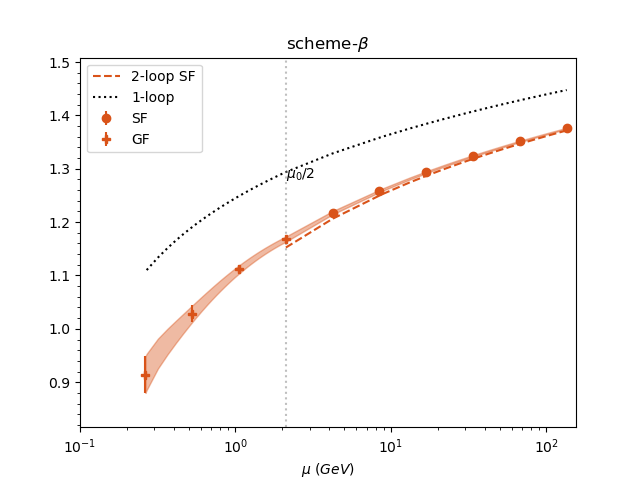}
\includegraphics[width=0.5\textwidth]{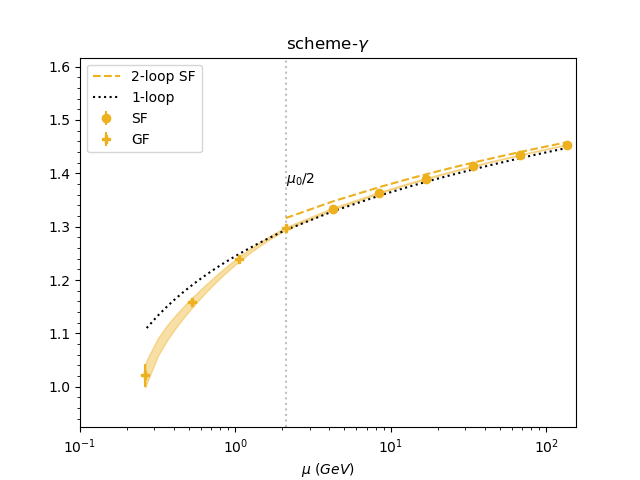}~\includegraphics[width=0.5\textwidth]{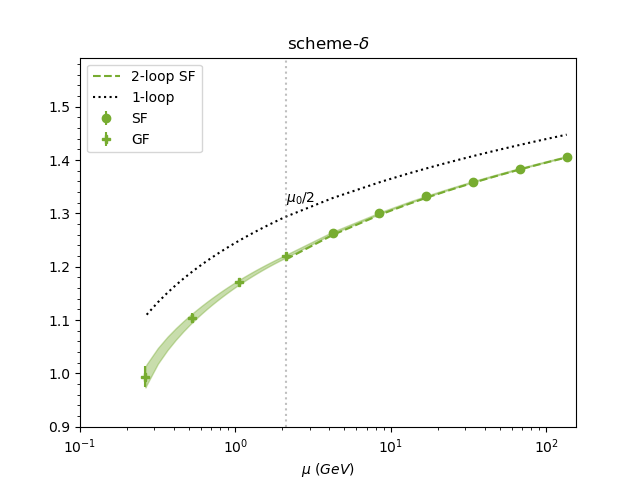}
\caption{Example of the running of the tensor current $T^{\mathrm{RGI}}/[ T(\mu) ]_{\scriptstyle \rm R}$ in the four schemes, as a function of $\mu$. In the high energy regime we show the $\gammaT$:{\it global} fit with $[n_\gamma, n_\sigma^\prime, j_{\mathrm{max}}, (L/a)_{\mathrm{min}}] = [2,2,4,6]$; in the low energy one, the $\gammaT$:{\it global} fit  with $[n_\gamma, n_\sigma^\prime, j_{\mathrm{max}}, (L/a)_{\mathrm{min}}] = [2,2,2,12]$. The non-perturbative results are
shown by a continuous band and datapoints . The vertical dotted line indicates the switching scale $\mu_0/2 \sim 2$GeV, which separates the high- and low-energy ranges. The universal LO perturbative prediction (``1-loop") is shown as a dotted line. The NLO perturbative prediction in the SF scheme (``2-loop SF") is shown as a dashed line. See main text for details.
}
\label{fig:op_runningl}
\end{figure*}

We can now obtain $T^{\mathrm{RGI}}/[ T(\mu_{\mathrm{had}}) ]_{\scriptstyle \rm R}$ from the product of the high-energy factor $T^{\mathrm{RGI}}/[T(\mu_0/2) ]_{\scriptstyle \rm R}$ and the low-energy one $[T(\mu_0/2) ]_{\scriptstyle \rm R}/[ T(\mu_{\mathrm{had}})]_{\scriptstyle \rm R}$, computed in Sections~\ref{sec:results-he} and \ref{sec:results-le} respectively. 
Before computing it in Section~\ref{sec:results-final}, we display in the present Section some graphic examples of the quality of the RG-running we have achieved for the tensor composite field. In Fig.~\ref{fig:op_runningl} we show an example of $T^{\mathrm{RGI}}/[ T(\mu)]_{\scriptstyle \rm R}$ for the four schemes $\alpha, \beta, \gamma, \delta$, as a function of the energy scale $\mu$ in physical units. It is obtained by using the results of ref.~\cite{Bruno:2017gxd}: in the notation of that work, a reference scale
$\mu^\ast_{\mathrm{ref}} = 478(7)$~MeV is related to a hadronic scale $\mu_{\mathrm{had,1}} = \mu^\ast_{\mathrm{ref}} /2.428(18)$, which corresponds to $u_{\mathrm{had,1}} \equiv \gbar^2_{\mathrm{GF}}(\mu_{\mathrm{had,1}} ) = 11.31$. From the pair 
$[\mu_{\mathrm{had,1}},u_{\mathrm{had,1}}]$
and the Callan-Symanzik $\beta$-function of our eqs.~(\ref{eq:betafunct-GF}),(\ref{eq:betafunct-GFp}) we obtain $\mu$ in physical units for any other $\gbar^2_{\mathrm{GF}}(\mu)$. For example, from eq.~(\ref{uGF-mu0/2}) we have $\mu_0/2 = 2.109(57)$~GeV.
Finally, eq.~(\ref{uSF-mu0/2}) for $\gbar^2_{\mathrm{SF}}(\mu_0/2)$ and the RG-running of the $\beta$-function of eqs.~(\ref{eq:beta}), (\ref{eq:betaPTa})-(\ref{eq:betaPTd}) gives us $\mu$ in physical units in the high-energy range.
These are the values of the abscissae of Figs.~\ref{fig:op_runningl}.

In the high-energy (SF) range, the curve and datapoints of Fig.~\ref{fig:op_runningl} correspond to fit parameters $[n_\gamma , n^\prime_\sigma, j_{\mathrm{max}} , (L/a)_{\mathrm{min}}] =[2,2,3,6]$, while in the low-energy (GF) range they correspond to fit parameters $[n_{\mathrm{f}}, n_\sigma^\prime, j_{\mathrm{max}}, (L/a)_{\mathrm{min}}] = [2,2,2,12]$. The continuum  extrapolation of these fits has been shown in Figs.~\ref{fig:sigmaT-global-SF} and \ref{fig:sigmaT-global-GF}. They are good representatives of the final results: as seen from the relevant entries of Tables~\ref{tab:SF_ris} (Appendix~\ref{app:TRGI/Tmu2-he})
and \ref{tab:GF_ris} (Appendix~\ref{app:Tmu2/Tmuhad-le}), these fits have high AIC weights and the results they produce for the running factors agree with those from all-fit-averages, both in the values and in the entity of the errors.

For high energies, once $\gammaT/\beta$ is extracted from the fit, the quantity $T^{\mathrm{RGI}}/[ T(\mu_0/2)]_{\scriptstyle \rm R}$ is known from eq.~(\ref{eq:T-barT-np}), whilst $[ T(\mu)]_{\scriptstyle \rm R}/[ T(\mu_0/2)]_{\scriptstyle \rm R}$ is known from eq.~(\ref{eq:Rk2}), computed at any scale $\mu$, rather than at $2^k \mu_0$. These two ratios are combined to give the non-perturbative continuous band with errors, shown in Fig.~\ref{fig:op_runningl} for scales $\mu \ge \mu_0/2$. In the low-energy range, again $\gammaT/\beta$ is extracted from the fit and the ratio $[ T(\mu_0/2)]_{\scriptstyle \rm R}/[ T(\mu)]_{\scriptstyle \rm R}$ can be worked out as in eq.~(\ref{eq:Tmu0half/Tmuhad}), for all scales $\mu$, rather than just for $\mu_{\mathrm{had}}$. This ratio, multiplied by the high-energy factor $T^{\mathrm{RGI}}/[ T(\mu_0/2)]_{\scriptstyle \rm R}$, gives the non-perturbative continuous band with errors at low energies $\mu < \mu_0/2$. We have also plotted solid points on top of the fit bands
at regular intervals of scales differing by a factor of 2; they are computed in equivalent fashion. 

The universal LO perturbative prediction, shown as a dotted line in Fig.~\ref{fig:op_runningl} and denoted as ``1-loop", is obtained from the coefficients $b_0, b_1$ and $\gammaT^{(0)}$. The NLO perturbative prediction in the SF scheme, shown as a dashed line and denoted as ``2-loop SF", is obtained from the coefficients $b_0, b_1, b_2, b_3^{\mathrm{eff}}$ and $\gammaT^{(0)},\gammaT^{(1)}$. See eqs.~(\ref{eq:betaPTa}-\ref{eq:betaPTd},\ref{eq:gamma0},\ref{eq:gamma1}) for the values of the perturbative coefficients. We see that for schemes$-\alpha$, $-\beta$ and $-\delta$ the 1-loop perturbative prediction clearly deviates from the non-perturbative results, while the 2-loop one follows them quite closely at high energies. In scheme-$\gamma$, while the 1-loop curve stays close to the non-perturbative band down to energies $\mu/\Lambda \sim 1$, the 2-loop one deviates halfway down the high-energy range. The impression is that perturbation theory performs poorly for this scheme and this is enforced by
the fact that $b_{30}$ for scheme$-\gamma$ is bigger than the others (cf. eq.~(\ref{eq:b30})). 
Also note that in scheme-$\gamma$ the coefficient $\gammaT^{(1)}$ is much smaller than those of the other schemes and of the opposite sign (cf. eq.~(\ref{eq:gamma1})).

A similar study for $\NF=0,2$ is presented in Figs.~6 and 10 of ref.~\cite{Pena:2017hct}. 

\begin{figure*}
\includegraphics[width=0.5\textwidth]{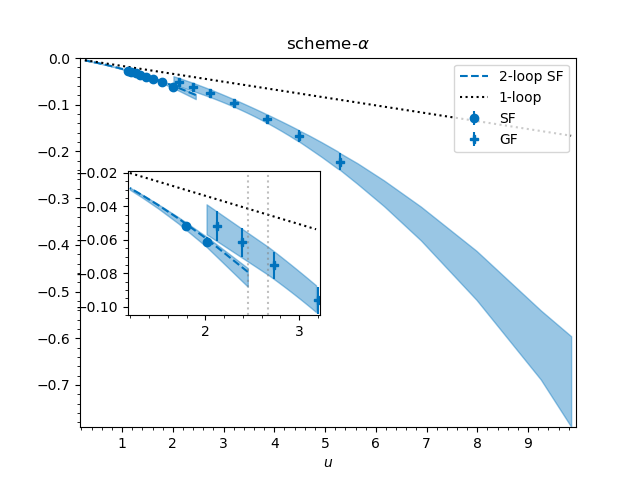}~\includegraphics[width=0.5\textwidth]{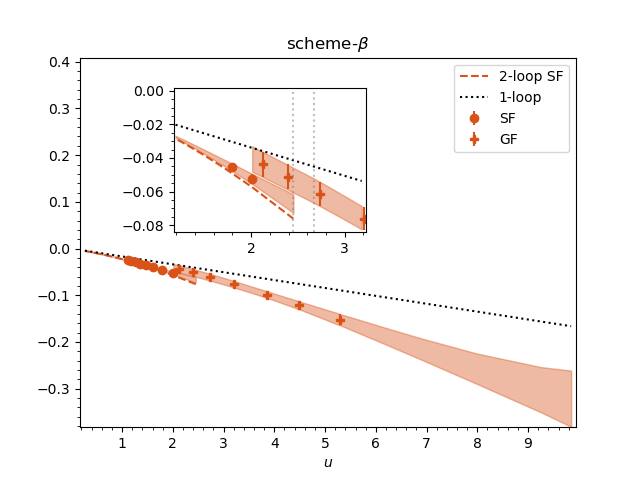}
\includegraphics[width=0.5\textwidth]{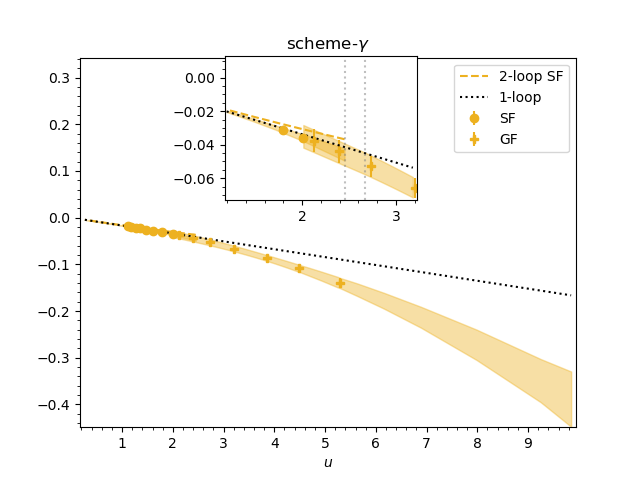}~\includegraphics[width=0.5\textwidth]{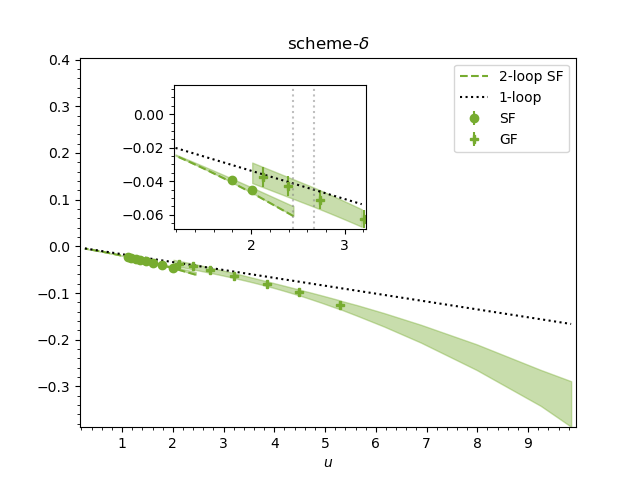}
\caption{Example of the tensor anomalous dimension $\gammaT$ in the four schemes, as a function of the squared renormalised coupling $u$.  In the high energy regime we show the $\gammaT$:{\it global} fit with $[n_\gamma, n_\sigma^\prime, j_{\mathrm{max}}, (L/a)_{\mathrm{min}}] = [2,2,4,6]$; in the low energy one, the $\gammaT$:{\it global} fit  with $[n_\gamma, n_\sigma^\prime, j_{\mathrm{max}}, (L/a)_{\mathrm{min}}] = [2,2,2,12]$. The non-perturbative results are shown as bands and datapoints (dots for high-energies (SF) and crosses for low-energies (GF)). The LO perturbative prediction (``1-loop") is shown as a dotted line and the NLO one (``2-loop") as a dashed line in the high-energy range. The insets zoom on the discontinuity in the bands, due to the fact that in the high- and low-energy ranges different definitions are used for the renormalised coupling (SF and GF respectively). The vertical dotted lines in the insets correspond to $u(\mu_0/2)$ in the SF and GF schemes  (left and right line respectively) . See main text for details.}
\label{fig:gamma_all}
\end{figure*}

In Fig.~\ref{fig:gamma_all} we show the variation of the tensor anomalous dimension with the renormalised coupling squared, for the same fit parameters as those used in Fig.~\ref{fig:op_runningl}. The bands, datapoints and perturbative curves have also been obtained like those of Fig.~\ref{fig:op_runningl}.
The insets in Fig.~\ref{fig:gamma_all} show that at scale $\mu_0/2$ (which corresponds to different $u$ values in the SF and GF schemes; see eqs.~(\ref{uSF-mu0/2}) and (\ref{uGF-mu0/2})) the SF and GF estimates of $\gammaT(u)$ agree within errors, in accordance to theoretical expectations.
Figure 8 of ref.~\cite{Chimirri:2023ovl} is a similar plot for the $\NF=3$ unitary setup.

\section{Switching schemes and final RG-running results}
\label{sec:results-final}
The final step of our analysis is the combination of the high- and low-energy running factors computed in Sections~\ref{sec:results-he} and \ref{sec:results-le} in order to obtain $T^{\mathrm{RGI}}/[ T(\mu_{\mathrm{had}}) ]_{\scriptstyle \rm R}$. As discussed in the previous Section, this is apparently a straightforward product of the high-energy factor $T^{\mathrm{RGI}}/[T(\mu_0/2) ]_{\scriptstyle \rm R}$ and the low-energy one $[T(\mu_0/2) ]_{\scriptstyle \rm R}/[ T(\mu_{\mathrm{had}})]_{\scriptstyle \rm R}$, {\it both computed in the same renormalisation scheme}. (Recall that at the switching scale the renormalisation condition of the coupling is changed from SF to GF, but the condition for the tensor operator remains the same).
We have seen however that the four renormalisation schemes employed have not given equally reliable results in both energy ranges. Thus it may be advantageous to combine high- and low-energy results in different schemes, for better accuracy. This is possible once we have computed the switching factor between two schemes at energy $\mu_0/2$. It is defined and computed as follows:
\begin{eqnarray}
&&\chi_{\mathrm T}^{\mathrm{s1,s2}}(\mu_0/2) \equiv \dfrac{[T(\mu_0/2,\gbar^2_{\mathrm{SF}}(\mu_0/2))]^{\mathrm{s1}}_{\scriptstyle \rm R}}{[T(\mu_0/2,\gbar^2_{\mathrm{SF}}(\mu_0/2))]^{\mathrm{s2}}_{\scriptstyle \rm R}} 
\nonumber \\
&=& \lim_{a/(2L) \rightarrow 0} \,\, \dfrac{\ZT^{\mathrm{s1}}(g_0^2,2L/a)}{\ZT^{\mathrm{s2}}(g_0^2,2L/a)} \Bigg \vert_{\gbar^2(L) = u} \,\,,
\label{eq:DeltaZ}
\end{eqnarray}
where $\mathrm{s1},\mathrm{s2}$ are two renormalisation schemes, e.g. $[\mathrm{s1},\mathrm{s2}] = [\alpha,\delta]$. Once the above factor is known,
it can be combined with the running factor of the high-energy region in scheme $\mathrm{s1}$ and the one of the low-energy region in scheme $\mathrm{s2}$, giving the total running factor in scheme $\mathrm{s2}$:
\begin{equation}
\label{eq:total-running-s2}
\dfrac{T^{\mathrm{RGI}}}{[ T(\mu_{\mathrm{had}})]^{\mathrm{s2}}_{\scriptstyle \rm R}} \,\, = \,\,
 \dfrac{T^{\mathrm{RGI}}}{[ T(\mu_0/2)]^{\mathrm{s1}}_{\scriptstyle \rm R}} \, \chi_{\mathrm T}^{\mathrm{s1,s2}}(\mu_0/2) \,
 \dfrac{[T(\mu_0/2)]^{\mathrm{s2}}_{\scriptstyle \rm R}}{[ T(\mu_{\mathrm{had}})]^{\mathrm{s2}}_{\scriptstyle \rm R}} \,\,.
\end{equation}

The first and third factor in the above expression are given in eqs.~(\ref{eq:final-he}) and (\ref{eq:final-le}) respectively. The second factor is computed from eq.~(\ref{eq:DeltaZ}), in the high-energy (SF) range: for the bare couplings corresponding to $\gbar^2(L) = u = 2.012$ (i.e. at energy $\mu_0 = 1/L \simeq 4$~GeV), we know the renormalisation parameters $\ZT(g_0^2,2L/a)$ at the ``doubled" lattices $2L/a=12,16,24,32$ (i.e. at energy $\mu_0/2 = 1/(2L) \simeq 2$~GeV). The ratios of these $\ZT$ computed in two different schemes, extrapolated to the continuum limit, give us the desired switching factor.
 
The datapoints are fit by the expression
\begin{equation}
\chi_{\mathrm T}^{\mathrm{s1,s2}}(\mu_0/2) \,\, = \,\, 1 + a_0 + a_2 \big(\dfrac{a}{2L}\big)^2 + a_{n_\mathrm{max}} \big(\dfrac{a}{2L}\big)^{n_\mathrm{max}} \,\, ,
\end{equation} 
with $n_\mathrm{max} =2,3,4$. By this we mean that when $n_\mathrm{max}=2$ the expression contains neither cubic nor quartic terms;
when $n_\mathrm{max}=3$ or $n_\mathrm{max}=4$, then either the cubic or the quartic term is added to the quadratic one.
The continuum limit of the switching factor is given by $\chi_{\mathrm T}^{\mathrm{s1,s2}} = 1 + a_0$. The results of our fits are collected in Table~\ref{tab:DeltaZ}. We see that they are insensitive to the choice of $n_\mathrm{max}$. Moreover, $\chi^2/{\mathrm{ d.o.f.}} \lesssim 1$ in all but a couple of cases (where it is less than 1.5, anyway). For each pair of schemes $[\mathrm{s1,s2}]$ we average with AIC weights the results from the three fits shown in Table~\ref{tab:DeltaZ} for varying $n_\mathrm{max}$ and $(2L/a)_{\mathrm{min}}$. This gives our final estimate of $\chi_{\mathrm T}^{\mathrm{s1,s2}}$, also displayed in Table~\ref{tab:DeltaZ}.

A crosscheck of the above result consists in computing the switching factor from the ratio of ratios
\begin{eqnarray}
\overline{\chi}_{\mathrm T}^{\mathrm{s1,s2}}(\mu_0/2) \equiv \dfrac{\dfrac{T^{\mathrm{RGI}}}{[ T(\mu_0/2)]^{\mathrm{s2}}_{\scriptstyle \rm R}}}{
\dfrac{T^{\mathrm{RGI}}}{[ T(\mu_0/2)]^{\mathrm{s1}}_{\scriptstyle \rm R}}} \,\, .
\label{eq:DeltaZprime}
\end{eqnarray}
Each ratio in the fraction on the rhs is known for each scheme; cf. eq.~(\ref{eq:final-he}). Since the scheme-independent quantity $T^{\mathrm{RGI}}$ cancels, $\overline{\chi}_{\mathrm T}^{\mathrm{s1,s2}}(\mu_0/2)$ is another estimate of the switching factor. Our results for $\overline{\chi}_{\mathrm T}^{\mathrm{s1,s2}}(\mu_0/2)$ are also shown in Table~\ref{tab:DeltaZ}. Compared to the ``direct'' determination $\chi_{\mathrm T}^{\mathrm{s1,s2}}$, the errors are bigger. This is not surprising, as $\overline{\chi}_{\mathrm T}^{\mathrm{s1,s2}}$ is a ratio of two factors, each computed iteratively, with accumulated error after each iteration. The agreement with the ``direct'' determination $\chi_{\mathrm T}^{\mathrm{s1,s2}}$ is within 1 or 1.5~$\sigma$ for the majority of results; only in a couple of cases there is a 2~$\sigma$ agreement. We therefore consider this comparison as a raw crosscheck.
\begin{table}
\begin{tabular}{c|c|c|c|c}
\hline 
$n_{{\rm max}}$ & $(2L/a)_{min}$ & $[\alpha,\beta]$ & $[\alpha,\gamma]$ & $[\alpha,\delta]$\tabularnewline
\hline 
\hline 
3 & 12 & 1.03524(93) & 1.1297(60) & 1.0865(36)\tabularnewline
4 & 12 & 1.03522(80) & 1.1287(51) & 1.0867(31)\tabularnewline
2 & 16 & 1.03519(61) & 1.1270(37) & 1.0870(23)\tabularnewline
\hline 
\hline 
 & $\chi_{\mathrm T}^{\mathrm{s1,s2}}(\mu_0/2)$ & 1.03522(77) & 1.1285(49) & 1.0867(30)\tabularnewline
\cline{2-5} \cline{3-5} \cline{4-5} \cline{5-5} 
 & $\overline{\chi}_{\mathrm T}^{\mathrm{s1,s2}}(\mu_0/2)$ & 1.0305(23) & 1.144(16) & 1.0767(94)\tabularnewline
\hline 
\end{tabular}
\begin{tabular}{c|c|c|c|c}
\hline 
$n_{{\rm max}}$ & $(2L/a)_{min}$ & $[\beta,\gamma]$ & $[\beta,\delta]$ & $[\gamma,\delta]$\tabularnewline
\hline 
\hline 
3 & 12 & 1.0915(53) & 1.0496(29) & 0.9618(44)\tabularnewline
4 & 12 & 1.0905(45) & 1.0498(25) & 0.9628(37)\tabularnewline
2 & 16 & 1.0889(32) & 1.0501(18) & 0.9645(27)\tabularnewline
\hline 
\hline 
 & $\chi_{\mathrm T}^{\mathrm{s1,s2}}(\mu_0/2)$ & 1.0903(43) & 1.0498(24) & 0.9630(36)\tabularnewline
\cline{2-5} \cline{3-5} \cline{4-5} \cline{5-5} 
 & $\overline{\chi}_{\mathrm T}^{\mathrm{s1,s2}}(\mu_0/2)$ & 1.110(14) & 1.0448(77) & 0.941(11)\tabularnewline
\hline 
\end{tabular}
\caption{Continuum limit of $\chi_{\mathrm T}^{\mathrm{s1,s2}}$ for various combinations of renormalisation schemes $[\mathrm{s1,s2}]$.
Results from three fits, differing by the choice of $[n_{{\rm max}}, (2L/a)_{min}]$ are shown, together with their weighted average $\chi_{\mathrm T}^{\mathrm{s1,s2}}(\mu_0/2)$. The cross-checking estimate $\overline{\chi}_{\mathrm T}^{\mathrm{s1,s2}}(\mu_0/2)$  is also shown.}\label{tab:DeltaZ}
\end{table}

We can combine the results of eqs.~(\ref{eq:final-he}) and (\ref{eq:final-le}) with the switching factors $\chi_{\mathrm T}
^{\mathrm{s1,s2}}$ for various scheme pairings $[\mathrm{s1,s2}]$, in order to obtain the overall running factor $T^{\mathrm{RGI}}/[ T(\mu_{\mathrm{had}})]^{\mathrm{s2}}_{\scriptstyle \rm R}$, which in principle is independent of scheme-$\mathrm{s1}$. Results are collected in Table~\ref{tab:overall_running_s2}. Moreover, there seems to be no substantial 
difference between results obtained in the same scheme throughout the whole energy range (i.e. with $\mathrm{s1} = \mathrm{s2} = \alpha, \beta, \gamma, \delta$ and $\chi_{\mathrm T}^{\mathrm{s1,s2}}=1$) and those obtained by combining different schemes (i.e. with $\mathrm{s1} \neq \mathrm{s2}$ and $\chi_{\mathrm T}^{\mathrm{s1,s2}} \neq 1$).

\begin{table}
\begin{adjustbox}{width=\columnwidth,center}
\begin{tabular}{| l ||c|c|c|c|}
\hline 
\backslashbox{$\mathrm{s1}$}{$\mathrm{s2}$}
 & $\alpha$ & $\beta$ & $\gamma$ & \textbf{$\delta$}\tabularnewline
\hline 
\hline 
$\alpha$ & \textcolor{black}{0.734(18)} & \textcolor{black}{0.882(16)} & \textcolor{black}{0.964(22)} & \textcolor{black}{0.963(15)}\tabularnewline
$\beta$ & \textcolor{black}{0.731(18)} & \textcolor{black}{0.878(15)} & \textcolor{black}{0.960(21)} & \textcolor{black}{0.959(14)}\tabularnewline
$\gamma$ & \textcolor{black}{0.744(17)} & \textcolor{black}{0.893(14)} & \textcolor{black}{0.977(21)} & \textcolor{black}{0.976(13)}\tabularnewline
\textbf{$\delta$} & \textcolor{black}{0.727(17)	} & \textcolor{black}{0.873(14)} & \textcolor{black}{0.955(20)} & \textcolor{black}{0.954(13)}\tabularnewline
\hline \hline
 \backslashbox{$\mathrm{s1-averages}$}{$\mathrm{s2}$} & 0.734(17) & 0.882(14)& 0.964(21) & 0.963(12)\tabularnewline
\hline 
\end{tabular}
\end{adjustbox}
\caption{The upper part of the table displays the overall running factors $T^{\mathrm{RGI}}/[ T(\mu_{\mathrm{had}})]^{\mathrm{s2}}_{\scriptstyle \rm R}$. Each result has been obtained as the product of the three factors in the rhs of eq.~(\ref{eq:total-running-s2}): The third factor is computed in scheme-$\mathrm{s2}$, indicated by the first row of the table; the first factor is computed in scheme-$\mathrm{s1}$, indicated by the first column; the second factor is listed in the $\chi_{\mathrm T}^{\mathrm{s1,s2}}(\mu_0/2)$ row of Table~\ref{tab:DeltaZ} for different schemes and is 1 for the same scheme. The lower part of the table (bottom line) shows the weighted average of the four results above it.
}
\label{tab:overall_running_s2}
\end{table}

The weighted average of the four results in each column of Table~\ref{tab:overall_running_s2}, with correlations properly taken into account, is our final estimate of $T^{\mathrm{RGI}}/[ T(\mu_{\mathrm{had}})]^{\mathrm{s}}_{\scriptstyle \rm R}$ for each scheme $s$ (previously denoted as $s2$):
\begin{equation}
\dfrac{T^{\mathrm{RGI}}}{[T(\mu_{\mathrm{had}})]^{\mathrm{s}}_{\scriptstyle \rm R}}=\begin{cases}
 0.734(17) & {\rm scheme-}\alpha\\
 0.882(14) & {\rm scheme-}\beta\\
0.964(21) & {\rm scheme-}\gamma\\
0.963(12)& {\rm scheme-}\delta
\end{cases}
\label{eq:final-he-le}
\end{equation}
We draw the reader's attention to Appendix~\ref{app:comparison}, where we compare our scheme-$\alpha$ and -$\beta$ results to the ones obtained in a unitary SF setup~\cite{Chimirri:2023ovl}, using the same gauge ensembles.

\section{Renormalisation at a hadronic scale}
\label{sec:hadronic-coupling}

So far we have computed the RG-running factor $T^{\mathrm{RGI}}/[ T(\mu_{\mathrm{had}})]^{\mathrm{s}}_{\scriptstyle \rm R}$ for four schemes $\mathrm{s} = \alpha, \beta, \gamma, \delta$. In order for the renormalisation programme to be complete, we
must still provide our estimates for the tensor renormalisation parameter $Z^{\mathrm{s}}_{\mathrm T}(g_0^2,a\mu_{\mathrm{had}})$, computed at fixed hadronic scale $\mu_{\mathrm{had}}$ for varying bare couplings $g_0^2$.

Following ref.~\cite{Capitani:1998mq} (see also the more recent refs.~\cite{Campos:2018ahf,Chimirri:2023ovl}),
we define the so-called renormalisation group invariant parameter  $Z^{\mathrm{RGI}}(g_0^2)$ 
\begin{align} 
\label{eq:ZTRGI}
Z^{\mathrm{RGI}}_{\mathrm T}(g_0^2) \equiv \dfrac{T^{\mathrm{RGI}}}{[ T(\mu_{\mathrm{had}})]^{\mathrm{s}}_{\scriptstyle \rm R}} \,\,\, Z^{\mathrm{s}}_{\mathrm T}(g_0^2,a\mu_{\mathrm{had}}) \, .
\end{align}
On the rhs, the ratio $T^{\mathrm{RGI}}/[ T(\mu_{\mathrm{had}})]^{\mathrm{s}}_{\scriptstyle \rm R}$ is a scheme\&scale dependent continuum quantity, independent 
of the lattice regularisation, whilst $Z^{\mathrm{s}}_{\mathrm T}(g_0^2, a\mu_{\mathrm{had}})$ depends on scheme\&scale as well as on the details of the lattice formulation. The scheme\&scale dependence cancels out in the product $Z^{\mathrm{RGI}}_{\mathrm T}$, which relates the bare tensor composite field $T(g_0^2)$ to its RGI counterpart and depends only on the lattice regularisation. Schematically we write this as:
\begin{subequations}
\begin{align} 
T^{\mathrm{RGI}} & = \lim_{g_0^2 \rightarrow 0} \big [ Z^{\mathrm{RGI}}_{\mathrm T}(g_0^2) \,\, T(g_0^2) \big ]
\label{eq:TRGI-a} \\
& = \dfrac{T^{\mathrm{RGI}}}{[ T(\mu_{\mathrm{had}})]^{\mathrm{s}}_{\scriptstyle \rm R}} \,\, \lim_{g_0^2 \rightarrow 0} \big [ Z^{\mathrm{s}}_{\mathrm T}(g_0^2,a\mu_{\mathrm{had}}) \,\, T(g_0^2) \big ] \, .
\label{eq:TRGI-b}
\end{align}
\end{subequations}
 In principle the value of $\mu_{\mathrm{had}}$ could  be chosen arbitrarily. In practice it is chosen in a range where large-volume simulations of the bare matrix element $\langle f \vert T(g_0^2) \vert i \rangle$ are performed, with $\vert i \rangle$ and $\vert f \rangle$ some initial and final hadronic states. 

Following for example refs.~\cite{Campos:2018ahf,Bruno:2019vup}, the bare couplings are tuned so as to span the interval $\beta = 6/g_0^2 \in [3.40, 3.85]$ of the CLS ensembles~\cite{Bruno:2014jqa,Mohler:2017wnb}.
The renormalised coupling is set to $u_{\mathrm{had}}=\gbar^2_{\mathrm{GF}}(\mu_{\mathrm{had}})= 9.25$, which corresponds to the energy value $\mu_{\mathrm{had}}=233(8) \MeV$, once the scale of the lattices is fixed by the method of ref.~\cite{Bruno:2016plf}.

The details and the results of the simulations at the hadronic scale are collected in Table~\ref{tab:ztabcd} of Appendix~\ref{app:zhad}. 
As in ref.~\cite{Chimirri:2023ovl}, the  data  are well represented by a polynomial fit
\begin{subequations}
\begin{align}
Z_\mathrm{T}(g_0^2,u,Lm) & = Z_{\mathrm T}(g_0^2,a\mu_{\mathrm{had}})
+ t_{10}\,(u-u_{\rm\scriptscriptstyle had}) +  t_{01}\,Lm \,, 
\label{eq:zhadfit-a}
\\
Z_{\mathrm T}(g_0^2,a\mu_{\mathrm{had}})
&= z_0 + z_1(\beta-\beta_0) + z_2(\beta-\beta_0)^2\,,
\label{eq:zhadfit-b}
\end{align}
\end{subequations}
where $\beta \equiv 6/g_0^2$, $\beta_0=3.79$, and $m$ is the residual PCAC quark mass computed in the SF setup. The terms with the parameters $ t_{10},\, t_{01}$ take into account 
deviations of the coupling and the mass 
from their reference values $(u_{\mathrm{had}},mL)=(9.25,0)$. We checked that higher powers in $(u-u_{\rm\scriptscriptstyle had})$ and $Lm$ are of hardly any consequence.
The fit results of the $z_i$ parameters and their covariances can be found in Appendix~\ref{app:zhad}, for all four schemes. 

\begin{figure*}
\includegraphics[width=0.5\textwidth]{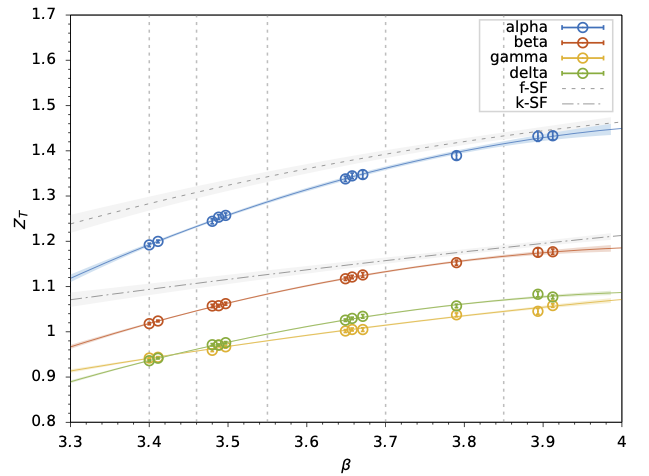}\includegraphics[width=0.5\textwidth]{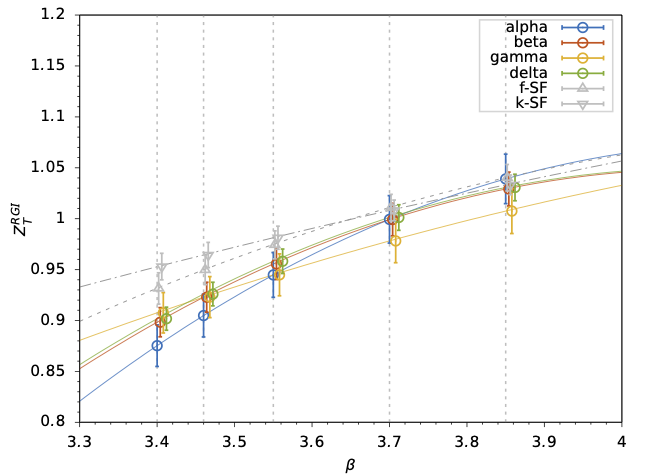}

\caption{Left panel: the datapoints are estimates of $Z_{\mathrm T}(g_0^2,a\mu_{\mathrm{had}})$ at the
$\beta$-values of our simulations, for schemes$-\alpha$, $-\beta$, $-\gamma$ and $-\delta$,
obtained from $Z_\mathrm{T}(g_0^2,u,Lm)$ as explained in the text. The four coloured curves are really very narrow bands
showing the fit results of eq.~(\ref{eq:zhadfit-b}) with errors. The two grey lines are the results of
ref.~\cite{Chimirri:2023ovl}.
Right panel: the parameter $Z^{\mathrm{RGI}}_{\mathrm T}(g_0^2)$, obtained as described in the text, from
data in our four renormalisation schemes, plotted against $\beta$. The four coloured curves are
the fit results. The grey points and lines are the results of
ref.~\cite{Chimirri:2023ovl}. Vertical dashed lines mark the $\beta$-values used in CLS simulations.}
\label{fig:zthad}
\end{figure*}

Once the fit parameters $t_{10}$ and $t_{01}$ have been determined, we can subtract the correction 
$[t_{10}\,(u-u_{\rm\scriptscriptstyle had}) +  t_{01}\,Lm]$ from
the data points of $Z_\mathrm{T}(g_0^2,u,Lm)$, obtaining estimates of $Z_{\mathrm T}(g_0^2,a\mu_{\mathrm{had}})$ at each $\beta$;
cf. eq.~(\ref{eq:zhadfit-a}). These points, together with the fit function of eq.~(\ref{eq:zhadfit-b}), are shown in Fig.~\ref{fig:zthad} (left panel).
In the same figure we also plot the results of ref.~\cite{Chimirri:2023ovl}, obtained with SF boundary conditions in the f- and k-schemes, which correspond to our schemes$-\alpha$ and $-\beta$ in the continuum. These data have been obtained from the same gauge ensembles, with the exception of those with $L/a=12,\,16$, which could not be recovered and had to be generated afresh.
Our $Z_{\mathrm T}(g_0^2,a\mu_{\mathrm{had}})$ results with their statistical errors are listed in Table~\ref{tab:zCLS}.

From eq.~(\ref{eq:final-he-le}) and Table~\ref{tab:zCLS},
we see that the most accurate results are those of schemes$-\beta$ and $-\delta$, with scheme$-\delta$ having slightly smaller errors. 

Combining  the running factor $T^{\mathrm{RGI}}/[ T(\mu_{\mathrm{had}})]^{\mathrm{s}}_{\scriptstyle \rm R}$ 
of eq.~(\ref{eq:final-he-le})
with the renormalisation constant $Z_{\mathrm T}(g_0^2,a\mu_{\mathrm{had}})$ of Table~\ref{tab:zCLS}, we can now
compute $Z^{\mathrm{RGI}}_{\mathrm T}(g_0^2)$; see eq.~(\ref{eq:ZTRGI}). 
Errors are added in quadrature.
Determinations from different schemes are expected to agree up to cutoff effects. This is confirmed in Fig.~\ref{fig:zthad} (right panel) for all four schemes $\alpha, \, \beta, \, \gamma,\, \delta$. In particular, we note the excellent agreement between schemes-$\beta$ and $-\delta$, which are superimposed over the whole range of bare couplings.
\begin{table}[h]
  \centering
  \small
\begin{tabular}{c |c |c |c |c  }
\hline 
   $\beta$ & ${Z}^{\alpha}_\mathrm{T}$ & ${Z}^{\beta}_\mathrm{T}$& ${Z}^{\gamma}_\mathrm{T}$& ${Z}^{\delta}_\mathrm{T}$ \\
\hline 
3.40  &    1.1925(29) &   1.0185(18) &   0.9415(14) &   0.9364(15) \\ 
 3.46  &   1.2327(20) &   1.0463(13) &   0.9575(10) &   0.9615(10) \\ 
 3.55  &   1.2870(26) &   1.0834(16) &   0.9803(14) &   0.9950(12) \\ 
 3.70  &   1.3613(32) &   1.1328(20) &   1.0148(17) &   1.0396(15) \\ 
 3.85  &   1.4156(41) &   1.1669(28) &   1.0451(21) &   1.0702(20) \\ 
   \hline 
  \end{tabular}
  \caption{Results for $Z^s_{\mathrm T}(g_0^2,a\mu_{\mathrm{had}})$, for schemes $s=\alpha, \beta, \gamma, \delta$, at the $\beta$ values of the CLS $\NF=2+1$ simulations.}
  \label{tab:zCLS}
\end{table}

We do not show our $Z^{\mathrm{RGI}}_{\mathrm T}(g_0^2)$ results in a table because 
the continuum extrapolation of the lattice data of the tensor matrix elements (i.e. the term 
$[ Z^{\mathrm{s}}_{\mathrm T}(g_0^2,a\mu_{\mathrm{had}}) \,\, T(g_0^2) ]$ on the rhs of eq.~(\ref{eq:TRGI-b})) 
must not include the error of the running factor $T^{\mathrm{RGI}}/[ T(\mu_{\mathrm{had}})]^{\mathrm{s}}_{\scriptstyle \rm R}$
(i.e. the first term on the rhs of eq.~(\ref{eq:TRGI-b})).
The latter factor can be taken into account after the continuum limit extrapolation has been performed, with its error safely added in quadrature. 
This was first pointed out in ref.~\cite{Capitani:1998mq} for the quark mass renormalisation factor $Z_\mathrm{M}$. 


%
\section*{Acknowledgements}
\label{sec:acknowl}
We wish to thank Alessandro Conigli, Patrick Fritzsch, Carlos Pena, David Preti, and Alberto Ramos for discussions and help. We thank Mattia Tamburrini for participating in a preliminary analysis of the hadronic data. 
MDB would like to thank the members of the Lattice QCD group at Instituto de F\'isica Te\'orica of Madrid
for their kind hospitality and support during the final stages of this work.
This work  is partially supported by INFN and CINECA, as part of research project of the QCDLAT INFN-initiative. 
We acknowledge the Santander Supercomputacion support group at the University of Cantabria which provided access to the supercomputer Altamira Supercomputer at the Institute of Physics of Cantabria (IFCA-CSIC). We also acknowledge the access granted by the Galician Supercomputing Center (CESGA) to the supercomputer FinisTerrae-III, and the support by the Poznan Supercomputing and Networking Center (PSNC) under the project with grant number 466.
AL acknowledges support by the U.S. Department of Energy under grant number DE-SC0015655.

\begin{appendix}
\section{Series expansions} 
\label{app:pert-coeffs}
In this Appendix we gather the polynomial expansions we use in our fits
and list the coefficients known from perturbative calculations.
 We also report as an example two results from the $u$-by-$u$ procedures.

In the SF scheme the Callan-Symanzik $\beta$ function is expressed as a truncated power series
\begin{equation}
\beta(g)=-g^3 \sum_{n=0}^3 b_{n} g^{2n} \,\, ,
\label{eq:beta}
\end{equation}
with
\begin{subequations}
\begin{align}
b_0 \,\, &= \dfrac{1}{(4\pi)^2} \Big (11 - \dfrac{2}{3} \NF \Big ) \,\, , \label{eq:betaPTa} \\
b_1 \,\, &= \dfrac{1}{(4\pi)^4} \Big (102 - \dfrac{38}{3}\NF \Big ) \,\, , \label{eq:betaPTb} \\
b_2^{\mathrm{SF}} \,\, &= \dfrac{1}{(4\pi)^3} \Big ( 0.483 - 0.275 \NF + 0.0361 \NF^2 - 0.00175 \NF^3 \Big ) \,\, ,  \label{eq:betaPTc} \\
b_3^{\mathrm{SF,eff}} \,\, &\stackrel{\NF=3}{=} \dfrac{1}{(4\pi)^4} \times 4(3)  \label{eq:betaPTd} \,\, .
\end{align}
\end{subequations}
The first three coefficients are calculated in perturbation theory; $b_0$ and $b_1$ are universal; $b_2$ is given in the Schr\"odinger Functional scheme~\cite{Bode:1999sm} and $b_3$ is a fit parameter obtained in ref.~\cite{DallaBrida:2018rfy,Campos:2018ahf}.

The truncated power series for the tensor operator is given by eq.~(\ref{eq:gamma}), repeated here for convenience,
\begin{equation}
\gammaT(g)=-g^{2} \sum_{n=0}^{n_\gamma} \gammaT^{(n)}g^{2n} \,\, ,
\label{eq:gamma-B}
\end{equation}
with the perturbative universal coefficient
\begin{equation}
\gammaT^{(0)}=\frac{2C_{F}}{(4\pi)^{2}}=0.016887
\label{eq:gamma0}
\end{equation}
and the scheme dependent one(s) given in $\chi$SF by
\begin{equation}
\gammaT^{(1)}=\begin{cases}
0.0062755(11) & {\rm scheme}-\alpha \,\, ,\\
0.0057956(11) & {\rm scheme}-\beta \,\, ,\\
-0.0007746(11) & {\rm scheme}-\gamma \,\, ,\\
0.0032320(11) & {\rm scheme}-\delta \,\, .
\end{cases}
\label{eq:gamma1}
\end{equation}
Analogously to ref.\cite{Pena:2017hct}, the above values are obtained by
matching to the $\MSbar$ renormalisation scheme, in which $\gammaT^{(1)}$
is known.
The starting point are the relations between two mass independent
renormalisation schemes~\cite{Sint:1998iq}:
\begin{equation}
\begin{array}{c}
\gbar'=\sqrt{\chi_{g}(\gbar)}\gbar \,\, ,\\
T_{\mathrm R}^\prime=\chi_{\mathrm T}(\gbar) T_{\mathrm R} \,\, .
\end{array}
\label{eq:change-scheme}
\end{equation}
(N.B. We simplify the notation in the last equation by writing
as $\chi_{\mathrm T}$ what had been defined as $\chi_{\mathrm T}^{\mathrm{s1,s2}}$
in eq.~(\ref{eq:DeltaZ}).)
The matching functions $\chi$ can be expanded as: 
\begin{equation}
\chi\underset{g\rightarrow0}{\sim}1+\underset{k}{\sum}\chi^{(k)}\gbar^{2k}\sim(1+\gbar^{2}\chi^{(1)}+\gbar^{4}\chi^{(2)}+...) \,\, .
\label{eq: the transform-function}
\end{equation}
Plugging Eqs.(\ref{eq:change-scheme},\ref{eq: the transform-function})
in the Callan-Symanzik equations, we find 
\begin{equation}
\gammaT^{'(1)}=\gammaT^{(1)}+2b_{0}\chi_{T}^{(1)}-\gammaT^{(0)}\chi_{g}^{(1)},\label{eq:gamma-transf}
\end{equation}
where $\gammaT^{(1)\MSbar}=-\frac{1}{(16\pi^{2})^{2}}\frac{4}{27}(26N_{f}-543)$
was computed in ref.\cite{Gracey:2003yr} and $\chi_{g}^{(1)}$ in refs.\cite{Luscher:1993gh,Sint:1995ch}.
The matching factor $\chi_{T}^{(1)}$ is determined from the difference 
$\chi_{T}^{(1)}=\chi_{T\{ \mathrm{SF},{\rm lat}\}}^{(1)}-\chi_{T\{\MSbar,{\rm lat}\}}^{(1)}$,
with $\chi_{T\{\MSbar,{\rm lat}\}}^{(1)}=-0.0939960952(3)$
from ref.\cite{Skouroupathis:2008mf}, while $\chi_{T\{\mathrm{SF},{\rm lat}\}}^{(1)}$ is
to be extracted from the 1-loop expansion of our $\ZT$ definitions:
\begin{equation}
\ZT(g_{0}^{2},L/a)=1+g_{0}^{2}\ZT^{(1)} (L/a)+... \,\, .
\end{equation}
$\ZT^{(1)}$ expected to have the following asymptotic behaviour for $a/L\rightarrow0$:
\begin{equation}
\ZT^{(1)}(L/a)\sim\underset{n=0}{\overset{n_{max}}{\sum}}[r_{n}+s_{n}\ln(L/a)]\Big (\frac{a}{L} \Big )^{n},
\label{eq:ZT^(1)}
\end{equation}
where $s_{0}$ corresponds to LO term $\gammaT^{(0)}$, while $r_{0}=\chi_{T\{\mathrm{SF},{\rm lat}\}}^{(1)}$.
In our setup, we expect $r_{1},s_{1}\sim0$.
The perturbative computations of the two-point $\chi SF$ correlation
functions have been obtained in ref.\cite{Brida:2016rmy} for a range of $L/a\in[6,48]$.
We used these results in order to perform two independent analyses to extract
$r_{0}$. Both exploited a variety of fits of the type (\ref{eq:ZT^(1)}),
varying $n_{{\rm max}}\in$ $[1,2,3]$, the range of data that were
used for the fit $(L/a)_{min}\in[6,26]$, $(L/a)_{max}\in[38,48]$
and the number of fixed terms in the fits ($s_{0}$, $r_{1}$, $s_{1}$).
One analysis made a average of the best fit results, for which
the expected values of $s_{0}$, $r_{1}$ and $s_{1}$ where obtained with a certain precision.
The second analysis chose, independently, a single fit.
Both agree on the following estimates for $r_{0}$:
\begin{equation}
\label{eq:r0-4 schemes}
r_{0}=\begin{cases}
-0.07939(1) & {\rm scheme}-\alpha \\
-0.08360(1) & {\rm scheme}-\beta \\
-0.14124(1) & {\rm scheme}-\gamma \\
-0.10609(1) & {\rm scheme}-\delta \,\, .
\end{cases}
\end{equation}
For scheme-$\alpha$ and -$\beta$ our results are comparable with
those of refs.\cite{Pena:2017hct,Brida:2016rmy}:
\begin{equation}
\begin{array}{cc}
r_{0}^{\alpha}=-0.05963(4)\times C_{F}=-0.07951(5) & \text{ref. \cite{Pena:2017hct}} \label{eq:r0-A-PP}   \,\, ,
\end{array}
\end{equation}
\begin{equation}
\begin{array}{cc}
r_{0}^{\beta}=-0.06279(4)\times C_{F}=-0.08372(5) & \text{ref. \cite{Pena:2017hct}} \label{eq:r0-B-PP}  \,\, ,
\end{array}
\end{equation}
\begin{equation}
\begin{array}{cc}
r_{0}^{\beta}=-0.06270(1)\times C_{F}=-0.08360(1) & \text{ref. \cite{Brida:2016rmy}} \label{eq:zt1-xsf}  \,\, ,
\end{array}
\end{equation}

Finally, for the tensor SSF we have eq.~(\ref{eq:sigma-ubyu}), repeated here for convenience: 
\begin{equation}
\sigmaT (u) \,\, = \,\, 1  \, + \, \sum_{i=1}^{n_\sigma} b_{i0}u^{i} \,\, .
\label{eq:sigma-ubyu-B}
\end{equation}
The coefficients $b_{i0}$ are defined in eq.~(\ref{eq:sigmaT_fit}) for the $\SigmaT$ double power series.
The universal coefficient $b_{10}$ and the perturbative one $b_{20}$ in $\chi$SF are:
\begin{eqnarray}
\label{eq:sigma-b10b20}
b_{10} \,\, &=& \gammaT^{(0)}\ln(2) \,\, , \\
b_{20} \,\, &=& \gammaT^{(1)}\ln(2)+ \Big [ \frac{(\gammaT^{(0)})^{2}}{2}+b_{0}\gammaT^{(0)} \Big ](\ln(2))^{2} \,\, . \nonumber
\end{eqnarray}

As an example, we quote $b_{30}$, obtained by
fitting our $\sigmaT^{\text{sub}}$ data with $n_\sigma = 4$ ($\sigmaT$:$u$-by-$u$ procedure, described in Sect.~\ref{sec:results-he}): 
\begin{equation}
b_{30}=\begin{cases}
0.0016(10) & \alpha-{\rm scheme}\\
0.0004(09) & \beta-{\rm scheme}\\
0.0022(07) & \gamma-{\rm scheme}\\
-0.0007(05) & \delta-{\rm scheme}
\end{cases} \,\, .
\label{eq:b30}
\end{equation}
Similarly, fitting our $\sigmaT^{\text{sub}}$ data with $n_\gamma = 3$ ($\gammaT$:$u$-by-$u$ procedure, described in Sect.~\ref{sec:results-he}), we obtain:
\begin{equation}
\gammaT^{(2)}\stackrel{\NF=3}{=}\begin{cases}
0.0014(10) & \alpha-{\rm scheme}\\
-0.0002(09) & \beta-{\rm scheme}\\
0.0027(07) & \gamma-{\rm scheme}\\
-0.0011(06) & \delta-{\rm scheme}
\end{cases}
\label{eq:gammaT2}
\end{equation}

We also quote the results from a $\sigmaT$-\emph{global} high-energy fit, $[n_{\sigma},j_{{\rm max}},(L/a)_{{\rm min}}]=[3,4,6]$ in Appendix~\ref{tab:fit-bi0-global-SF}.

We conclude this Appendix by giving the fit coefficients for the anomalous dimension $\gammaT$ displayed 
in Fig.~\ref{fig:op_runningl}. Recall that in the high-energy region, the result in the figure 
was obtained with a fit having $[n_\gamma, n_\sigma^\prime, j_{\mathrm{max}}, (L/a)_{\mathrm{min}}] = [2,2,4,6]$, whereas in
the low-energy region, the result was obtained with a fit having $[n_\gamma, n_\sigma^\prime, j_{\mathrm{max}}, (L/a)_{\mathrm{min}}] = [2,2,2,12]$.
The coefficients $\gammaT^{(n)}(\gbar_{\SF})$ and $\gammaT^{(n)}(\gbar_{\GF})$ for $n=0,1,2$ are shown in Tables~\ref{tab:fit-gamma-coeffs-SF} and~\ref{tab:fit-gamma-coeffs-GF} respectively.
\begin{table} [h]
\begin{tabular}{c|c|c|c|c}
\hline 
$n$ & $\alpha$ & $\beta$ & $\gamma$ & $\delta$\tabularnewline
\hline 
0 & 0.0168869 & 0.0168869 & 0.0168869 & 0.0168869\tabularnewline
1 & 0.0062755(11) & 0.0057956(11) & -0.00077460(11) & 0.0032320(11)\tabularnewline
2 & 0.00022(37) & -0.00056(32) & 0.00062(25) & -0.00020(19)\tabularnewline
\hline 
\end{tabular}
\caption{Results of the $\gammaT^{(n)}(\gbar_{\SF})$ coefficients in the high-energy region for the four
renormalisation schemes for the fit $[n_\gamma, n_\sigma^\prime, j_{\mathrm{max}}, (L/a)_{\mathrm{min}}] = [2,2,4,6]$. The $n=0,1$ results are the perturbative ones of eqs.~ (\ref{eq:gamma0})
and (\ref{eq:gamma1}) respectively.}
\label{tab:fit-gamma-coeffs-SF}
\end{table}

\begin{table} [h]
\begin{tabular}{c|c|c|c|c}
\hline 
$n$ & $\alpha$ & $\beta$ & $\gamma$ & $\delta$\tabularnewline
\hline 
0 & 0.23(18) & 0.21(14) & 0.20(12) & 0.24(11)\tabularnewline
1 & 0.076(65) & 0.069(48) & 0.042(42) & 0.023(38)\tabularnewline
2 & 0.0021(52) & -0.0035(38) & 0.0006(33) & 0.0012(28)\tabularnewline
\hline 
\end{tabular}
\caption{Results of the $\gammaT^{(n)}(\gbar_{\GF})$ coefficients in the low-energy region for the four
renormalisation schemes for the fit $[n_\gamma, n_\sigma^\prime, j_{\mathrm{max}}, (L/a)_{\mathrm{min}}] = [2,2,2,12]$.}
\label{tab:fit-gamma-coeffs-GF}
\end{table}

\begin{table}
\begin{tabular}{|c|c|c|c|c|}
\hline 
$b_{i0}$ & $b_{i0}^{\alpha}$ & $b_{i0}^{\beta}$ & $b_{i0}^{\gamma}$ & $b_{i0}^{\delta}$\tabularnewline
\hline 
\hline 
$b_{10}$ & 0.011705082 & 0.011705082 & 0.011705082 & 0.011705082\tabularnewline
\hline 
$b_{20}$ & 0.004880755 & 0.004548113 & -6.00213E-06 & 0.002771161\tabularnewline
\hline 
$b_{30}$ & 0.00093(37) & 0.00010(31) & 0.00066(24) & 0.00012(19)\tabularnewline
\hline 
\end{tabular}
\caption{Results of the  $b_{i0}$ coefficients in the $(u^{i},(a/L)^{j})$
double series expansion of $\Sigma_{T}(u,a/L)$ in the high-energy region for the four
renormalisation schemes for the fit $[n_{\sigma},j_{{\rm max}},(L/a)_{{\rm min}}]=[3,4,6]$. The $i=1, 2$ results are perturbative.}
\label{tab:fit-bi0-global-SF}
\end{table}

\cleardoublepage

\section{Results for renormalisation parameters and SSF's in the high-energy regime} 
\label{app:Z-Sigma-he}
  
\providecommand{\tabularnewline}{\\}

\begin{table*}[h]
\begin{tabular}{c|c|c|c|c|c|c|c|c}
\hline 
$u_{\mathrm {SF}}$ & $L/a$ & $\beta$ & $\hopc$ & $z_{{\rm f}}$ & $Z_{\mathrm T}(g_{0}^{2},L/a)$ & $Z_{\mathrm T}(g_{0}^{2},2L/a)$ & $\Sigma_{\mathrm T}(g_{0}^{2},L/a)$ & $\Sigma_{\mathrm T}^{{\rm sub}}(g_{0}^{2},L/a)$\tabularnewline
\hline 
1.11 & 6  & 8.5403   & 0.1323361   & 1.185887 & 0.97242(37) & 0.98797(47) & 1.01599(62) & 1.01942(62)\tabularnewline
1.11 & 8  & 8.7325   & 0.1321338   & 1.183747 & 0.97722(34) & 0.99397(82) & 1.01715(91) & 1.01862(91)\tabularnewline
1.11 & 12 & 8.995    & 0.1318621   & 1.176246 & 0.98610(62) & 1.00593(87) & 1.0201(11) & 1.0206(11)\tabularnewline
\hline 
1.1844 & 6  & 8.217  & 0.1326903   & 1.194279 & 0.97307(40) & 0.98713(47) & 1.01445(64) & 1.01811(64)\tabularnewline
1.1844 & 8  & 8.4044 & 0.1324767   & 1.192242 & 0.97760(38) & 0.9970(11) & 1.0199(12) & 1.0214(12)\tabularnewline
1.1844 & 12 & 8.6769 & 0.13217153  & 1.184179 & 0.98701(60) & 1.00865(89) & 1.0219(11) & 1.0224(11)\tabularnewline
\hline 
1.2656 & 6  & 7.9091 & 0.1330572   & 1.202121 & 0.97286(38) & 0.98877(54) & 1.01636(68) & 1.02028(69)\tabularnewline
1.2656 & 8  & 8.0929 & 0.1328312   & 1.200268 & 0.97706(45) & 0.99671(88) & 1.0201(10) & 1.0218(10)\tabularnewline
1.2656 & 12 & 8.373  & 0.13249231  & 1.191639 & 0.98931(65) & 1.0111(12) & 1.0220(14) & 1.0225(14)\tabularnewline
\hline 
1.3627 & 6  & 7.5909 & 0.1334693   & 1.211572 & 0.97239(40) & 0.99185(62) & 1.02002(76) & 1.02426(76)\tabularnewline
1.3627 & 8  & 7.7723 & 0.1332283   & 1.210094 & 0.97753(42) & 1.0022(14) & 1.0253(15) & 1.0271(15)\tabularnewline
1.3627 & 12 & 8.0578 & 0.13285365  & 1.201006 & 0.98769(75) & 1.0153(14) & 1.0280(16) & 1.0285(16)\tabularnewline
\hline 
1.4808 & 6  & 7.2618 & 0.1339337   & 1.221192 & 0.97183(47) & 0.99367(71) & 1.02247(88) & 1.02709(89)\tabularnewline
1.4808 & 8  & 7.4424 & 0.1336745   & 1.220256 & 0.97797(45) & 1.00578(78) & 1.02844(93) & 1.03042(93)\tabularnewline
1.4808 & 12 & 7.7299 & 0.13326353  & 1.210493 & 0.99004(89) & 1.0205(11) & 1.0307(15) & 1.0313(15)\tabularnewline
\hline 
1.6173 & 6  & 6.9433 & 0.134422    & 1.231196 & 0.97337(51) & 0.99863(73) & 1.02595(92) & 1.03102(93)\tabularnewline
1.6173 & 8  & 7.1254 & 0.1341418   & 1.230529 & 0.97862(48) & 1.0088(14) & 1.0308(15) & 1.0330(15)\tabularnewline
1.6173 & 12 & 7.4107 & 0.13369922  & 1.220678 & 0.99398(92) & 1.0254(18) & 1.0316(21) & 1.0323(21)\tabularnewline
\hline 
1.7943 & 6  & 6.605  & 0.1349829   & 1.241693 & 0.97368(58) & 1.00215(91) & 1.0292(11) & 1.0349(11)\tabularnewline
1.7943 & 8  & 6.7915 & 0.1346765   & 1.242185 & 0.98107(57) & 1.0167(20) & 1.0363(21) & 1.0388(21)\tabularnewline
1.7943 & 12 & 7.0688 & 0.13420891  & 1.231700 & 0.9966(11) & 1.0360(17) & 1.0394(20) & 1.0402(20)\tabularnewline
\hline 
2.012 & 6  & 6.2735  & 0.13565774  & 1.264242 & 0.97543(64) & 1.0100(12) & 1.0354(14) & 1.0418(14)\tabularnewline
2.012 & 8  & 6.468   & 0.13521177  & 1.248867 & 0.98442(66) & 1.0280(13) & 1.0443(15) & 1.0470(15)\tabularnewline
2.012 & 12 & 6.72995 & 0.13474560  & 1.240103 & 1.0003(10) & 1.0500(18) & 1.0497(21) & 1.0505(21)\tabularnewline
2.012 & 16 & 6.9346  & 0.13440568  & 1.232916 & 1.01572(97) & 1.0661(31) & 1.0496(32) & 1.0500(32)\tabularnewline
\end{tabular}
\caption{Results for the renormalisation parameters $Z_{\mathrm T}$ and the SSFs $\SigmaT$ and $\SigmaT^{\mathrm{sub}}$ in the $\alpha$-scheme, for all squared couplings $u_{\mathrm {SF}}$, lattice resolutions $L/a$, inverse bare couplings $\beta$, and critical hopping parameters $\hopc$.}
\label{tab:SF-Z-Sigma-alpha}
\end{table*}


\begin{table*}
\begin{tabular}{c|c|c|c|c|c|c|c|c}
\hline 
$u_{\mathrm {SF}}$ & $L/a$ & $\beta$ & $\hopc$ & $z_{{\rm f}}$ & $Z_{\mathrm T}(g_{0}^{2},L/a)$ & $Z_{\mathrm T}(g_{0}^{2},2L/a)$ & $\Sigma_{\mathrm T}(g_{0}^{2},L/a)$ & $\Sigma_{\mathrm T}^{{\rm sub}}(g_{0}^{2},L/a)$\tabularnewline
\hline 
1.11 & 6  & 8.5403   & 0.1323361   & 1.185887 & 0.95937(34) & 0.97429(42) & 1.01555(56) & 1.01823(56)\tabularnewline
1.11 & 8  & 8.7325   & 0.1321338   & 1.183747 & 0.96504(30) & 0.98055(70) & 1.01607(79) & 1.01730(79)\tabularnewline
1.11 & 12 & 8.995    & 0.1318621   & 1.176246 & 0.97437(55) & 0.99246(74) & 1.01856(96) & 1.01899(96)\tabularnewline
\hline 
1.1844 & 6  & 8.217  & 0.1326903   & 1.194279 & 0.95860(36) & 0.97219(41) & 1.01418(57) & 1.01704(58)\tabularnewline
1.1844 & 8  & 8.4044 & 0.1324767   & 1.192242 & 0.96424(34) & 0.98215(91) & 1.0186(10) & 1.0199(10)\tabularnewline
1.1844 & 12 & 8.6769 & 0.13217153  & 1.184179 & 0.97420(51) & 0.99343(76) & 1.01974(94) & 1.02019(95)\tabularnewline
\hline 
1.2656 & 6  & 7.9091 & 0.1330572   & 1.202121 & 0.95696(35) & 0.97224(47) & 1.01596(62) & 1.01902(62)\tabularnewline
1.2656 & 8  & 8.0929 & 0.1328312   & 1.200268 & 0.96235(40) & 0.98050(77) & 1.01886(90) & 1.02026(90)\tabularnewline
1.2656 & 12 & 8.373  & 0.13249231  & 1.191639 & 0.97475(57) & 0.9944(11) & 1.0202(13) & 1.0207(13)\tabularnewline
\hline 
1.3627 & 6  & 7.5909 & 0.1334693   & 1.211572 & 0.95469(36) & 0.97317(53) & 1.01935(68) & 1.02265(68)\tabularnewline
1.3627 & 8  & 7.7723 & 0.1332283   & 1.210094 & 0.96115(37) & 0.9832(13) & 1.0230(14) & 1.0245(14)\tabularnewline
1.3627 & 12 & 8.0578 & 0.13285365  & 1.201006 & 0.97224(63) & 0.9965(11) & 1.0249(14) & 1.0254(14)\tabularnewline
\hline 
1.4808 & 6  & 7.2618 & 0.1339337   & 1.221192 & 0.95218(42) & 0.97252(61) & 1.02136(78) & 1.02496(78)\tabularnewline
1.4808 & 8  & 7.4424 & 0.1336745   & 1.220256 & 0.95960(39) & 0.98420(66) & 1.02563(81) & 1.02728(81)\tabularnewline
1.4808 & 12 & 7.7299 & 0.13326353  & 1.210493 & 0.97255(72) & 0.99894(94) & 1.0271(12) & 1.0277(12)\tabularnewline
\hline 
1.6173 & 6  & 6.9433 & 0.134422    & 1.231196 & 0.95080(46) & 0.97400(62) & 1.02440(82) & 1.02834(82)\tabularnewline
1.6173 & 8  & 7.1254 & 0.1341418   & 1.230529 & 0.95785(42) & 0.9840(12) & 1.0273(13) & 1.0291(13)\tabularnewline
1.6173 & 12 & 7.4107 & 0.13369922  & 1.220678 & 0.97361(78) & 1.0008(15) & 1.0279(17) & 1.0285(17)\tabularnewline
\hline 
1.7943 & 6  & 6.605  & 0.1349829   & 1.241693 & 0.94772(51) & 0.97331(75) & 1.02700(96) & 1.03139(97)\tabularnewline
1.7943 & 8  & 6.7915 & 0.1346765   & 1.242185 & 0.95688(50) & 0.9876(16) & 1.0321(17) & 1.0341(17)\tabularnewline
1.7943 & 12 & 7.0688 & 0.13420891  & 1.231700 & 0.97355(90) & 1.0069(14) & 1.0342(17) & 1.0349(17)\tabularnewline
\hline 
2.012 & 6  & 6.2735  & 0.13565774  & 1.264242 & 0.94448(55) & 0.97465(97) & 1.0319(12) & 1.0369(12)\tabularnewline
2.012 & 8  & 6.468   & 0.13521177  & 1.248867 & 0.95630(56) & 0.9925(11) & 1.0378(13) & 1.0401(13)\tabularnewline
2.012 & 12 & 6.72995 & 0.13474560  & 1.240103 & 0.97329(86) & 1.0142(14) & 1.0421(17) & 1.0429(17)\tabularnewline
2.012 & 16 & 6.9346  & 0.13440568  & 1.232916 & 0.98839(80) & 1.0296(25) & 1.0417(26) & 1.0421(26)\tabularnewline
\end{tabular}
\caption{Results for the renormalisation parameters $Z_{\mathrm T}$ and the SSFs $\SigmaT$ and $\SigmaT^{\mathrm{sub}}$ in the $\beta$-scheme, for all squared couplings $u_{\mathrm {SF}}$, lattice resolutions $L/a$, inverse bare couplings $\beta$, and critical hopping parameters $\hopc$.}
\label{tab:SF-Z-Sigma-beta}
\end{table*}


\begin{table*}
\begin{tabular}{c|c|c|c|c|c|c|c|c}
\hline 
$u_{\mathrm {SF}}$ & $L/a$ & $\beta$ & $\hopc$ & $z_{{\rm f}}$ & $Z_{\mathrm T}(g_{0}^{2},L/a)$ & $Z_{\mathrm T}(g_{0}^{2},2L/a)$ & $\Sigma_{\mathrm T}(g_{0}^{2},L/a)$ & $\Sigma_{\mathrm T}^{{\rm sub}}(g_{0}^{2},L/a)$\tabularnewline
\hline 
1.11 & 6  & 8.5403   & 0.1323361   & 1.185887 & 0.91668(24) & 0.92419(31) & 1.00819(42) & 1.01528(43)\tabularnewline
1.11 & 8  & 8.7325   & 0.1321338   & 1.183747 & 0.92091(26) & 0.93113(49) & 1.01110(61) & 1.01449(61)\tabularnewline
1.11 & 12 & 8.995    & 0.1318621   & 1.176246 & 0.92856(36) & 0.94074(64) & 1.01312(79) & 1.01437(79)\tabularnewline
\hline 
1.1844 & 6  & 8.217  & 0.1326903   & 1.194279 & 0.91306(26) & 0.92174(32) & 1.00951(45) & 1.01710(45)\tabularnewline
1.1844 & 8  & 8.4044 & 0.1324767   & 1.192242 & 0.91786(23) & 0.92956(58) & 1.01275(68) & 1.01638(68)\tabularnewline
1.1844 & 12 & 8.6769 & 0.13217153  & 1.184179 & 0.92653(46) & 0.94013(75) & 1.01468(96) & 1.01601(96)\tabularnewline
\hline 
1.2656 & 6  & 7.9091 & 0.1330572   & 1.202121 & 0.90930(26) & 0.91835(31) & 1.00995(45) & 1.01806(46)\tabularnewline
1.2656 & 8  & 8.0929 & 0.1328312   & 1.200268 & 0.91338(26) & 0.92609(66) & 1.01391(78) & 1.01779(78)\tabularnewline
1.2656 & 12 & 8.373  & 0.13249231  & 1.191639 & 0.92407(42) & 0.93955(92) & 1.0168(11) & 1.0182(11)\tabularnewline
\hline 
1.3627 & 6  & 7.5909 & 0.1334693   & 1.211572 & 0.90547(27) & 0.91514(38) & 1.01068(52) & 1.01942(52)\tabularnewline
1.3627 & 8  & 7.7723 & 0.1332283   & 1.210094 & 0.91005(29) & 0.92410(74) & 1.01543(88) & 1.01961(88)\tabularnewline
1.3627 & 12 & 8.0578 & 0.13285365  & 1.201006 & 0.92049(46) & 0.93718(93) & 1.0181(11) & 1.0197(11)\tabularnewline
\hline 
1.4808 & 6  & 7.2618 & 0.1339337   & 1.221192 & 0.90066(27) & 0.91178(46) & 1.01234(60) & 1.02186(61)\tabularnewline
1.4808 & 8  & 7.4424 & 0.1336745   & 1.220256 & 0.90508(30) & 0.92149(54) & 1.01813(69) & 1.02269(69)\tabularnewline
1.4808 & 12 & 7.7299 & 0.13326353  & 1.210493 & 0.91829(43) & 0.93585(65) & 1.01913(85) & 1.02080(85)\tabularnewline
\hline 
1.6173 & 6  & 6.9433 & 0.134422    & 1.231196 & 0.89568(32) & 0.90857(44) & 1.01439(61) & 1.02482(61)\tabularnewline
1.6173 & 8  & 7.1254 & 0.1341418   & 1.230529 & 0.90159(32) & 0.92030(87) & 1.0207(10) & 1.0257(10)\tabularnewline
1.6173 & 12 & 7.4107 & 0.13369922  & 1.220678 & 0.91461(63) & 0.9339(10) & 1.0211(13) & 1.0229(13)\tabularnewline
\hline 
1.7943 & 6  & 6.605  & 0.1349829   & 1.241693 & 0.89087(34) & 0.90437(52) & 1.01516(70) & 1.02675(71)\tabularnewline
1.7943 & 8  & 6.7915 & 0.1346765   & 1.242185 & 0.89678(36) & 0.9175(14) & 1.0231(16) & 1.0286(16)\tabularnewline
1.7943 & 12 & 7.0688 & 0.13420891  & 1.231700 & 0.90991(59) & 0.9331(10) & 1.0255(13) & 1.0275(13)\tabularnewline
\hline 
2.012 & 6  & 6.2735  & 0.13565774  & 1.264242 & 0.88202(39) & 0.90048(72) & 1.02093(93) & 1.03402(95)\tabularnewline
2.012 & 8  & 6.468   & 0.13521177  & 1.248867 & 0.89297(38) & 0.91615(74) & 1.02596(94) & 1.03221(94)\tabularnewline
2.012 & 12 & 6.72995 & 0.13474560  & 1.240103 & 0.90845(60) & 0.9347(10) & 1.0289(13) & 1.0312(13)\tabularnewline
2.012 & 16 & 6.9346  & 0.13440568  & 1.232916 & 0.91842(48) & 0.9445(17) & 1.0284(19) & 1.0295(19)\tabularnewline
\end{tabular}
\caption{Results for the renormalisation parameters $Z_{\mathrm T}$ and the SSFs $\SigmaT$ and $\SigmaT^{\mathrm{sub}}$ in the $\gamma$-scheme, for all squared couplings $u_{\mathrm {SF}}$, lattice resolutions $L/a$, inverse bare couplings $\beta$, and critical hopping parameters $\hopc$.}
\label{tab:SF-Z-Sigma-gamma}
\end{table*}


\begin{table*}
\begin{tabular}{c|c|c|c|c|c|c|c|c}
\hline 
$u_{\mathrm {SF}}$ & $L/a$ & $\beta$ & $\hopc$ & $z_{{\rm f}}$ & $Z_{\mathrm T}(g_{0}^{2},L/a)$ & $Z_{\mathrm T}(g_{0}^{2},2L/a)$ & $\Sigma_{\mathrm T}(g_{0}^{2},L/a)$ & $\Sigma_{\mathrm T}^{{\rm sub}}(g_{0}^{2},L/a)$\tabularnewline
\hline 
1.11 & 6  & 8.5403   & 0.1323361   & 1.185887 & 0.93231(20) & 0.94927(24) & 1.01820(34) & 1.01616(34)\tabularnewline
1.11 & 8  & 8.7325   & 0.1321338   & 1.183747 & 0.94077(21) & 0.95690(42) & 1.01715(50) & 1.01625(50)\tabularnewline
1.11 & 12 & 8.995    & 0.1318621   & 1.176246 & 0.95173(31) & 0.96792(51) & 1.01701(63) & 1.01662(63)\tabularnewline
\hline 
1.1844 & 6  & 8.217  & 0.1326903   & 1.194279 & 0.92919(21) & 0.94653(25) & 1.01866(36) & 1.01649(36)\tabularnewline
1.1844 & 8  & 8.4044 & 0.1324767   & 1.192242 & 0.93780(19) & 0.95451(53) & 1.01782(60) & 1.01686(60)\tabularnewline
1.1844 & 12 & 8.6769 & 0.13217153  & 1.184179 & 0.94947(37) & 0.96642(59) & 1.01785(74) & 1.01743(74)\tabularnewline
\hline 
1.2656 & 6  & 7.9091 & 0.1330572   & 1.202121 & 0.92578(21) & 0.94449(26) & 1.02020(37) & 1.01788(37)\tabularnewline
1.2656 & 8  & 8.0929 & 0.1328312   & 1.200268 & 0.93546(21) & 0.95345(51) & 1.01923(59) & 1.01820(59)\tabularnewline
1.2656 & 12 & 8.373  & 0.13249231  & 1.191639 & 0.94770(35) & 0.96493(70) & 1.01818(83) & 1.01773(83)\tabularnewline
\hline 
1.3627 & 6  & 7.5909 & 0.1334693   & 1.211572 & 0.92136(21) & 0.94234(31) & 1.02277(41) & 1.02026(41)\tabularnewline
1.3627 & 8  & 7.7723 & 0.1332283   & 1.210094 & 0.93210(23) & 0.95210(70) & 1.02145(79) & 1.02034(79)\tabularnewline
1.3627 & 12 & 8.0578 & 0.13285365  & 1.201006 & 0.94533(37) & 0.96490(72) & 1.02071(86) & 1.02022(86)\tabularnewline
\hline 
1.4808 & 6  & 7.2618 & 0.1339337   & 1.221192 & 0.91677(22) & 0.93959(39) & 1.02490(50) & 1.02216(50)\tabularnewline
1.4808 & 8  & 7.4424 & 0.1336745   & 1.220256 & 0.92888(25) & 0.95038(40) & 1.02314(51) & 1.02193(51)\tabularnewline
1.4808 & 12 & 7.7299 & 0.13326353  & 1.210493 & 0.94204(33) & 0.96434(53) & 1.02367(67) & 1.02315(66)\tabularnewline
\hline 
1.6173 & 6  & 6.9433 & 0.134422    & 1.231196 & 0.91215(25) & 0.93691(35) & 1.02715(47) & 1.02416(47)\tabularnewline
1.6173 & 8  & 7.1254 & 0.1341418   & 1.230529 & 0.92436(25) & 0.94771(75) & 1.02526(86) & 1.02394(86)\tabularnewline
1.6173 & 12 & 7.4107 & 0.13369922  & 1.220678 & 0.94041(49) & 0.96508(86) & 1.0262(11) & 1.0257(11)\tabularnewline
\hline 
1.7943 & 6  & 6.605  & 0.1349829   & 1.241693 & 0.90559(26) & 0.93479(42) & 1.03224(55) & 1.02890(55)\tabularnewline
1.7943 & 8  & 6.7915 & 0.1346765   & 1.242185 & 0.92030(29) & 0.9469(11) & 1.0289(12) & 1.0275(12)\tabularnewline
1.7943 & 12 & 7.0688 & 0.13420891  & 1.231700 & 0.93825(48) & 0.96545(81) & 1.0290(10) & 1.0283(10)\tabularnewline
\hline 
2.012 & 6  & 6.2735  & 0.13565774  & 1.264242 & 0.89982(30) & 0.93153(55) & 1.03524(70) & 1.03149(70)\tabularnewline
2.012 & 8  & 6.468   & 0.13521177  & 1.248867 & 0.91525(30) & 0.94670(55) & 1.03436(69) & 1.03270(69)\tabularnewline
2.012 & 12 & 6.72995 & 0.13474560  & 1.240103 & 0.93368(47) & 0.96626(77) & 1.03489(98) & 1.03417(98)\tabularnewline
2.012 & 16 & 6.9346  & 0.13440568  & 1.232916 & 0.94802(39) & 0.9813(15) & 1.0351(16) & 1.0347(16)\tabularnewline
\end{tabular}
\caption{Results for the renormalisation parameters $Z_{\mathrm T}$ and the SSFs $\SigmaT$ and $\SigmaT^{\mathrm{sub}}$ in the $\delta$-scheme, for all squared couplings $u_{\mathrm {SF}}$, lattice resolutions $L/a$, inverse bare couplings $\beta$, and critical hopping parameters $\hopc$.}
\label{tab:SF-Z-Sigma-delta}
\end{table*}

\cleardoublepage

\section{Continuum extrapolation results from the $\sigmaT$:$u$-by-$u$ and $\gammaT$:$u$-by-$u$ methods} 
\label{app:sigma-rho-he}

\providecommand{\tabularnewline}{\\}

\begin{table}[h]
{\footnotesize
\begin{tabular}{|c||c|c|c||c|c|c|}
$u$ & $\sigmaT$ & $\rho_{\mathrm T}$ & $\dfrac{\chi^{2}}{\mathrm{d.o.f.}}$ & $\sigmaT^{\mathrm{sub}}$ & $\rho_{\mathrm T}^{\mathrm{sub}}$ & $\dfrac{\chi^{2}}{\mathrm{d.o.f.}}$ \tabularnewline
& & & & & & \tabularnewline
\hline 
1.11 & 1.0206(13) & -0.173(56) & 1.2 & 1.0200(13) & -0.026(56) & 1.65\tabularnewline
1.1844 & 1.0249(13) & -0.374(58) & 0.62 & 1.0242(13) & -0.217(58) & 0.38\tabularnewline
1.2656 & 1.0243(15) & -0.282(66) & 0.13 & 1.0234(15) & -0.113(66) & 0.03\tabularnewline
1.3627 & 1.0310(19) & -0.392(78) & 0.11 & 1.0301(19) & -0.210(79) & 0.04\tabularnewline
1.4808 & 1.0345(16) & -0.426(75) & 0.77 & 1.0335(16) & -0.226(75) & 0.43\tabularnewline
1.6173 & 1.0348(23) & -0.313(96) & 0.62 & 1.0337(23) & -0.093(96) & 0.39\tabularnewline
1.7943 & 1.0434(24) & -0.51(10) & 0.22 & 1.0423(24) & -0.27(11) & 0.1\tabularnewline
2.012 & 1.0540(20) & -0.66(10) & 0.26 & 1.0529(20) & -0.40(10) & 0.14\tabularnewline
\end{tabular}
\caption{Results of the fit parameters $\sigmaT$ and $\rho_{\mathrm T}$ (both non-subtracted and subtracted) in the $\alpha$-scheme.}
\label{tab:sigma-rho-alpha}
}
\end{table}

\begin{table}[h]
{\footnotesize
\begin{tabular}{|c||c|c|c||c|c|c|}
$u$ & $\sigmaT$ & $\rho_{\mathrm T}$ & $\dfrac{\chi^{2}}{\mathrm{d.o.f.}}$ & $\sigmaT^{\mathrm{sub}}$ & $\rho_{\mathrm T}^{\mathrm{sub}}$ & $\dfrac{\chi^{2}}{\mathrm{d.o.f.}}$ \tabularnewline
& & & & & & \tabularnewline
\hline 
1.11 & 1.0187(11) & -0.121(49) & 1.55 & 1.0183(11) & -0.010(50) & 1.91\tabularnewline
1.1844 & 1.0221(11) & -0.282(51) & 0.98 & 1.0217(11) & -0.165(51) & 0.75\tabularnewline
1.2656 & 1.0220(14) & -0.214(60) & 0.13 & 1.0215(14) & -0.087(60) & 0.06\tabularnewline
1.3627 & 1.0270(16) & -0.273(68) & 0.06 & 1.0265(16) & -0.137(68) & 0.02\tabularnewline
1.4808 & 1.0298(13) & -0.299(65) & 0.64 & 1.0292(13) & -0.149(65) & 0.4\tabularnewline
1.6173 & 1.0297(19) & -0.189(81) & 0.26 & 1.0291(19) & -0.026(81) & 0.15\tabularnewline
1.7943 & 1.0371(21) & -0.361(90) & 0.19 & 1.0365(21) & -0.181(90) & 0.11\tabularnewline
2.012 & 1.0448(17) & -0.459(84) & 0.18 & 1.0442(17) & -0.260(85) & 0.13\tabularnewline
\end{tabular}
\caption{Results of the fit parameters $\sigmaT$ and $\rho_{\mathrm T}$ (both non-subtracted and subtracted) in the $\beta$-scheme.}
\label{tab:sigma-rho-beta}
}
\end{table}

\vskip 1.0 cm

\begin{table}[h]
{\footnotesize
\begin{tabular}{|c||c|c|c||c|c|c|}
$u$ & $\sigmaT$ & $\rho_{\mathrm T}$ & $\dfrac{\chi^{2}}{\mathrm{d.o.f.}}$ & $\sigmaT^{\mathrm{sub}}$ & $\rho_{\mathrm T}^{\mathrm{sub}}$ & $\dfrac{\chi^{2}}{\mathrm{d.o.f.}}$ \tabularnewline
& & & & & & \tabularnewline
\hline 
1.11 & 1.01479(89) & -0.238(39) & 0.0 & 1.01387(89) & 0.050(39) & 0.11\tabularnewline
1.1844 & 1.0166(10) & -0.254(45) & 0.06 & 1.0156(10) & 0.054(45) & 0.01\tabularnewline
1.2656 & 1.0190(12) & -0.326(49) & 0.0 & 1.0179(12) & 0.004(49) & 0.11\tabularnewline
1.3627 & 1.0209(13) & -0.368(53) & 0.13 & 1.0198(13) & -0.013(53) & 0.0\tabularnewline
1.4808 & 1.02260(100) & -0.357(48) & 4.28 & 1.02139(100) & 0.027(48) & 2.65\tabularnewline
1.6173 & 1.0252(15) & -0.380(63) & 3.49 & 1.0238(15) & 0.043(63) & 2.39\tabularnewline
1.7943 & 1.0297(16) & -0.519(68) & 1.12 & 1.0284(16) & -0.055(68) & 0.63\tabularnewline
2.012 & 1.0311(13) & -0.359(65) & 0.4 & 1.0297(13) & 0.157(65) & 0.12\tabularnewline
\end{tabular}
\caption{Results of the fit parameters $\sigmaT$ and $\rho_{\mathrm T}$ (both non-subtracted and subtracted) in the $\gamma$-scheme.}
}
\label{tab:sigma-rho-gamma}
\end{table}

\begin{table}
{\footnotesize
\begin{tabular}{|c||c|c|c||c|c|c|}
$u$ & $\sigmaT$ & $\rho_{\mathrm T}$ & $\dfrac{\chi^{2}}{\mathrm{d.o.f.}}$ & $\sigmaT^{\mathrm{sub}}$ & $\rho_{\mathrm T}^{\mathrm{sub}}$ & $\dfrac{\chi^{2}}{\mathrm{d.o.f.}}$ \tabularnewline
& & & & & & \tabularnewline
\hline 
1.11 & 1.01635(71) & 0.065(31) & 0.31 & 1.01664(71) & -0.018(31) & 0.08\tabularnewline
1.1844 & 1.01731(83) & 0.047(36) & 0.24 & 1.01762(83) & -0.041(36) & 0.06\tabularnewline
1.2656 & 1.01768(89) & 0.092(38) & 0.07 & 1.01803(89) & -0.004(38) & 0.28\tabularnewline
1.3627 & 1.01995(100) & 0.101(42) & 0.01 & 1.02028(100) & -0.000(42) & 0.01\tabularnewline
1.4808 & 1.02255(78) & 0.076(38) & 2.4 & 1.02293(78) & -0.034(38) & 1.43\tabularnewline
1.6173 & 1.0249(12) & 0.075(50) & 1.57 & 1.0254(12) & -0.047(50) & 1.03\tabularnewline
1.7943 & 1.0273(12) & 0.173(53) & 1.04 & 1.0277(12) & 0.040(53) & 0.66\tabularnewline
2.012 & 1.03448(97) & 0.020(50) & 0.34 & 1.03491(97) & -0.127(49) & 0.09\tabularnewline
\end{tabular}
\caption{Results of the fit parameters $\sigmaT$ and $\rho_{\mathrm T}$ (both non-subtracted and subtracted) in the $\delta$-scheme.}
\label{tab:sigma-rho-delta}
}
\end{table}

\cleardoublepage

\section{Results for the tensor RG evolution in the high-energy regime}
\label{app:TRGI/Tmu2-he}

\providecommand{\tabularnewline}{\\}

\begin{table*}[h]
\scalebox{0.9}{
\begin{tabular}{cc||c|c|c|c|c|c|c|c}
 &  & {\Large{}$\Big[\frac{T^{{\rm RGI}}}{[T(\mu_{0}/2)]_{\mathrm R}}\Big]^{\alpha}$} & $w_{{\rm AIC}}^{\alpha}$ & {\Large{}$\Big[\frac{T^{{\rm RGI}}}{[T(\mu_{0}/2)]_{\mathrm R}}\Big]^{\beta}$} & $w_{{\rm AIC}}^{\beta}$ & {\Large{}$\Big[\frac{T^{{\rm RGI}}}{[T(\mu_{0}/2)]_{\mathrm R}}\Big]^{\gamma}$} & $w_{{\rm AIC}}^{\gamma}$ & {\Large{}$\Big[\frac{T^{{\rm RGI}}}{[T(\mu_{0}/2)]_{\mathrm R}}\Big]^{\delta}$} & $w_{{\rm AIC}}^{\delta}$\tabularnewline
 &  &  &  &  &  &  &  &  & \tabularnewline
\hline 
$[n_{\gamma}]$ & {[}2{]} & 1.1248(50) &  & 1.1623(44) &  & 1.2823(38) &  & 1.2283(29) & \tabularnewline
 & {[}3{]} & 1.1210(69) &  & 1.1614(62) &  & 1.2731(55) &  & 1.2319(42) & \tabularnewline
\hline 
$[n_{\sigma}]$ & {[}3{]} & 1.1270(53) &  & 1.1658(47) &  & 1.2845(40) &  & 1.2325(31) & \tabularnewline
 & {[}4{]} & 1.1251(72) &  & 1.1652(65) &  & 1.2775(57) &  & 1.2358(44) & \tabularnewline
\hline 
$[n_{\gamma},n'_{\sigma},j_{\mathrm{max}},(L/a)_{\mathrm{min}}]$ & {[}2,2,3,6{]} & 1.138(12) & 0.2212 & 1.170(11) & 0.2321 & 1.3005(92) & 0.1722 & 1.2177(68) & 0.232\tabularnewline
 & {[}2,2,4,6{]} & 1.135(10) & 0.214 & 1.1681(90) & 0.2302 & 1.2967(78) & 0.1566 & 1.2198(58) & 0.2264\tabularnewline
 & {[}2,2,2,8{]} & 1.1305(77) & 0.0141 & 1.1655(67) & 0.0172 & 1.2910(57) & 0.0012 & 1.2231(43) & 0.0052\tabularnewline
 & {[}3,2,3,6{]} & 1.129(16) & 0.1116 & 1.165(14) & 0.1055 & 1.293(12) & 0.0862 & 1.2195(88) & 0.0969\tabularnewline
 & {[}3,2,4,6{]} & 1.127(14) & 0.1124 & 1.164(12) & 0.106 & 1.289(10) & 0.0854 & 1.2219(76) & 0.0964\tabularnewline
 & {[}3,2,2,8{]} & 1.123(10) & 0.0072 & 1.1622(93) & 0.0072 & 1.2773(80) & 0.0017 & 1.2281(60) & 0.0028\tabularnewline
 & {[}2,3,3,6{]} & 1.134(13) & 0.0473 & 1.167(11) & 0.0457 & 1.2963(96) & 0.0868 & 1.2191(71) & 0.054\tabularnewline
 & {[}2,3,4,6{]} & 1.132(11) & 0.0476 & 1.1664(95) & 0.0462 & 1.2937(81) & 0.0906 & 1.2208(60) & 0.0547\tabularnewline
 & {[}2,3,2,8{]} & 1.1291(78) & 0.0073 & 1.1646(67) & 0.0079 & 1.2892(57) & 0.002 & 1.2237(43) & 0.0031\tabularnewline
 & {[}3,3,3,6{]} & 1.128(20) & 0.0214 & 1.167(18) & 0.0199 & 1.299(16) & 0.0344 & 1.214(12) & 0.0233\tabularnewline
 & {[}3,3,4,6{]} & 1.126(17) & 0.0217 & 1.165(15) & 0.0201 & 1.293(13) & 0.0356 & 1.2180(98) & 0.0229\tabularnewline
 & {[}3,3,2,8{]} & 1.126(12) & 0.0031 & 1.165(10) & 0.0031 & 1.2831(92) & 0.0009 & 1.2255(69) & 0.0012\tabularnewline
 & {[}2,4,3,6{]} & 1.134(13) & 0.017 & 1.168(11) & 0.0149 & 1.2960(96) & 0.0182 & 1.2195(71) & 0.0184\tabularnewline
 & {[}2,4,4,6{]} & 1.132(11) & 0.017 & 1.1664(95) & 0.015 & 1.2936(81) & 0.0191 & 1.2210(60) & 0.0188\tabularnewline
 & {[}2,4,2,8{]} & 1.1290(78) & 0.0032 & 1.1645(67) & 0.0034 & 1.2892(57) & 0.0008 & 1.2237(43) & 0.0012\tabularnewline
 & {[}3,4,3,6{]} & 1.131(21) & 0.0074 & 1.169(19) & 0.0064 & 1.295(17) & 0.0071 & 1.219(12) & 0.0074\tabularnewline
 & {[}3,4,4,6{]} & 1.129(17) & 0.0075 & 1.167(15) & 0.0065 & 1.290(14) & 0.0076 & 1.221(10) & 0.0075\tabularnewline
 & {[}3,4,2,8{]} & 1.125(12) & 0.0013 & 1.164(10) & 0.0014 & 1.2831(93) & 0.0003 & 1.2254(69) & 0.0005\tabularnewline
 & {[}2,2,2,6{]} & 1.1254(49) &  & 1.1626(43) &  & 1.2821(37) &  & 1.2284(28) & \tabularnewline
 & {[}3,2,2,6{]} & 1.1198(64) &  & 1.1598(58) &  & 1.2750(51) &  & 1.2309(39) & \tabularnewline
 & {[}2,3,2,6{]} & 1.1245(50) &  & 1.1621(44) &  & 1.2816(38) &  & 1.2287(29) & \tabularnewline
 & {[}3,3,2,6{]} & 1.1199(69) &  & 1.1603(62) &  & 1.2734(55) &  & 1.2312(42) & \tabularnewline
 & {[}2,4,2,6{]} & 1.1241(50) &  & 1.1618(44) &  & 1.2819(38) &  & 1.2283(29) & \tabularnewline
 & {[}3,4,2,6{]} & 1.1204(69) &  & 1.1608(62) &  & 1.2728(55) &  & 1.2317(42) & \tabularnewline
\hline 
$[n_{\sigma},j_{\mathrm{max}},(L/a)_{\mathrm{min}}]$ & {[}3,3,6{]} & 1.132(14) & 0.0481 & 1.167(12) & 0.0458 & 1.295(10) & 0.0858 & 1.2185(75) & 0.0546\tabularnewline
 & {[}3,4,6{]} & 1.130(11) & 0.0488 & 1.1660(100) & 0.0463 & 1.2926(85) & 0.0905 & 1.2204(63) & 0.0551\tabularnewline
 & {[}3,2,8{]} & 1.1273(82) & 0.0049 & 1.1642(71) & 0.0055 & 1.2877(60) & 0.0019 & 1.2237(45) & 0.0025\tabularnewline
 & {[}4,3,6{]} & 1.129(21) & 0.0075 & 1.167(19) & 0.0064 & 1.294(17) & 0.0071 & 1.219(13) & 0.0074\tabularnewline
 & {[}4,4,6{]} & 1.127(18) & 0.0076 & 1.166(16) & 0.0065 & 1.290(14) & 0.0076 & 1.221(10) & 0.0075\tabularnewline
 & {[}4,2,8{]} & 1.125(12) & 0.0008 & 1.164(11) & 0.0009 & 1.2831(96) & 0.0003 & 1.2254(72) & 0.0004\tabularnewline
 & {[}3,2,6{]} & 1.1227(53) &  & 1.1615(46) &  & 1.2798(39) &  & 1.2289(30) & \tabularnewline
 & {[}4,2,6{]} & 1.1204(71) &  & 1.1605(64) &  & 1.2732(57) &  & 1.2315(44) & \tabularnewline
\end{tabular}}

\caption{Fit results for $T^{{\rm RGI}}/[T(\mu_{0}/2)]_{\mathrm R}$ for the four schemes $\alpha, \cdots, \delta$ in the high-energy (SF) range. The fit procedures are, from top to bottom: $\gammaT$:$u$-\emph{by-}$u$, $\sigmaT$:$u$-\emph{by-}$u$, $\gammaT$:\emph{global, }$\sigmaT$:\emph{global}. The fit parameters are shown in the second column. When the AIC weights are missing, the corresponding fit is not included in the final average of eq.(\ref{eq:final-he}).}
\label{tab:SF_ris}

\end{table*}

\cleardoublepage

\section{Results for renormalisation parameters and SSF's in the low-energy regime} 
\label{app:Z-Sigma-le}
  
\providecommand{\tabularnewline}{\\}

\begin{table*}[h]
\begin{tabular}{c|c|c|c|c|c|c|c}
\hline 
$u_{\mathrm{GF}}$ & $L/a$ & $\beta$ & $\hopc$ & $z_{{\rm f}}$ & $Z_{\mathrm T}(g_{0}^{2},L/a)$ & $Z_{\mathrm T}(g_{0}^{2},2L/a)$ & $\Sigma_{\mathrm T}(g_{0}^{2},L/a)$\tabularnewline
\hline 
2.1275(15) & 8  & 5.3715    & 0.13362120 & 1.226543 & 0.99567(59) & 1.0352(19) & 1.0397(20)\tabularnewline
2.1231(12) & 12 & 5.54307   & 0.13331407 & 1.219858 & 1.01342(69) & 1.0592(41) & 1.0452(42)\tabularnewline
2.1257(25) & 16 & 5.7       & 0.13304840 & 1.213896 & 1.02623(87) & 1.0710(60) & 1.0436(59)\tabularnewline
\hline 
2.3898(15) & 8  & 5.071     & 0.13421678 & 1.242454 & 0.99924(68) & 1.0463(22) & 1.0470(24)\tabularnewline
2.3913(10) & 12 & 5.242465  & 0.13387635 & 1.234761 & 1.01865(86) & 1.0770(44) & 1.0573(44)\tabularnewline
2.3900(31) & 16 & 5.4       & 0.13357851 & 1.227585 & 1.0343(12) & 1.0904(29) & 1.0542(31)\tabularnewline
\hline 
2.7363(14) & 8  & 4.7649    & 0.13488555 & 1.259474 & 1.00326(66) & 1.0608(25) & 1.0573(26)\tabularnewline
2.7389(14) & 12 & 4.938726  & 0.13450761 & 1.251937 & 1.0277(14) & 1.0941(35) & 1.0646(37)\tabularnewline
2.7359(36) & 16 & 5.1       & 0.13416889 & 1.243518 & 1.0447(12) & 1.1180(72) & 1.0702(70)\tabularnewline
\hline 
3.2040(18) & 8  & 4.4576    & 0.13560675 & 1.278015 & 1.01044(82) & 1.0880(47) & 1.0767(48)\tabularnewline
3.2051(17) & 12 & 4.634654  & 0.13519986 & 1.270631 & 1.0364(14) & 1.1265(53) & 1.0869(53)\tabularnewline
3.2029(51) & 16 & 4.8       & 0.13482139 & 1.261490 & 1.0609(15) & 1.1470(45) & 1.0812(45)\tabularnewline
\hline 
3.8636(21) & 8  & 4.1519    & 0.13632589 & 1.297229 & 1.0188(10) & 1.1291(53) & 1.1083(53)\tabularnewline
3.8633(21) & 12 & 4.33166   & 0.13592664 & 1.291442 & 1.0542(20) & 1.1743(79) & 1.1139(77)\tabularnewline
3.8643(66) & 16 & 4.5       & 0.13552582 & 1.281991 & 1.0804(18) & 1.2309(91) & 1.1393(86)\tabularnewline
\hline 
4.4848(24) & 8  & 3.9479    & 0.13674684 & 1.307866 & 1.0315(12) & 1.1744(56) & 1.1385(56)\tabularnewline
4.4864(27) & 12 & 4.128217  & 0.13640300 & 1.305825 & 1.0738(22) & 1.2575(83) & 1.1710(81)\tabularnewline
4.4901(75) & 16 & 4.3       & 0.13600821 & 1.296227 & 1.1074(20) & 1.289(12) & 1.164(11)\tabularnewline
\hline 
5.3009(32) & 8  & 3.75489   & 0.13701929 & 1.314612 & 1.0474(17) & 1.277(14) & 1.220(13)\tabularnewline
5.2970(36) & 12 & 3.936816  & 0.13679805 & 1.317703 & 1.1036(26) & 1.371(16) & 1.242(15)\tabularnewline
5.301(14)  & 16  & 4.1      & 0.13647301 & 1.310054 & 1.1402(40) & 1.456(25) & 1.277(22)\tabularnewline
\end{tabular}
\caption{Results for the renormalisation parameters $Z_{\mathrm T}$ and the SSFs $\SigmaT$ in the $\alpha$-scheme, for
all squared couplings $u_{\mathrm{GF}}$, lattice resolutions $L/a$, inverse bare couplings $\beta$, and critical hopping parameters $\hopc$.}
\label{tab:Z-Sigma-alpha}
\end{table*}


\begin{table*}
\begin{tabular}{c|c|c|c|c|c|c|c}
\hline 
$u_{\mathrm{GF}}$ & $L/a$ & $\beta$ & $\hopc$ & $z_{{\rm f}}$ & $Z_{\mathrm T}(g_{0}^{2},L/a)$ & $Z_{\mathrm T}(g_{0}^{2},2L/a)$ & $\Sigma_{\mathrm T}(g_{0}^{2},L/a)$\tabularnewline
\hline 
2.1275(15) & 8  & 5.3715    & 0.13362120 & 1.226543  & 0.97229(50) & 1.0050(16) & 1.0336(17)\tabularnewline
2.1231(12) & 12 & 5.54307   & 0.13331407 & 1.219858  & 0.98920(56) & 1.0263(34) & 1.0375(35)\tabularnewline
2.1257(25) & 16 & 5.7       & 0.13304840 & 1.213896  & 1.00112(72) & 1.0393(45) & 1.0381(46)\tabularnewline
\hline 
2.3898(15) & 8  & 5.071     & 0.13421678 & 1.242454  & 0.97216(56) & 1.0101(18) & 1.0390(20)\tabularnewline
2.3913(10) & 12 & 5.242465  & 0.13387635 & 1.234761  & 0.99001(70) & 1.0369(33) & 1.0473(35)\tabularnewline
2.3900(31) & 16 & 5.4       & 0.13357851 & 1.227585  & 1.0050(10) & 1.0486(28) & 1.0434(30)\tabularnewline
\hline 
2.7363(14) & 8  & 4.7649    & 0.13488555 & 1.259474  & 0.97111(56) & 1.0158(19) & 1.0461(20)\tabularnewline
2.7389(14) & 12 & 4.938726  & 0.13450761 & 1.251937  & 0.9938(11) & 1.0455(31) & 1.0519(33)\tabularnewline
2.7359(36) & 16 & 5.1       & 0.13416889 & 1.243518  & 1.00949(93) & 1.0661(60) & 1.0560(61)\tabularnewline
\hline 
3.2040(18) & 8  & 4.4576    & 0.13560675 & 1.278015  & 0.97085(66) & 1.0301(37) & 1.0610(39)\tabularnewline
3.2051(17) & 12 & 4.634654  & 0.13519986 & 1.270631  & 0.9956(11) & 1.0617(38) & 1.0663(40)\tabularnewline
3.2029(51) & 16 & 4.8       & 0.13482139 & 1.261490  & 1.0175(12) & 1.0797(34) & 1.0612(36)\tabularnewline
\hline 
3.8636(21) & 8  & 4.1519    & 0.13632589 & 1.297229  & 0.96983(84) & 1.0487(34) & 1.0814(37)\tabularnewline
3.8633(21) & 12 & 4.33166   & 0.13592664 & 1.291442  & 1.0015(15) & 1.0801(62) & 1.0785(64)\tabularnewline
3.8643(66) & 16 & 4.5       & 0.13552582 & 1.281991  & 1.0252(14) & 1.1293(56) & 1.1015(57)\tabularnewline
\hline 
4.4848(24) & 8  & 3.9479    & 0.13674684 & 1.307866  & 0.97209(93) & 1.0650(39) & 1.0956(41)\tabularnewline
4.4864(27) & 12 & 4.128217  & 0.13640300 & 1.305825  & 1.0093(16) & 1.1230(60) & 1.1126(62)\tabularnewline
4.4901(75) & 16 & 4.3       & 0.13600821 & 1.296227  & 1.0389(15) & 1.1486(72) & 1.1056(72)\tabularnewline
\hline 
5.3009(32) & 8  & 3.75489   & 0.13701929 & 1.314612  & 0.9742(13) & 1.1099(73) & 1.1393(76)\tabularnewline
5.2970(36) & 12 & 3.936816  & 0.13679805 & 1.317703  & 1.0219(19) & 1.1599(94) & 1.1351(95)\tabularnewline
5.301(14)  & 16  & 4.1      & 0.13647301 & 1.310054  & 1.0523(29) & 1.2066(95) & 1.1466(96)\tabularnewline
\end{tabular}
\caption{Results for the renormalisation parameters $Z_{\mathrm T}$ and the SSFs $\SigmaT$ in the $\beta$-scheme, for
all squared couplings $u_{\mathrm{GF}}$, lattice resolutions $L/a$, inverse bare couplings $\beta$, and critical hopping parameters $\hopc$.}
\label{tab:Z-Sigma-beta}
\end{table*}


\begin{table*}
\begin{tabular}{c|c|c|c|c|c|c|c}
\hline 
$u_{\mathrm{GF}}$ & $L/a$ & $\beta$ & $\hopc$ & $z_{{\rm f}}$ & $Z_{\mathrm T}(g_{0}^{2},L/a)$ & $Z_{\mathrm T}(g_{0}^{2},2L/a)$ & $\Sigma_{\mathrm T}(g_{0}^{2},L/a)$\tabularnewline
\hline 
2.1275(15) & 8  & 5.3715    & 0.13362120 & 1.226543   & 0.91090(36) & 0.9336(12) & 1.0249(14)\tabularnewline
2.1231(12) & 12 & 5.54307   & 0.13331407 & 1.219858   & 0.92346(49) & 0.9472(16) & 1.0257(19)\tabularnewline
2.1257(25) & 16 & 5.7       & 0.13304840 & 1.213896   & 0.93489(48) & 0.9555(31) & 1.0221(33)\tabularnewline
\hline 
2.3898(15) & 8  & 5.071     & 0.13421678 & 1.242454   & 0.90554(39) & 0.9296(12) & 1.0266(14)\tabularnewline
2.3913(10) & 12 & 5.242465  & 0.13387635 & 1.234761   & 0.92133(54) & 0.9496(17) & 1.0307(19)\tabularnewline
2.3900(31) & 16 & 5.4       & 0.13357851 & 1.227585   & 0.93178(66) & 0.9656(34) & 1.0363(37)\tabularnewline
\hline 
2.7363(14) & 8  & 4.7649    & 0.13488555 & 1.259474   & 0.90010(42) & 0.9309(13) & 1.0342(15)\tabularnewline
2.7389(14) & 12 & 4.938726  & 0.13450761 & 1.251937   & 0.91617(69) & 0.9489(19) & 1.0357(23)\tabularnewline
2.7359(36) & 16 & 5.1       & 0.13416889 & 1.243518   & 0.92979(79) & 0.9701(34) & 1.0433(38)\tabularnewline
\hline 
3.2040(18) & 8  & 4.4576    & 0.13560675 & 1.278015   & 0.89580(49) & 0.9309(16) & 1.0391(19)\tabularnewline
3.2051(17) & 12 & 4.634654  & 0.13519986 & 1.270631   & 0.91371(88) & 0.9601(31) & 1.0508(35)\tabularnewline
3.2029(51) & 16 & 4.8       & 0.13482139 & 1.261490   & 0.9293(10) & 0.9799(35) & 1.0545(39)\tabularnewline
\hline 
3.8636(21) & 8  & 4.1519    & 0.13632589 & 1.297229   & 0.89176(59) & 0.9423(21) & 1.0567(24)\tabularnewline
3.8633(21) & 12 & 4.33166   & 0.13592664 & 1.291442   & 0.91488(90) & 0.9771(42) & 1.0680(47)\tabularnewline
3.8643(66) & 16 & 4.5       & 0.13552582 & 1.281991   & 0.9320(11) & 0.9973(35) & 1.0701(40)\tabularnewline
\hline 
4.4848(24) & 8  & 3.9479    & 0.13674684 & 1.307866   & 0.88812(63) & 0.9540(28) & 1.0742(33)\tabularnewline
4.4864(27) & 12 & 4.128217  & 0.13640300 & 1.305825   & 0.9177(14) & 0.9947(55) & 1.0839(62)\tabularnewline
4.4901(75) & 16 & 4.3       & 0.13600821 & 1.296227   & 0.9349(12) & 1.0215(62) & 1.0926(68)\tabularnewline
\hline 
5.3009(32) & 8  & 3.75489   & 0.13701929 & 1.314612   & 0.88951(88) & 0.9772(42) & 1.0986(48)\tabularnewline
5.2970(36) & 12 & 3.936816  & 0.13679805 & 1.317703   & 0.9195(15) & 1.0366(75) & 1.1273(84)\tabularnewline
5.301(14)  & 16  & 4.1      & 0.13647301 & 1.310054   & 0.9431(17) & 1.089(10) & 1.155(11)\tabularnewline
\end{tabular}
\caption{Results for the renormalisation parameters $Z_{\mathrm T}$ and the SSFs $\SigmaT$ in the $\gamma$-scheme, for
all squared couplings $u_{\mathrm{GF}}$, lattice resolutions $L/a$, inverse bare couplings $\beta$, and critical hopping parameters $\hopc$.}
\label{tab:Z-Sigma-gamma}
\end{table*}


\begin{table*}
\begin{tabular}{c|c|c|c|c|c|c|c}
\hline 
$u_{\mathrm{GF}}$ & $L/a$ & $\beta$ & $\hopc$ & $z_{{\rm f}}$ & $Z_{\mathrm T}(g_{0}^{2},L/a)$ & $Z_{\mathrm T}(g_{0}^{2},2L/a)$ & $\Sigma_{\mathrm T}(g_{0}^{2},L/a)$\tabularnewline
\hline 
2.1275(15) & 8  & 5.3715    & 0.13362120 & 1.226543    & 0.93304(28) & 0.96117(97) & 1.0302(11)\tabularnewline
2.1231(12) & 12 & 5.54307   & 0.13331407 & 1.219858    & 0.95047(38) & 0.9821(12) & 1.0332(13)\tabularnewline
2.1257(25) & 16 & 5.7       & 0.13304840 & 1.213896    & 0.96205(39) & 0.9961(39) & 1.0353(41)\tabularnewline
\hline 
2.3898(15) & 8  & 5.071     & 0.13421678 & 1.242454    & 0.92894(32) & 0.9625(10) & 1.0361(12)\tabularnewline
2.3913(10) & 12 & 5.242465  & 0.13387635 & 1.234761    & 0.94741(44) & 0.9831(14) & 1.0376(15)\tabularnewline
2.3900(31) & 16 & 5.4       & 0.13357851 & 1.227585    & 0.96173(54) & 0.9965(32) & 1.0362(33)\tabularnewline
\hline 
2.7363(14) & 8  & 4.7649    & 0.13488555 & 1.259474    & 0.92404(33) & 0.9620(11) & 1.0411(12)\tabularnewline
2.7389(14) & 12 & 4.938726  & 0.13450761 & 1.251937    & 0.94664(54) & 0.9893(15) & 1.0451(17)\tabularnewline
2.7359(36) & 16 & 5.1       & 0.13416889 & 1.243518    & 0.96195(60) & 1.0032(27) & 1.0429(29)\tabularnewline
\hline 
3.2040(18) & 8  & 4.4576    & 0.13560675 & 1.278015    & 0.91818(37) & 0.9663(17) & 1.0524(19)\tabularnewline
3.2051(17) & 12 & 4.634654  & 0.13519986 & 1.270631    & 0.94398(74) & 0.9939(25) & 1.0529(28)\tabularnewline
3.2029(51) & 16 & 4.8       & 0.13482139 & 1.261490    & 0.96276(75) & 1.0140(23) & 1.0532(25)\tabularnewline
\hline 
3.8636(21) & 8  & 4.1519    & 0.13632589 & 1.297229    & 0.91183(43) & 0.9716(15) & 1.0655(17)\tabularnewline
3.8633(21) & 12 & 4.33166   & 0.13592664 & 1.291442    & 0.94292(82) & 1.0096(32) & 1.0708(35)\tabularnewline
3.8643(66) & 16 & 4.5       & 0.13552582 & 1.281991    & 0.96436(79) & 1.0367(30) & 1.0750(32)\tabularnewline
\hline 
4.4848(24) & 8  & 3.9479    & 0.13674684 & 1.307866    & 0.91011(50) & 0.9857(24) & 1.0830(27)\tabularnewline
4.4864(27) & 12 & 4.128217  & 0.13640300 & 1.305825    & 0.9429(11) & 1.0356(39) & 1.0983(43)\tabularnewline
4.4901(75) & 16 & 4.3       & 0.13600821 & 1.296227    & 0.96971(93) & 1.0605(57) & 1.0936(60)\tabularnewline
\hline 
5.3009(32) & 8  & 3.75489   & 0.13701929 & 1.314612    & 0.90681(74) & 1.0098(55) & 1.1136(61)\tabularnewline
5.2970(36) & 12 & 3.936816  & 0.13679805 & 1.317703    & 0.9490(11) & 1.0674(61) & 1.1247(66)\tabularnewline
5.301(14)  & 16  & 4.1      & 0.13647301 & 1.310054    & 0.9757(13) & 1.1054(91) & 1.1329(95)\tabularnewline
\end{tabular}

\caption{Results for the renormalisation parameters $Z_{\mathrm T}$ and the SSFs $\SigmaT$ in the $\delta$-scheme, for
all squared couplings $u_{\mathrm{GF}}$, lattice resolutions $L/a$, inverse bare couplings $\beta$, and critical hopping parameters $\hopc$.}
\label{tab:Z-Sigma-delta}
\end{table*}

\cleardoublepage

\section{Results for the tensor RG evolution in the low-energy regime}
\label{app:Tmu2/Tmuhad-le}
  
\providecommand{\tabularnewline}{\\}

\begin{table*}[h]
\begin{tabular}{c||c|c|c|c|c|c|c|c}
$[n_{{\rm f}},n'_{\sigma},j_{\mathrm{max}},(L/a)_{\mathrm{min}}]$ & {\Large{}$\Big[\frac{[T(\mu_{0}/2)]_{\mathrm R}}{[T(\mu_{\mathrm{had}})]_{\mathrm R}}\Big]^{\alpha}$} & $w_{{\rm AIC}}^{\alpha}$ & {\Large{}$\Big[\frac{[T(\mu_{0}/2)]_{\mathrm R}}{[T(\mu_{\mathrm{had}})]_{\mathrm R}}\Big]^{\beta}$} & $w_{{\rm AIC}}^{\beta}$ & {\Large{}$\Big[\frac{[T(\mu_{0}/2)]_{\mathrm R}}{[T(\mu_{\mathrm{had}})]_{\mathrm R}}\Big]^{\gamma}$} & $w_{{\rm AIC}}^{\gamma}$ & {\Large{}$\Big[\frac{[T(\mu_{0}/2)]_{\mathrm R}}{[T(\mu_{\mathrm{had}})]_{\mathrm R}}\Big]^{\delta}$} & $w_{{\rm AIC}}^{\delta}$\tabularnewline
 &  &  &  &  &  &  &  & \tabularnewline
\hline 
{[}2,2,2,12{]} & 0.647(20) & 0.1975 & 0.750(15) & 0.1757 & 0.761(15) & 0.2125 & 0.782(12) & 0.188\tabularnewline
{[}2,2,2,8{]} & 0.6488(82) & 0.1813 & 0.7528(65) & 0.1842 & 0.7741(61) &  & 0.7791(50) & 0.1623\tabularnewline
{[}2,2,3,8{]} & 0.663(33) & 0.0377 & 0.755(26) & 0.042 & 0.758(24) & 0.0346 & 0.791(21) & 0.0355\tabularnewline
{[}2,2,4,8{]} & 0.658(27) & 0.0376 & 0.753(22) & 0.0422 & 0.760(20) & 0.0351 & 0.788(17) & 0.0351\tabularnewline
{[}2,3,2,12{]} & 0.647(22) & 0.078 & 0.752(16) & 0.078 & 0.753(16) & 0.1369 & 0.787(14) & 0.0885\tabularnewline
{[}2,3,2,8{]} & 0.6459(90) & 0.084 & 0.7535(68) & 0.0823 & 0.7734(66) &  & 0.7773(55) & 0.0713\tabularnewline
{[}2,3,3,8{]} & 0.654(35) & 0.0074 & 0.755(27) & 0.0084 & 0.755(26) & 0.0066 & 0.792(23) & 0.0064\tabularnewline
{[}2,3,4,8{]} & 0.651(30) & 0.0074 & 0.753(23) & 0.0084 & 0.757(22) & 0.0067 & 0.789(19) & 0.0063\tabularnewline
{[}2,4,2,12{]} & 0.643(23) & 0.0328 & 0.751(16) & 0.0324 & 0.754(16) & 0.056 & 0.782(15) & 0.0464\tabularnewline
{[}2,4,2,8{]} & 0.6458(93) & 0.0349 & 0.7539(69) & 0.0368 & 0.7724(67) &  & 0.7766(57) & 0.0297\tabularnewline
{[}2,4,3,8{]} & 0.638(42) & 0.0015 & 0.754(29) & 0.0019 & 0.742(28) & 0.0023 & 0.787(26) & 0.0013\tabularnewline
{[}2,4,4,8{]} & 0.641(35) & 0.0014 & 0.753(24) & 0.0019 & 0.747(23) & 0.0022 & 0.785(21) & 0.0012\tabularnewline
{[}3,2,2,12{]} & 0.647(20) & 0.0781 & 0.750(15) & 0.0747 & 0.754(15) & 0.2454 & 0.785(13) & 0.1181\tabularnewline
{[}3,2,2,8{]} & 0.6475(84) & 0.0815 & 0.7524(67) & 0.0812 & 0.7736(63) &  & 0.7795(51) & 0.0671\tabularnewline
{[}3,2,3,8{]} & 0.657(34) & 0.0173 & 0.753(27) & 0.0192 & 0.756(25) & 0.0153 & 0.792(21) & 0.0141\tabularnewline
{[}3,2,4,8{]} & 0.654(28) & 0.0172 & 0.752(22) & 0.0192 & 0.758(21) & 0.0156 & 0.789(18) & 0.014\tabularnewline
{[}3,3,2,12{]} & 0.646(22) & 0.0312 & 0.752(16) & 0.032 & 0.754(16) & 0.0959 & 0.785(14) & 0.0449\tabularnewline
{[}3,3,2,8{]} & 0.6459(90) & 0.0355 & 0.7533(69) & 0.0378 & 0.7733(66) &  & 0.7772(55) & 0.0326\tabularnewline
{[}3,3,3,8{]} & 0.660(37) & 0.0032 & 0.758(28) & 0.0044 & 0.756(26) & 0.0029 & 0.793(24) & 0.0025\tabularnewline
{[}3,3,4,8{]} & 0.655(31) & 0.0032 & 0.756(23) & 0.0043 & 0.758(22) & 0.0029 & 0.790(20) & 0.0025\tabularnewline
{[}3,4,2,12{]} & 0.639(24) & 0.014 & 0.750(17) & 0.0136 & 0.741(17) & 0.1258 & 0.783(15) & 0.0177\tabularnewline
{[}3,4,2,8{]} & 0.643(10) & 0.0156 & 0.7528(74) & 0.0167 & 0.7658(72) &  & 0.7785(62) & 0.0134\tabularnewline
{[}3,4,3,8{]} & 0.630(38) & 0.0009 & 0.741(28) & 0.0015 & 0.732(27) & 0.0016 & 0.783(25) & 0.0005\tabularnewline
{[}3,4,4,8{]} & 0.630(31) & 0.0009 & 0.742(23) & 0.0014 & 0.737(22) & 0.0017 & 0.782(21) & 0.0005\tabularnewline
\end{tabular}

\caption{Fit results for $[T(\mu_{0}/2)]_{\mathrm R}/[T(\mu_{\mathrm{had}})]_{\mathrm R}$ for the four schemes $\alpha, \cdots, \delta$ in the low-energy (GF) range. First column: fit parameters. When the AIC weights are missing, the corresponding fit is not included in the final average of eq.(\ref{eq:final-le}).}
\label{tab:GF_ris}

\end{table*}

\cleardoublepage

\section{Comparison of our RG-running results and those of ref.~\cite{Chimirri:2023ovl}}
\label{app:comparison}

In this Appendix we compare our results to those of ref.~\cite{Chimirri:2023ovl}, which have been obtained
on practically the same gauge ensembles used by us. The principal difference is that ref.~\cite{Chimirri:2023ovl}
uses f- and k-schemes with SF boundary conditions (see eqs.~(\ref{eq:ZT-SFf}) and (\ref{eq:ZT-SFk})), 
whereas our schemes-$\alpha$ and -$\beta$ involve $\chi$SF boundary conditions (see eq.~(\ref{eq:ZT-SFf}) and (\ref{eq:ZT-SFk})). 
Otherwise, the renormalised continuum ratios of correlation functions 
defining the f- and k-schemes are the same as those defining schemes-$\alpha$ and -$\beta$ respectively.
Another important difference is that the authors of ref.~\cite{Chimirri:2023ovl}, having tried out several fit Ans\"atze,
single out the result of one given Ansatz, while we prefer to quote as best result the average a several fit Ans\"atze.

We discuss results in the high-energy range first. The raw data for $\ZT(g_0^2,L/a), \ZT(g_0^2,2L/a)$, and $\SigmaT(g_0^2,L/a)$
in Tables~8 and 9 of ref.~\cite{Chimirri:2023ovl} are very similar to the ones in Tables~\ref{tab:SF-Z-Sigma-alpha} and~\ref{tab:SF-Z-Sigma-beta} of our Appendix~\ref{app:Z-Sigma-he}; also the errors are comparable. The best result of ref.~\cite{Chimirri:2023ovl} is reportedly obtained from fitting all datapoints (i.e. including resolution $L/a =6$) with the tensor anomalous dimension approximated by
$\gammaT(x) = -x^2 [t_0 + t_1 x^2 + t_2 x^4]$. The cutoff effects are only quadratic in this fit and have the form $[\rho_2 u^2 + \rho_3 u^3] (a/L)^2$. They find (see their eq.(4.15)):
\begin{subequations}
\begin{alignat}{2}
    \frac{T^{\mathrm{RGI}}}{[T (\mu_{0}/2)]_{\mathrm R}}  &=& 1.1213(74) &\qquad {\rm f-scheme} \, ,
   \label{eq:result-f-scheme-he}\\
       \frac{T^{\mathrm{RGI}}}{[T (\mu_{0}/2)]_{\mathrm R}}  &=&1.1586(63) &\qquad {\rm k-scheme} \, .
   \label{eq:result-k-scheme-he}
\end{alignat}
\end{subequations}
This closely corresponds to our result from $[n_\gamma, n_\sigma^\prime, j_{\mathrm{max}}, (L/a)_{\mathrm{min}}] = [2,3,2,6]$ which we may read off from Table~\ref{tab:SF_ris} of Appendix~\ref{app:TRGI/Tmu2-he}:
\begin{subequations}
\begin{alignat}{2}
    \frac{T^{\mathrm{RGI}}}{[T (\mu_{0}/2)]_{\mathrm R}}  &=& 1.1245(50) &\qquad {\rm scheme}-\alpha \, ,
   \label{eq:result-alpha-1fit-he}\\
       \frac{T^{\mathrm{RGI}}}{[T (\mu_{0}/2)]_{\mathrm R}}  &=&1.1621(44) &\qquad {\rm scheme}-\beta \, .
   \label{eq:result-beta-1fit-he}
\end{alignat}
\end{subequations}
Results are compatible pairwise (f-scheme with scheme-$\alpha$; k-scheme with scheme-$\beta$), with those of the present work having somewhat smaller errors. Note that in our case, we have allowed for a contribution to the quadratic discretisation effects which is linear in $u$ (i.e. $[b_{12} u + b_{22} u^2 + b_{32} u^3 ](a/L)^2$), in order to account for the effects discussed right after eq.~(\ref{eq:SigmaT_sub}).

Next we compare results in the low-energy region. The raw data for $\ZT(g_0^2,L/a), \ZT(g_0^2,2L/a)$, and $\SigmaT(g_0^2,L/a)$ are displayed in Tables~10 and 11 of ref.~\cite{Chimirri:2023ovl}. Upon comparing them to the results in our Tables~\ref{tab:Z-Sigma-alpha} and \ref{tab:Z-Sigma-beta} of Appendix~\ref{app:Z-Sigma-le}, we see that the central values are in the same ballpark and the errors of $\ZT(g_0^2,L/a)$ are very similar. 
However, the errors of $\ZT(g_0^2,2L/a)$ (and consequently of 
$\SigmaT(g_0^2,L/a)$) are significantly bigger in our data for the bigger lattice resolutions. Given the fact that the results of both groups were
largely computed on the same configuration ensembles, these differences can only be explained by the fact that our choice of parameter for the estimation of the autocorrelation times in the error analysis lead to more conservative errors.

The authors of ref.~\cite{Chimirri:2023ovl} approximate the anomalous dimension ratio by the polynomial $f(x) \equiv \gammaT(x)/\beta(x) = (1/x)[f_0 + f_1 x^2 + f_2 x^4]$. The cutoff effects are only quadratic in this fit and have the form $[\rho_1 u + \rho_2 u^2] (a/L)^2$. They find (see their eq.(4.24)):
\begin{subequations}
\begin{alignat}{2}
    \frac{[T (\mu_{0}/2)]_{\mathrm R}}{[T(\mu_{\mathrm{had}})]_{\mathrm R}}  &=& 0.6475(59) &\qquad {\rm f-scheme} \, ,
   \label{eq:result-f-scheme-le}\\
     \frac{[T (\mu_{0}/2)]_{\mathrm R}}{[T(\mu_{\mathrm{had}})]_{\mathrm R}}  &=&0.7519(45) &\qquad {\rm k-scheme} \, .
   \label{eq:result-k-scheme-le}
\end{alignat}
\end{subequations}
This corresponds to our result from $[n_\gamma, n_\sigma^\prime, j_{\mathrm{max}}, (L/a)_{\mathrm{min}}] = [2,2,2,8]$ which we may read off from Table~\ref{tab:GF_ris} of Appendix~\ref{app:Tmu2/Tmuhad-le}:
\begin{subequations}
\begin{alignat}{2}
    \frac{[T (\mu_{0}/2)]_{\mathrm R}}{[T(\mu_{\mathrm{had}})]_{\mathrm R}}  &=& 0.6488(82) &\qquad {\rm scheme}-\alpha \, ,
   \label{eq:result-f-scheme-1fit-le}\\
     \frac{[T (\mu_{0}/2)]_{\mathrm R}}{[T(\mu_{\mathrm{had}})]_{\mathrm R}}  &=&0.7528(65) &\qquad {\rm scheme}-\beta \, .
   \label{eq:result-k-scheme-1fit-le}
\end{alignat}
\end{subequations}
Once more the central values agree pairwise (f-scheme with scheme-$\alpha$; k-scheme with scheme-$\beta$), but the errors of the present work are 30\%-50\% larger than those estimated in ref.~\cite{Chimirri:2023ovl}.

Combining their high- and low-energy running factors, the authors of ref.~\cite{Chimirri:2023ovl} quote in their eq.~(4.31):
\begin{subequations}
\begin{alignat}{2}
    \frac{T^{\mathrm{RGI}}}{[T(\mu_{\mathrm{had}})]_{\mathrm R}} &=& 0.7260(81)(14) &\qquad {\rm f-scheme} \, ,
   \label{eq:result-f-scheme-all}\\
     \frac{T^{\mathrm{RGI}}}{[T(\mu_{\mathrm{had}})]_{\mathrm R}} &=&0.8711(70)(11) &\qquad {\rm k-scheme} \, .
   \label{eq:result-k-scheme-all}
\end{alignat}
\end{subequations}
These are to be compared to the scheme-$\alpha$ and -$\beta$ results of eq.~(\ref{eq:final-he-le}). It is hardly surprising that our errors are twice as big, since they incorporate the uncertainty arising from the choice of fitting Ansatz. We consider this a more conservative approach which properly takes into account a significant, if habitually overlooked, source of systematic error.

\cleardoublepage

\section{Details and results of the simulations at the hadronic scale}
\label{app:zhad}
\begin{table}[h]
  \centering
  \small
\begin{tabular}{c |c |c |c |c |c |c |c |c |c |c }
\hline 
     $L/a$   &  $\beta$ & $\hopc$ &  $\zf$  & $u$ & $Lm$ & ${Z}^\alpha_\mathrm{T}(g_0^2,u,Lm)$ & ${Z}^\beta_\mathrm{T}(g_0^2,u,Lm)$  & ${Z}^\gamma_\mathrm{T}(g_0^2,u,Lm)$ & ${Z}^\delta_\mathrm{T}(g_0^2,u,Lm)$ &  $N_\mathrm{ms}$  \\
\hline 
10 & 3.400000 & 0.136804 & 1.360627 &    9.282(39) &  -0.0237(33)  &1.1989(52)  & 1.0208(31)  &  0.9392(25)  & 0.9426(18) &   2489 \\ 
10 & 3.411000 & 0.136765 & 1.312881 &    9.290(32) &  +0.0178(23) & 1.2010(37) &  1.0235(20) &   0.9472(18) & 0.9400(14) &   4624 \\ 
12 & 3.480000 & 0.137039 & 1.356632 &    9.486(42) &  -0.0095(26)  &1.2695(70)  & 1.0645(39)  &  0.9631(33)  & 0.9852(28) &   2400 \\ 
12 & 3.488000 & 0.137021 & 1.334415 &    9.374(41) &  +0.0018(25) & 1.2656(66) & 1.0610(31)  &  0.9736(29) & 0.9766(19) &   3200 \\ 
12 & 3.497000 & 0.137063 & 1.354166 &    9.171(40) &  -0.0106(24)  &1.2507(64)  & 1.0609(34)  &  0.9633(29)  & 0.9740(25) &   2400 \\ 
16 & 3.649000 & 0.137158 & 1.342554 &    9.368(32) &  -0.0041(13)  &1.3505(66)  & 1.1208(36)  &  1.0038(33)  & 1.0325(28) &   4650 \\ 
16 & 3.657600 & 0.137154 & 1.342971 &    9.163(33) &  -0.0042(14)  &1.3365(70)  & 1.1190(38)  &  1.0025(34)  & 1.0261(28) &   3719 \\ 
16 & 3.671000 & 0.137148 & 1.344328 &    8.940(42) &  -0.0078(23)  &1.3174(106) & 1.1175(58) &  0.9966(49)  & 1.0200(43) &   1600 \\ 
20 & 3.790000 & 0.137048 & 1.331205 &    9.256(41) &  +0.0002(11) & 1.3898(83)  & 1.1529(48) &  1.0377(39) & 1.0573(35) &   4305 \\ 
24 & 3.893412 & 0.136894 & 1.325811 &    9.396(49) &  +0.0008(12) & 1.4467(110)& 1.1796(76) &  1.0488(63) & 1.0902(50) &   3008 \\ 
24 & 3.912248 & 0.136862 & 1.325102 &    9.176(37) &  +0.0008(8)   & 1.4257(85) &  1.1747(50) &  1.0564(39)  & 1.0729(33)  &  5086 \\ 
   \hline 
  \end{tabular}
  \caption{Results for ${Z}^\beta_\mathrm{T}(g_0^2,u,Lm)$ in the hadronic matching region for all four renormalisation schemes.}
  \label{tab:ztabcd}
\end{table}

Fit coefficients of eq.~(\ref{eq:zhadfit-b}) and covariance matrices for all four renormalisation schemes.
\begin{align}
  & z^\alpha_0=+1.396300 \,, \quad z^\alpha_1=+0.348406 \,, \quad z^\alpha_2=-0.446360 \qquad\qquad \chi^2/\mathrm{d.o.f.}=0.217\,, 
 \nonumber \\ 
 &{\rm cov}(z^\alpha_i,z^\alpha_j) = \left( \begin{array}{rrr} +1.359453 \times 10^{-5}    & +2.406576 \times 10^{-5}    & -4.387235 \times 10^{-5} \\ +2.406576 \times 10^{-5}    & +9.634172 \times 10^{-4}    & +2.458513 \times 10^{-3} \\ -4.387235 \times 10^{-5}    & +2.458513 \times 10^{-3}    & +7.515493 \times 10^{-3} \\ \end{array}\right)
\end{align}
%
%
\begin{align}
 &z^\beta_0=+1.155089 \,, \quad z^\beta_1=+0.216761 \,, \quad z^\beta_2=-0.342213 \,, \qquad\qquad \chi^2/\mathrm{d.o.f.}=0.122\,,
  \nonumber \\ 
 & {\rm cov}(z^\beta_i,z^\beta_j) = \left( \begin{array}{rrr} +4.677974 \times 10^{-6}    & +1.268042 \times 10^{-5}    & -3.578771 \times 10^{-6} \\ +1.268042 \times 10^{-5}    & +3.389808 \times 10^{-4}    & +8.367065 \times 10^{-4} \\ -3.578771 \times 10^{-6}    & +8.367065 \times 10^{-4}    & +2.472715 \times 10^{-3} \\ \end{array}\right)
\end{align}
\begin{align}
 & z^\gamma_0=+1.033467 \,, \quad z^\gamma_1=+0.198771 \,, \quad z^\gamma_2=-0.095224\,, \qquad\qquad \chi^2/\mathrm{d.o.f.}=1.513\,, 
  \nonumber \\ 
 & {\rm cov}(z^\gamma_i,z^\gamma_j) = \left( \begin{array}{rrr} +3.204288 \times 10^{-6}    & +5.808756 \times 10^{-6}    & -1.047494 \times 10^{-5} \\ +5.808756 \times 10^{-6}    & +2.234198 \times 10^{-4}    & +5.695031 \times 10^{-4} \\ -1.047494 \times 10^{-5}    & +5.695031 \times 10^{-4}    & +1.752155 \times 10^{-3} \\ \end{array}\right)
\end{align}
\begin{align}
 &z^\delta_0=+1.059607 \,, \quad z^\delta_1=+0.194894 \,, \quad z^\delta_2=-0.310301\,, \qquad\qquad \chi^2/\mathrm{d.o.f.}=0.749\,, 
  \nonumber \\ 
 & {\rm cov}(z^\delta_i,z^\delta_j) = \left( \begin{array}{rrr} +2.244394 \times 10^{-6}    & +4.692534 \times 10^{-6}    & -5.179732 \times 10^{-6} \\ +4.692534 \times 10^{-6}    & +1.503983 \times 10^{-4}    & +3.742533 \times 10^{-4} \\ -5.179732 \times 10^{-6}    & +3.742533 \times 10^{-4}    & +1.118873 \times 10^{-3} \\ \end{array}\right)
\end{align}

%
\cleardoublepage

\end{appendix}

\bibliography{paper}
\end{document}